\documentclass{cimento}

\usepackage{graphicx}  % got figures? uncomment this
\usepackage{epigraph}
\usepackage[cbgreek]{textgreek}
\usepackage[position=top,labelfont=normal]{subcaption}
\usepackage{amsmath}
\usepackage{amssymb}
\usepackage{multirow}
\usepackage{nicefrac}

\newcommand{\dmskedit}[2]{{\color{blue}#2}}

\newcommand{\psip}{\hat{\psi}_{1}}
\newcommand{\psipd}{\hat{\psi}^\dagger_{1}}
\newcommand{\psiz}{\hat{\psi}_{0}}
\newcommand{\psizd}{\hat{\psi}^\dagger_{0}}
\newcommand{\psim}{\hat{\psi}_{-1}}
\newcommand{\psimd}{\hat{\psi}^\dagger_{-1}}

\title{Spinor Bose-Einstein gases}
\author{G.\ Edward Marti\from{ins:x}\from{ins:x2} \atque
Dan M.\ Stamper-Kurn\from{ins:y}\from{ins:z}
}
\instlist{\inst{ins:x} JILA, National Institute of Standards and Technology and University of Colorado, Boulder, Colorado 80309-0440, USA
\inst{ins:x2} Department of Physics, University of Colorado, Boulder, Colorado 80309-0390, USA
  \inst{ins:y}Department of Physics, University of California, Berkeley, California 94720, USA
  \inst{ins:z} Materials Sciences Division, Lawrence Berkeley National Laboratory, Berkeley, California 94720, USA }
\begin{document}

\maketitle

\begin{abstract}
In a spinor Bose-Einstein gas, the non-zero hyperfine spin of the gas becomes an accessible degree of freedom.  At low temperature, such a gas shows both magnetic and superfluid order, and undergoes both density and spin dynamics.  These lecture notes present a general overview of the properties of spinor Bose-Einstein gases.  The notes are divided in five sections.  In the first, we summarize basic properties of multi-component quantum fluids, focusing on the specific case of spinor Bose-Einstein gases and the role of rotational symmetry in defining their properties.  Second, we consider the magnetic state of a spinor Bose-Einstein gas, highlighting effects of thermodynamics and Bose-Einstein statistics and also of spin-dependent interactions between atoms.  In the third section, we discuss methods for measuring the properties of magnetically ordered quantum gases and present newly developed schemes for spin-dependent imaging.  We then discuss the dynamics of spin mixing in which the spin composition of the gas evolves through the spin-dependent interactions within the gas.  We discuss spin mixing first from a microscopic perspective, and then advance to discussing collective and beyond-mean-field dynamics.  The fifth section reviews recent studies of the magnetic excitations of quantum-degenerate spinor Bose gases.  We conclude with some perspectives on future directions for research.
\end{abstract}

Back in 1998, an ``Enrico Fermi'' summer school on \emph{Bose-Einstein Condensation in Atomic Gases} was organized and held at Varenna.  One of the topics touched lightly upon was the novelty of optically trapped Bose-Einstein condensates, which allowed one to study ultracold atoms occupying several spin states, including those that were not amenable to magnetic trapping.  And one of the applications of that new capability was the creation of spinor Bose-Einstein condensates.  Results from the first experiment on spinor Bose-Einstein gases \cite{sten98spin} were presented by Wolfgang Ketterle in one short segment of a several lecture series \cite{kett99var}.  In the intervening 16 years, the study of spinor Bose-Einstein gases has broadened to the point that the topic of just a few overhead transparencies grew into the focus of a three-lecture series given at the 2014 ``Enrico Fermi'' summer school on \emph{Quantum Matter at Ultralow Temperatures} by one of the authors of the present manuscript.

The present document and the lectures they summarize are meant to share with newcomers to the cold-atom community some of the central concepts that spinor Bose-Einstein gases bring to the fore, and some of the experimental techniques by which these concepts have been, and continue to be, studied.  A lack of depth in parts of our presentation does not reflect a shortage of interesting ideas and work by the community exploring these topics, but rather reflects our attempt to present material in an accessible, pedagogical style.  We hope those interested will turn their attention to several review articles \cite{kawa12review,ueda12annrev,stam13rmp} and book chapters \cite{peth08book,ueda10book,stam15seeing} that have been written about spinor Bose-Einstein gases.  We have chosen to make two of the sections in these notes --  Sec.\ \ref{sec:imaging} on the topic of imaging spinor gases, and Sec.\ \ref{sec:excitations} magnetic excitations -- more technical and detailed than the others, for the reason that the material in those sections has not been covered deeply in previous reviews. 
\newcommand{\calFpair}{\mathcal{F}_\mathrm{pair}}
\newcommand{\Lpair}{L_\mathrm{pair}}
\newcommand{\bfLpair}{\mathbf{L}_\mathrm{pair}}
\newcommand{\hatbfFpair}{\hat{\mathbf{F}}_\mathrm{pair}}

\section{Basic properties}

\subsection{The quantum fluids landscape}

Before 1995, the list of quantum fluids that were available to study experimentally was short.  Superfluid $^4$He is the oldest item on that list, first created when Kamerlingh Onnes liquified helium in 1908. However, at the time Onnes did not recognize that he had produced a new state of matter. It was not until 1937 that it was  ``discovered'' that, at low temperatures, the $^4$He liquid was indeed a superfluid \cite{kapi38,alle38flow}.   This fluid is an example of a ``scalar superfluid'' because its order parameter is a single complex number: the ground state is described by a single complex number everywhere, while its dynamics are described by a complex number assigned to each point in space (a scalar field).  The simplicity of this order parameter reflects the fact that $^4$He is a spinless atom: its electrons are paired into a spin singlet state, and so are the neutrons and protons within its nucleus.  %Helium atoms can acquire non-zero spin in states where the electrons are promoted to different electronic orbitals and assume a spin-triplet state, as is done in modern-day experiments with trapped gases of metastable helium; however, at equilibrium at low temperatures such excitations are too high in energy to be thermodynamically relevant.

Electronic superconductors are the second oldest on the list.  These \emph{were} discovered by Onnes.  The simplest and best understood superconductors are described by the theory of Bardeen, Cooper and Schrieffer as having an s-wave order parameter.  This order parameter can be understood as describing the center-of-mass wavefunction of pairs of electrons, the Cooper pairs, that have formed a spinless composite boson.  So, again, the quantum fluid is described by a scalar complex field, though one that it is quite different from liquid $^4$He owing to the long-range Coulomb interaction among Cooper pairs.  There are examples of superconductors in which the crystal structure environment of the electrons favors their adopting an electron-pair state of different symmetry.  In p-wave superconductors, Cooper pairs are formed between electrons of like spin \cite{mack03pwave}.  In this case, the electron pair carries both orbital and spin angular momentum.  In d-wave superconductors, the Cooper pairs are spinless but have a non-zero orbital angular momentum.  However, the possible states of both the p-wave and d-wave superconductors are highly constrained by the broken rotational symmetry of the crystal lattice.  Thus very few -- usually just one -- of the conceivable angular momentum states of the Cooper pairs are accessible, and the order parameter is greatly simplified.

The last quantum fluid to be discovered before 1995 is superfluid $^3$He.  The description of this superfluid is far from simple.  Like the electrons in a superconductor, $^3$He atoms are fermions.  At temperatures down to a few mK and pressures below 30 atmospheres, $^3$He forms a Fermi liquid.  Like electrons in a metal, the distribution of helium atoms is characterized by a Fermi surface. Low-energy excitations involve the correlated motion of helium atoms near the Fermi surface and their neighbors.  Below 3 mK, these quasiparticles undergo Cooper pairing.  It is found that the short-range repulsion among helium atoms causes p-wave pairing to be favored, similar to the case of p-wave superconductors.  However, unlike the electronic superconductor, here the helium fluid is free of the anisotropy of an underlying crystal.  The physics that ensues is fascinating  \cite{voll90he3,dobb00,volo03uni}.

%Two different types of superfluid form at low magnetic field.  At lower pressures and temperatures, the superfluid adopts the ``B-phase,'' in which the (triplet-state) spin composition of the Cooper-paired electrons varies with their relative momentum.  The state is ``quasi-isotropic'' in that the superfluid has no overall spin order, when integrated over all relative momenta, and also in that the Cooper pairing gap has constant magnitude across the Fermi surface.  At higher temperatures and pressures, a small sliver appears in the phase diagram where the ``A-phase'' is favored.  In this phase, the Cooper-pair wavefunction is a product of a spin-triplet state and a p-wave orbital state.  The spin-triplet state is the state with zero angular momentum projection along some axis $\mathbf{d}$, while the orbital state is the plus-1 angular momentum projection along an axis $\mathbf{l}$.  The relative orientation of these two vectors $\mathbf{d}$ and $\mathbf{l}$ is sensitive to effects that couple spin and orbital degrees of freedom, most notably the nuclear dipole-dipole interaction.  The orientation of each vector is also affected by conditions such as physical boundaries (affecting the orbital wavefunction) and magnetic fields (favoring $\mathbf{d}$ to lie perpendicular to the field).  Under certain conditions, both phases may exist within the superfluid, abutting one another in a variety of topological structures.  Altogether, the phenomenology of this fluid is extremely rich (codeword for ``complicated'').

The realization of Bose-Einstein condensation in atomic gases has produced a panoply of new quantum fluids.  Quantum fluids have already been formed of more than a dozen atomic species, in one-, two-, or many-electron atoms, some in their electronic ground state and others in metastable states.  Molecular quantum gases have been produced by associating ultracold atoms into high lying molecular states, and efforts are underway to repeat the feat with molecules in their internal ground state as well.  Some of these are Bose fluids while other are Cooper-paired Fermi gases; the distinction between these two fluids is blurred by experiments span the crossover from a Cooper-paired Fermi gas to a Bose fluid of bound molecules.  The strength of interactions among atoms, and thereby of correlations among them, is broadly tunable \cite{inou98,poll09extreme}.  So, rather than studying only specific instances of broad theories of possible quantum fluids, we can now study many of them experimentally.  The quantum fluids landscape is newly beckoning, inviting our exploration.

One of the appealing areas for investigation is the behavior that comes about when several quantum fluids are mixed together.  Such a system will be represented by a multicomponent order parameter, for example, with one complex field for each component of the quantum fluids.  In choosing which mixtures to study, it is clear that some are going to be more interesting than others.  For example, if we immerse a superconductor in a cup of superfluid $^4$He, we have indeed mixed quantum fluids, but we haven't produced something that seems very interesting; there is no motion or mixing between the fluids, and the only exchange is the (boring) thermodynamic exchange of energy. The same perhaps can be said of mixtures of $^3$He, $^4$He, and $^6$He that were considered previously \cite{volo75mixture,cols78}.

To refine our sense of when something interesting will arise, we might look for three things (these are personal preferences, not strict rules):
\begin{enumerate}
\item{} Cases where the components of the quantum fluids can be transformed into one another, where we're not considering apples and oranges (or $^4$He and $^6$He liquids), but rather situations where the experimentalist can drive transitions that convert one component of the quantum fluids into the other.  %The interesting case of an atom/molecule Bose-Einstein condensate exposed to light that drives a photoassociation resonance illustrates this point \cite{wink05dark}.
\item{} Cases where the system is not ``trivially'' ordered, but rather where order is established from a set of degenerate or at least nearly degenerate possibilities.  The degeneracy of the ground state can be brought about either by fine tuning, or else by symmetries, continuous or discrete, possessed by the system Hamiltonian.  For a bulk fluid (outside of a crystal lattice), the symmetries most likely at play are spatial and/or spin-space rotations.
\item{} Cases where the system can undergo dynamics that explore the different choices of order parameter implied above.
\end{enumerate}

Spinor Bose-Einstein gases, which are ones comprised of atoms (and potentially, some day, molecules too) whose internal state is free to lie anywhere within a single manifold of angular momentum states, satisfy all three of these criteria.  (1) Converting one component of the gas into another is accomplished, for example,  by driving magnetic dipole transitions between Zeeman levels.  (2) In the absence of conditions that explicitly break rotational symmetry (such as anisotropic trapping containers or applied fields), we are certain that any many-body state with non-zero angular momentum is perforce degenerate.  A ground state with angular momentum zero would not necessarily be degenerate, but we assert that such a state (discussed in Sec.\ \ref{sec:manybody}) would already be interesting enough to satisfy the ``non-trivial'' requirement.   (3) As we will see, interactions quite generically allow for ``spin-mixing'' dynamics where the composition of the gas among accessible Zeeman states can vary.

We would be remiss if we did not mention another class of quantum fluids that is appearing on the landscape, namely an assortment of Bose-Einstein condensates of quasi-particle excitations in non-equilibrium settings.  This class includes Bose-Einstein condensation of excitons, exciton-polaritons, photons, and magnons (probably others as well).  These quasi-particle condensates occur in systems that are pumped with a population of excitations that is larger than would exist at thermal equilibrium.  Under favorable conditions, this excess number of quasi-particles decays over a time that is sufficiently long that the quasi-particles can undergo enough collisions amongst themselves or with the medium in which they propagate in order to reach a near-equilibrium distribution in position and momentum space.   There are fundamental questions about whether these driven and decaying quasi-particle gases can form quantum fluids that are the same as those achieved in thermal equilibrium systems in which the particle number is truly conserved.  For lack of space, we do not elaborate on the properties of such quasi-particle quantum gases here and simply refer the interested reader to the literature for further reading \cite{bene13book}.

\subsection{Atomic species}

Experiments on spinor Bose gases have taken place using either alkali atoms (notably sodium and rubidium) or high-spin atoms (notably the transition metal chromium).  The selections are summarized in Table \ref{tab:spinorlist}.  In all cases, the spin manifold that is occupied by the atoms of the spinor gas is one of the hyperfine spin manifolds of the electronic ground state.

\renewcommand{\multirowsetup}{\centering}
\begin{table}[tb]
\begin{center}
\begin{tabular}{ccc}
\hline
\multicolumn{2}{c}{Stable} & \multirow{2}{2cm}{Unstable} \\
\cline{1-2} $\langle F_z \rangle$ conserved & $\langle F_z \rangle$  not conserved &  \\
\hline
$^7$Li, $F=1$ (f) & $^{52}$Cr, $F = 3$ (not f) & $^7$Li, $F=2$ \\
$^{23}$Na, $F=1$ (af) & Dy, $F=8$ (unknown) & $^{23}$Na, $F=2$ \\
$^{41}$K, $F=1$ (f) & Er, $F=6$ (unknown) & $^{39}$K \\
$^{87}$Rb, $F=1$ (f)&  &  $^{85}$Rb \\
$^{87}$Rb, $F=2$ (af or cyc) & & $^{133}$Cs \\
\hline
$^{87}$Rb pseudo-spin: & \multicolumn{2}{c}{Tm, $F=4$ (unknown) } \\
%$|F=1, m_F=0\rangle$, $|F=2, m_F = 0\rangle$ & \multicolumn{2}{c}{Ho, $F=4$ (unknown) } \\
%$|F=1, m_F = \pm 1 \rangle$, $|F=2, m_F = \mp 1\rangle $ &  & \\
$|1,0\rangle$, $|2,0\rangle$ & \multicolumn{2}{c}{Ho, $F=4$ (unknown) }  \\
$|1, \pm 1\rangle$, $|2, \mp 1\rangle$ & \multicolumn{2}{c}{} \\
\hline
\end{tabular}
\caption{Experimental candidates for the study of ultracold spinor Bose gases.  Species are divided according to whether they are stable at zero magnetic field (information on thulium and holmium is lacking), and whether the dipolar relaxation rate is small enough to allow the longitudinal magnetization ($\langle F_z\rangle$) to be conserved in an experiment.  The nature of the spin-dependent contact interactions is indicated in parentheses (f: ferromagnetic, af: antiferromagnetic, cyc:cyclic or tetrahedral).  Stable pseudo-spin-1/2 gases of $^{87}$Rb are indicated, with states labeled with quantum numbers $|F, m_F\rangle$ having the same low-field magnetic moment.  Table taken from Ref.\ \cite{stam13rmp}}
\label{tab:spinorlist}
\end{center}
\end{table}

\subsubsection{Alkali atoms}

To identify atomic properties that come up later in this document, let us remind the reader briefly of the structure of an alkali atom.  The alkali atom is a one-electron atom in the sense that, in the electronic ground state and accessible excited states, all but one of the electrons in the atom remain in closed atomic shells while only the state of the last electron varies.  Focusing on the electronic ground state, that one electron is unpaired in an s-orbital, so that the atom has no internal orbital angular momentum.  The atom does have two sources of spin angular momentum in this electronic ground state: the 1/2 spin of the unpaired electron, and the spin of the nucleus denoted by the angular momentum quantum number $I$.  These two spins are energetically coupled by the hyperfine interaction.

At zero magnetic field, the atomic energy eigenstates are also eigenstates of the total angular momentum of the atom, denoted by the vector operator $\hat{\mathbf{F}}$ and known as the hyperfine spin\footnote{for notational convenience, we will make the spin vector and spin quadrupole operators dimensionless, and include factors of $\hbar$ explicitly where needed}.  The angular momentum quantum number for the hyperfine spin, $F$, takes one of two values: $I \pm 1/2$.  We observe that for bosonic alkali atoms $I$ is a half-integer number\footnote{We note that the quantum statistics of a neutral atom, i.e.\ whether it is a boson or fermion, depend on whether the number of neutrons in the nucleus is even (boson) or odd (fermion), because the sum of the number of electrons and protons is always even.  The nucleus of a bosonic alkali atom will contain an even number of neutrons and an odd number of protons; hence the nuclear spin will be non-zero and half-integer.  As a counter-example, for alkali-earth atoms, the bosonic isotopes have an even number of nucleons.  Nuclear stability most often then implies that the nuclear spin is $I = 0$, so that the bosonic atom will have zero net spin in its ground state and will not be a candidate for spinor physics.}. These two hyperfine spin states are separated by an energy on the order of $h \times 1$ GHz.  Each of these energy states is $(2F + 1)$-fold degenerate.

Thus, an alkali atom supports, in principle, two different species of spinor Bose-Einstein gases: an $F = I+1/2$ spinor gas, and an $F = I - 1/2$ spinor gas.  In reality, one finds often that the gas of atoms in the higher-energy hyperfine spin manifold is unstable against collisions that relax the hyperfine energy, limiting the amount of time one can allow such a gas to evolve; $^{87}$Rb is an exception in that both its hyperfine manifolds are long lived.  The remaining choices are listed in Table \ref{tab:spinorlist}.

The identification of an atomic state with a hyperfine spin quantum number $F$ is strictly valid only at zero magnetic field.  A magnetic field breaks rotation symmetry, and introduces coupling and energy repulsion between the two states of the same magnetic quantum number $m_F$ that characterizes the spin projection along the field axis.  Strong mixing between states is obtained when the magnetic field is strong enough so that the magnetic energy, roughly $\mu_B B$, is greater than the zero-field hyperfine energy splitting.  With $\mu_B = e \hbar /2 m_e h \times 1.4$ MHz/G being the Bohr magneton ($e$ is the electron charge, $m_e$ is the electron mass) and using a typical $h \times \mbox{1 GHz}$ hyperfine energy, this strong-field regime applies for $B \gtrsim 700$ G.  So, if we restrict the magnetic field to be less than 100 G or so in strength, we can still take the atomic states as comprising spin manifolds of well-defined spin.

\subsubsection{High-spin atoms}

More recent additions to the family of atoms that can be laser-cooled and brought to quantum degeneracy are elements in the transition metal and lanthanide series.  These are many-electron atoms, with several electrons lying in the valence shell, and for which the optical spectrum is rich with different series of states corresponding to different arrangements of these many electrons.  It came as a surprise that laser cooling and trapping techniques work so well for these complex atoms and their complex spectra \cite{mccl06erbium,berg07erbium}.

The valence shell for these atoms has a high angular momentum.  When these shells are only partly filled, the electrons arrange themselves to maximize the total electron spin $S$ that can thus become large.  The partly filled shells can also have a non-zero orbital angular momentum $J$.  Both $S$ and $J$ contribute substantially to the magnetic moment of the atom.   Add in the possible contribution of nuclear spin, and one has the potential for creating spinor Bose-Einstein gases of very high spin $F$.

The ground state angular momentum and the magnetic moment of these atoms can be determined by applying Hund's rules for atomic structure, an exercise that is described in standard atomic and condensed-matter physics textbooks.  For bosonic chromium \cite{grie05crbec}, with the configuration $[\mbox{Ar}] 4s^1 3d^5$, one finds the quantum numbers $S=3, L=0; J=3, I=0; F = 3$ and magnetic moment $\mu = 6 \, \mu_B$.  For bosonic dysprosium \cite{lu11dybec}, with the configuration $[\mbox{Xe}] 6s^2 4f^{12}$, one finds the quantum numbers $S=2, L=6; J=8, I=0; F=8$ and magnetic moment $\mu = 10 \, \mu_B$.  For bosonic erbium \cite{aika12erbium}, with two extra electrons in the $4f$ orbital, we find $S=1, L=5; J=6, I=0; F=6$ and $\mu = 7 \, \mu_B$.  The point here is not only that the spin quantum number of gases formed of these atoms is large, which, for example, leads to an interesting variety of magnetically ordered states that might be realized in such gases \cite[for example]{sant06spin3,dien06crspinor,make07cr,yip07inert,lian12}, but also that the effects of dipolar interactions, scaling in strength as $\mu^2$, become increasingly dominant.

\subsubsection{Stability against dipolar relaxation}

One consequence of the large magnetic moment is that the high-spin gases undergo rapid dipolar relaxation whereas the alkali gases do not.  Let us distinguish here between two types of collisions: s-wave contact collisions and dipolar relaxation collisions (see Ref.\ \cite{bloc08rmp} for a nice review of quantum scattering in the context of ultracold atomic gases).  The s-wave contact interaction describes low-energy collisions where atoms interact with one another via a short-range molecular potential, where ``short-range'' means that the range of the potential is small compared with the deBroglie wavelength defined by the small relative velocity of the incident colliding atoms.  In an s-wave contact collision, two atoms enter into and emerge from the collision with no relative, center-of-mass orbital angular momentum.  As we discuss further below, this situation requires that the total spin angular momentum of the two colliding bodies be conserved in the collision.  In contrast, in a dipolar relaxation collision, the colliding atoms emerge with a non-zero value of the center-of-mass angular momentum.  This change in orbital angular momentum implies a change also in the total hyperfine spin of the colliding pair.

In alkali spinor gases, dipolar relaxation collisions occur (typically) much less frequently than s-wave contact interactions.  Consider the evolution of such a gas in the presence of a small, uniform applied magnetic field.  Single-particle dynamics in the presence of this field conserve the populations in each of the Zeeman sublevels of the spinor gas.  The s-wave contact interaction between two atoms conserves their total hyperfine spin, but not necessarily the spin of each atom.  Generically, then, one finds that such collisions redistribute atoms among the Zeeman sublevels of the gas.  However, the conservation of the total hyperfine spin implies that the net hyperfine spin along the field axis remains constant.

Remarkably, the conservation of the longitudinal (along the field axis) total hyperfine spin implies that, to first order, the Zeeman energy shifts imposed by applied magnetic fields are irrelevant to the evolution of the spinor gas.  If the gas were completely free to change its spin composition, then the linear Zeeman energy from magnetic fields of typical strength -- say 10's of mG -- would completely overwhelm all other energy scales of the gas (the temperature, the chemical potential, the spin-dependent s-wave interaction strength, the magnetic dipole-dipole interaction, etc.).  The ground state of such a spinor gas would trivially be one in which the gas is fully magnetized along the magnetic field direction.  However, absent dipolar relaxation collisions, this Zeeman-energy equilibration does not occur.  Rather, the gas evolves as if these Zeeman energy shifts did not exist at all, i.e.\ as if it were in a zero-field environment (except for residual effects of the magnetic dipole interaction).  The s-wave contact interactions then provide a means for the gas to evolve to an equilibrium constrained to a constant longitudinal magnetization, the value of which is established by the experimentalist by the application of radio or microwave frequency fields that can drive atoms between Zeeman states.

In contrast, in the high-spin spinor gases, dipolar relaxation collisions occur more rapidly, and the constraint of constant longitudinal magnetization is lifted.  In an applied magnetic field of moderate strength, the high-spin spinor gas will indeed evolve toward the trivial magnetically ordered state where all atomic spins are aligned with the field.  The resulting quantum fluid only occupies a single magnetic  state ($m_F = \pm F$), and so the physics is described by only a single complex number, just like a scalar fluid.  If one wants to access non-trivial magnetic order in such gases, one must reduce the applied magnetic field to very small values.  For example, one may want to make the applied field smaller than the field generated by the magnetization of the gas in order to realize the chiral spin textures proposed in Refs.\ \cite{sait06chiral,yi06texture}.  We can estimate this field as $B = \mu_0 \mu n$ where $n$ is the atomic density (say $10^{13} \, \mbox{cm}^{-3}$) and $\mu$ is the atomic magnetic moment (say dysprosium's $10 \, \mu_B$).  We arrive at a field of $B \sim 10 \, \mu\mbox{G}$.  So experiments on these high-spin spinor gases require that the magnetic field be controlled with great care.

This distinction between ``small'' and ``large'' magnetic moments can be made quantitative by comparing two energies: the van der Waals energy, and the magnetic dipole energy.  The van der Waals interaction leads to a potential of the form $-C_6/ r^6$ where $r$ is the distance between two atoms and $C_6$ quantifies the dc electric polarizability of the atoms.  We can turn this potential into an energy scale $E_a$ by calculating the distance $l$, known as the van der Waals length, where the confinement energy $E_a = \hbar^2/ 2 m l^2$ equals the magnitude of the van der Waals potential.  We obtain $l = (2 m C_6 / \hbar^2)^{1/4}$, where $m$ is the atomic mass.  We can now assess the magnetic dipole energy as the energy of two atomic dipoles separated by a distance $l$; this energy depends on the orientation of the two dipoles relative to the separation between them, but let's just define the energy as being $E_d = \mu_0 \mu^2/ (4 \pi l^3)$ where $\mu_0$ is the vacuum permeability and $\mu$ is the atomic magnetic moment.  So now the importance of the dipolar interactions in the collisions between atoms is quantified by the ratio
\begin{equation}
\frac{E_d}{E_a} \sim \frac{\mu_0 \mu^2 m}{2 \pi l \hbar^2} = \frac{\mu_0 \mu^2 m^{3/4}}{2^{5/4} \pi C_6^{1/4} \hbar^{3/2}}
\end{equation}

A similar ratio is obtained by taking the ratio of the mean-field contact interaction energy and the mean-field magnetic dipole energy of a spherical ball of atoms with uniform density $n$.  The mean-field contact interaction energy is given as $E_a = (4 \pi \hbar^2 a / m) n$, where $a$ is the s-wave scattering length, while the magnetic dipole energy is $E_d = (1/3) \mu_0 \mu^2 n$.  So we obtain the ratio
\begin{equation}
\frac{E_d}{E_a} = \frac{\mu_0 \mu^2 m}{12 \pi a \hbar^2} = 8 \times 10^{-3} \times \left( \frac{\mu}{\mu_B}\right)^2 \times \left( \frac{m}{\mbox{amu}}\right) \times \left( \frac{a_0}{a}\right)
\end{equation}
Recall that the van der Waals potential plays a dominant role in low-energy atomic collisions, so that $a$ is indeed on the order of $l$ except near a collisional resonance \cite{hein99var}.  Thus, our two expressions for $E_d/E_a$ are the same apart for a numerical factor of order unity.

On the right hand side of the expression above, we quantify the ratio $E_d/E_a$ with the magnetic moment being measured in units of the Bohr magneton, $m$ being measured in amu, and the $a$ being measured in units of the Bohr radius $a_0$.  With this expression we can compare the relevance of dipolar interactions in the collisions of various gases.  For example, for $^{87}$Rb, with $\mu = 1/2 \, \mu_B$ and $a \sim 100 \, a_0$, we have $E_d/E_a = 2 \times 10^{-3}$, confirming the fact that s-wave interactions dominate the collisional physics of alkali gases.  In contrast, for $^{192}$Dy, with $\mu = 10 \, \mu_B$ and $a = 92 \, a_0$ \cite{tang15dys}, we have $E_d/E_a = 1.4$ so that dipolar interactions are highly significant.

\subsection{Rotationally symmetric interactions}

%\epigraph{Break a vase, and the love that reassembles the fragments is stronger than that love which took its symmetry for granted when it was whole.}{--- \textup{Derek Walcott, in his lecture receiving the 1992 Nobel Prize for Literature}}

Spinor Bose-Einstein gases are not just multicomponent quantum fluids, but specifically ones in which the components of the fluid are simply different orientations of the same atomic spin.  The spin operator is a vector operator, one that transforms as a vector under physical rotations. The implications of this statement for the properties of spinor gases are profound.  Just as we find the notion of continuous symmetry groups powerful in explaining the nature of quantum field theories or condensed-matter systems, so too do we find the notion of rotational symmetry a powerful one in simplifying and constraining the physics of spinor gases.

Consider the constraints that rotational symmetry places on the possible interactions among atoms in the spinor gas.  In the absence of applied fields, and in a spherically symmetric trapping volume, the interaction between two atoms must conserve their total angular momentum, the quantum number for which we will denote by the symbol $\calFpair$.  This total angular momentum includes both the internal angular momenta of the two atoms, and also the orbital angular momentum of their relative motion.

Let us assume that the collision between two particles occurs only in the s-wave channel, where the relative center-of-mass orbital angular momentum of the two atoms, with quantum number $\Lpair$, is zero.  This assumption is justified if we consider low-energy collisions between two particles with a short-range two-body potential.  As discussed below, this assumption is questionable in the case of dipolar interactions.

Under this assumption, the total angular momentum of the incoming state is just the sum of the two atomic hyperfine spins, $\hatbfFpair  = \hat{\mathbf{F}}_1 + \hat{\mathbf{F}}_2$ with total spin quantum number $F_\mathrm{pair}$.  As we are considering the collision between two identical bosons, exchange symmetry implies that the total angular momentum quantum number of the colliding pair must be an even integer; spin combinations with an odd-integer value of the total pair angular momentum simply do not collide in the s-wave regime.  So the total atomic spin $F_\mathrm{pair}$ (in this case also the total atomic angular momentum $\calFpair$) can take $F + 1$ different values: $\{0, 2,\ldots 2F\}$.  The states of the colliding incident particles can thus be written in the following basis:
\begin{equation}
\{ |F_\mathrm{pair}, \Lpair=0; \calFpair = F_\mathrm{pair}, m_{F_\mathrm{pair}}\rangle \}
\end{equation}
where $m_{F_\mathrm{pair}}$ is the projection of the total (dimensionless) angular momentum onto a selected axis.

The conservation of angular momentum under collisions now implies that the collisional interaction between two particles can be written in the form
\begin{equation}
\hat{V}_\mathrm{int} = \sum_{\calFpair \in \{0, 2, \ldots 2F\}} \hat{V}_{\calFpair} \hat{P}_{\calFpair}
\end{equation}
where $\hat{P}_{\calFpair}$ projects the two-atom state onto the subspace with total angular momentum $\calFpair$, and $\hat{V}_{\calFpair}$ is a scalar two-body operator.

The effects of interactions are now fairly well characterized.  Consider collisions where the outgoing atom pair also has zero center-of-mass orbital angular momentum, and hence can be written in the same basis as above (denoted by primed symbols).  Since $\hat{V}_\mathrm{int}$ is a scalar operator, and recalling the Wigner-Eckart theorem, non-zero matrix elements for the two-body interaction are only obtained when the outgoing total spin angular momentum is the same as the incident total spin angular momentum:
\begin{multline}
\langle F_\mathrm{pair}^\prime, \Lpair =0; \calFpair^\prime = F_\mathrm{pair}^\prime, m_{F_\mathrm{pair}}^\prime | \hat{V}_\mathrm{int} | F_\mathrm{pair}, \Lpair=0; \calFpair  = F_\mathrm{pair}, m_{F_\mathrm{pair}} \rangle \\ = g_{F_\mathrm{pair}} \delta_{\calFpair^\prime, \calFpair} \delta_{m_{F_\mathrm{pair}}^\prime, m_{F_\mathrm{pair}}}
\end{multline}
The simple dependence of the matrix element on $m_{F_\mathrm{pair}}$ and $m_{F_\mathrm{pair}}^\prime$ (diagonal, with only one overall constant $g_{F_\mathrm{pair}}$) again comes from the fact that $\hat{V}_\mathrm{int}$ is a scalar operator.  The constants $g_{F_\mathrm{pair}}$ are determined from $F+1$ different scattering lengths, $a_{F_\mathrm{pair}}$.  To capture this part of the interaction, we can adopt the pseudo-potential approach and replace the collisional interaction with the operator
\begin{equation}
\hat{{V}}_\mathrm{int} =  \frac{1}{2} \sum_{F_\mathrm{pair} \in \{0, 2, \ldots 2F\}} \frac{4 \pi \hbar^2 a_{F_\mathrm{pair}}}{m} \, \delta^3(\mathbf{r}) \hat{P}_{F_\mathrm{pair}} \label{eq:swaveprojection}
\end{equation}
where $\mathbf{r}$ is the interatomic separation.  The factor of $1/2$ is inserted to account for double-counting in a summation over atom pairs.

%\subsubsection{Rotational symmetry and dipolar interactions}

It remains to consider the effects of dipolar interactions that can couple the incident s-wave collision channel to an outgoing channel with non-zero center-of-mass orbital angular momentum with total angular momentum quantum number $\Lpair = 2$ (coupling to $\Lpair=1$ is forbidden by parity).  In addition to the treatment above, we now have three more outgoing basis sets to consider:
\begin{equation}
\begin{array}{l}
 \{ | F_\mathrm{pair} = \calFpair-2, \Lpair=2; \calFpair, m_{F_\mathrm{pair}}\rangle \}  \\
 \{ | F_\mathrm{pair} = \calFpair, \Lpair=2; \calFpair, m_{F_\mathrm{pair}}\rangle \}  \\
 \{ | F_\mathrm{pair} = \calFpair+2, \Lpair=2; \calFpair, m_{F_\mathrm{pair}}\rangle \}\end{array}
\end{equation}
Matrix elements to each of these final basis states are again independent of $m_{F_\mathrm{pair}}$.  The dipolar interaction does not conserve the hyperfine spin; it leads to a ``spin-orbit'' interaction through which spin angular momentum can be converted into orbital angular momentum, a phenomenon known as the Einstein-de Haas effect \cite{eins15dehaas,kawa06dehaas,sant06spin3,garw07res}.  This coupling also leads to dipolar relaxation among Zeeman levels, allowing the Zeeman energy of a spinor gas be exchanged with kinetic energy.

It turns out to be advisable to treat the magnetic dipolar interactions as an explicit long-range interaction, rather than adopting a pseudo-potential approach and replacing the dipolar interaction with a zero-range contact interaction \cite{laha09review}.  The pseudo-potential method is appropriate for describing the non-dipolar molecular potential between two atoms, which falls off as $1/r^6$ at long distance.  This rapid fall-off ensures that only s-wave interactions will persist at low energies of the incident particles.  In contrast, the slower $1/r^3$ fall-off of the dipole-dipole interaction allows many partial waves to remain relevant to low-energy collisions. 
\section{Magnetic order of spinor Bose-Einstein condensates}

Having established some of the basic properties of spinor Bose-Einstein gases, we ask how they behave at low temperature.  Our expectations are set by the example of scalar Bose-Einstein gases: At low temperature, these gases undergo Bose-Einstein condensation, a phase transition driven by Bose-Einstein statistics.  This phase transition is observable by the growth to a non-zero value of a local order parameter, the complex field $\Psi(\mathbf{r})$ which is the superfluid order parameter.  This non-zero order parameter represents a spontaneous breaking of a symmetry obeyed by the system's Hamiltonian.  In this particular case, the broken symmetry is the gauge symmetry for which the conserved quantity is the particle number.

Following this example, we expect that order will be established in low-temperature spinor Bose-Einstein gases.  We expect that this order may come about through the effects of Bose-Einstein statistics, and that it may represent a spontaneous breaking of symmetries of the many-body Hamiltonian, in particular the symmetry under rotation discussed in the previous sections.

We analyze the subject of magnetic ordering in spinor gases in three stages.  For simplicity, each will focus on the spin-1 spinor Bose-Einstein gas.  First, we present the simple case of a non-interacting spinor gas that equilibrates in the presence of an applied magnetic field.  This case illustrates the statistical origin of magnetic ordering in a degenerate spinor Bose gas, and also exemplifies some basic features we should expect of the thermodynamic properties of a spinor gas.  However, we will show that this non-interacting model provides little guidance as to the exact nature of magnetic ordering, particularly for highly degenerate gases at very low temperature.  So, second, we discuss the role of those rotationally symmetric interactions outlined in the previous section.  We analyze what type of magnetic order is favored under different conditions, often through the spontaneous breaking of rotational symmetry, and summarize experiments that have revealed such order and such symmetry breaking.  Finally, we find an exact solution to the spin-dependent many-body Hamiltonian -- exact at least under the assumption of there being no spatial dynamics and no dipolar interactions.  We find that the ground state of one type of spin-1 spinor Bose-Einstein gas may be a highly correlated, non-condensed, rotationally symmetric spin state, providing a theoretical counter-example to our prediction of local mean-field ordering and spontaneous rotational symmetry breaking.

\subsection{Bose-Einstein magnetism in a non-interacting spinor gas}
\label{sec:nonint}

We consider a simple model of a non-interacting spinor Bose-Einstein gas in a uniform box of volume $V$, and exposed to a uniform magnetic field $\mathbf{B} = B \mathbf{z}$ \cite{yama82}.  We assume the gas comes to thermal equilibrium, constrained only by the constant particle number $N$.  The magnetic field shifts the single-atom energy levels according to the Hamiltonian
\begin{equation}
H = - \boldsymbol{\mu} \cdot \mathbf{B} \equiv - \frac{\mu}{F} B \hat{F}_z
\end{equation}
where $\mu$ is the atomic magnetic moment (assumed positive), and where we include only the linear Zeeman energy.  This Zeeman energy causes the chemical potentials \textmu$_{m_F}$ for atoms in the different magnetic sublevels to differ, i.e.\
\begin{equation}
\mbox{\textmu}_{m_F} = \mbox{\textmu}_0 + \frac{\mu}{F} B m_F.
\end{equation}

Let us pause and ask whether the assumptions of this model are appropriate.  For high-spin atoms, rapid dipolar relaxation collisions allow the gas atoms to redistribute themselves among the Zeeman sublevels, and thus the model should apply directly.  Experiments on the thermodynamics of condensed chromium atoms in moderate-strength magnetic fields are described well by our model (though with higher $F=3$ spin) \cite{pasq12thermo}.

For the low-spin, alkali metal spinor gases, the model we have presented is still valid so long as we regard the applied magnetic field as being an effective, fictitious magnetic field, the value of which is determined by the magnetization of the gas.  In these gases, collisions among atoms will redistribute the kinetic (and trap potential) energy among them, but will not change the magnetization.  The linear Zeeman energy determined by the applied magnetic field is inaccessible to the gas, and is thus irrelevant to its dynamics.  That is, without dipolar interactions, the gas is subject to collisions of the form
\begin{equation}
|m_F = m_1\rangle + |m_F = m_2\rangle \leftrightarrow |m_F = m_3\rangle + |m_F = m_4\rangle
\end{equation}
only so long as $m_1 + m_2 = m_3 + m_4$.  These processes evolve independent of the linear Zeeman shift, and thus such shifts cannot enter into the dynamics of a spinor gas.  However, spin-mixing collisions do generally occur, allowing the Zeeman-state populations to redistribute and come to chemical equilibrium.  For example, a spin mixing collision within a spin-1 gas can be viewed as the following reaction:
\begin{equation}
|m_F = 0\rangle + |m_F = 0\rangle \leftrightarrow |m_F = -1\rangle + |m_F = +1 \rangle.
\label{eq:spinmixingreaction}
\end{equation}
Chemical equilibrium for this reaction (i.e.\ where the forward and backward reaction rates are equal) implies the chemical potentials of the Zeeman sublevels are related as follows:
\begin{equation}
\mbox{\textmu}_0 + \mbox{\textmu}_0 = \mbox{\textmu}_{-1} + \mbox{\textmu}_{+1}.
\end{equation}
Rearranging this equation, and considering all other possible spin-mixing collisions (in case of a higher spin), we see that the chemical-potential difference $\mbox{\textmu}_i - \mbox{\textmu}_{i-1} \equiv \mu B_\mathrm{eff}/F$ is independent of $i$.  In other words, spin-mixing collisions in the absence of dipolar relaxation allow an equilibration at constant magnetization that is equivalent to the equilibration in an applied magnetic field, where now this effective magnetic field $B_\mathrm{eff}$ is not the actual magnetic field present in the laboratory, but rather a Lagrange multiplier that serves to enforce the constraint of constant magnetization.

We continue to explore our simple model.  The number of non-condensed atoms in the gas in each of the magnetic sublevels is given by the Bose-Einstein distribution,
\begin{equation}
N_{\mathrm{th},m_F} = g_{3/2}(z_{m_F}) \lambda_T^{-3} V
\end{equation}
where $g_{3/2}(z) = \sum_{i=1}^{\infty} z^i / i^{3/2}$, $z_{m_F} = \exp\left(\mbox{\textmu}_{m_F} / k_B T\right)$ are fugacities and $\lambda_T = \sqrt{2 \pi \hbar^2 / m k_B T}$ is the thermal deBroglie wavelength.  The Bose-Einstein distribution is defined for fugacities being in the range $0 \leq z \leq 1$, i.e.\ for chemical potentials $\mbox{\textmu}\leq 0$.

\begin{figure}[t]
\centering
\includegraphics[width=\textwidth]{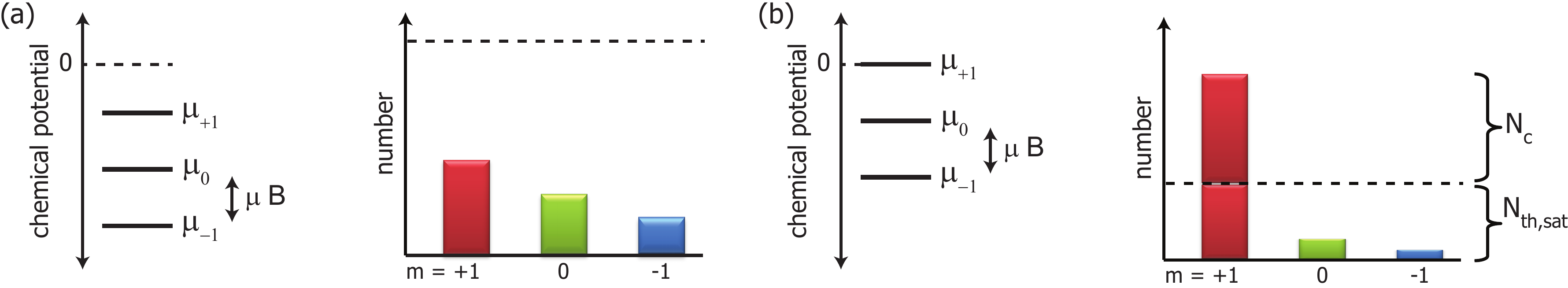}
\caption{Composition of a non-interacting $F=1$ spinor Bose gas at thermal equilibrium.  An applied magnetic field of strength $B$ separates the chemical potentials $\mbox{\textmu}_{m_F}$ for the three Zeeman sublevels, labeled by magnetic quantum number $m_F$, by an amount $\mu B$.  (a) Above the critical temperature for Bose-Einstein condensation, the chemical potentials are all below zero, and the gas is non-degenerate.  The number of atoms in each sublevel, shown in the bar chart, is below the critical number for Bose-Einstein condensation (dashed line).  This gas is paramagnetic, with a finite magnetic susceptibility that lends it a magnetization under an applied magnetic field.  (b) Upon cooling below the critical temperature for Bose-Einstein condensation, the chemical potential of the lowest-energy sublevel, here with $m_F = 1$, reaches zero, and a Bose-Einstein condensate forms in this sublevel.  The chemical potentials for the remaining sublevels are negative, and the populations in these sublevels remain non-degenerate.  Here, the gas is ferromagnetic, acquiring a large magnetization even in an infinitesimal magnetic field, implying an infinite magnetic susceptibility.\label{fig:barchart}}
\end{figure}

We consider three generic scenarios.  First, at high temperature, all $N$ atoms in the gas can be accommodated within the thermal Bose-Einstein distribution of the Zeeman components, i.e.\
\begin{equation}
N_{\mathrm{th}} = \sum_{m_F} N_{\mathrm{th},m_F} = N
\end{equation}
In this non-degenerate gas, the chemical potentials of all Zeeman components are negative.  In an applied magnetic field (assumed positive, without loss of generality), the populations in the three magnetic sublevels of a spin-1 gas will differ, with the largest population being in the $|m_F = +1\rangle$ state (Fig.\ \ref{fig:barchart}a).  The non-degenerate gas thus acquires a non-zero magnetization.

At lower temperature, the gas undergoes Bose-Einstein condensation, marked by the fact that the chemical potential of at least one of the magnetic sublevels goes to zero.  In an applied, positive magnetic field, the linear Zeeman shift implies that $\mbox{\textmu}_{+1} = 0$ because the chemical potential of the non-interacting Bose gas cannot be positive.  Therefore, the remaining chemical potentials, $\mbox{\textmu}_{0}$ and  $\mbox{\textmu}_{-1}$, are both negative.  The number of non-condensed atoms saturates at the value
\begin{equation}
N_{\mathrm{th},\mathrm{sat}} = \lambda_T^{-3} V \left[ g_{3/2}(1) + g_{3/2}\left(e^{-x}\right) + g_{3/2}\left(e^{-2 x}\right) \right]
\end{equation}
with $x = \mu B / k_B T$.  The Bose-Einstein condensation transition occurs when $N_{\mathrm{th},\mathrm{sat}} = N$.  The transition temperature, defined implicitly by the above equation, has the following limiting values
\begin{equation}
T_\mathrm{c}(x) = \left(\frac{n}{\zeta(3/2)} \right)^{2/3} \frac{2 \pi \hbar^2}{m k_B} \left\{\begin{array}{c r} 1 & x \rightarrow \pm \infty \\ (2F + 1)^{-2/3} & x = 0 \end{array} \right.
\end{equation}
with $\zeta(x)$ being the Riemann zeta function.

Bose-Einstein condensation occurs in just one of the magnetic sublevels (Fig.\ \ref{fig:barchart}b).  The Bose-Einstein condensate, containing $N_\mathrm{c} = N - N_{\mathrm{th},\mathrm{sat}}$ atoms, is completely spin polarized along the direction of the applied field, while the non-degenerate fraction of the gas is only partly polarized.

A third scenario to consider is Bose-Einstein condensation at zero magnetic field.  In this case, the chemical potentials of all Zeeman sublevels go to zero simultaneously.  The condensate number is defined as before.  However, the state of the Bose-Einstein condensate is not determined within this model, requiring instead that we begin accounting for the effects of spin-dependent interactions.

\begin{figure}[t]
\centering
\includegraphics[width=\textwidth]{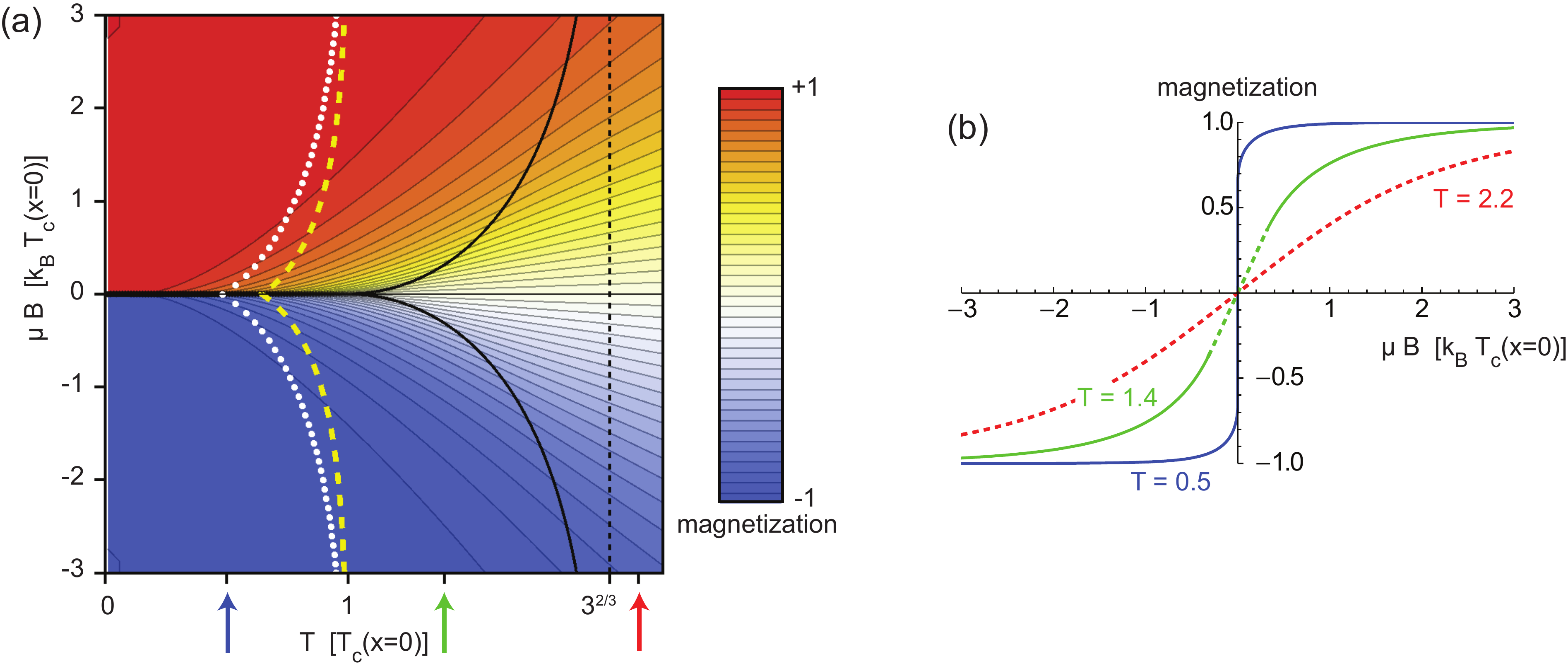}
\caption{Magnetization of a non-interacting $F=1$ spinor Bose gas at thermal equilibrium in an applied field.  (a) A contour plot shows the magnetization as a function of applied field strength and temperature.  The Zeeman energy $\mu B$ and temperature $T$ are shown scaled by the energy and temperature defined by the critical temperature for Bose-Einstein condensation at zero magnetic field, $T_\mathrm{c}(x=0)$, defined in the text.  The solid black line shows the Bose-Einstein condensation transition. The maximum transition temperature $T_\mathrm{c}(x \rightarrow \pm \infty) = 3^{2/3} \, T_\mathrm{c}(x=0)$ is indicated by the dashed black line.  The dashed yellow line demarcates a line of constant thermal energy, while the dotted white line demarcates a line of constant entropy, both calculated using expressions in Eqs.\ \ref{eq:energy} and \ref{eq:entropy}.  (b) The magnetization vs.\ magnetic field is plotted at several representative temperatures, indicated by the arrows in (a).  The lines here are dashed in regimes where the gas is non-degenerate, and solid where the gas is degenerate.  At temperatures $T < T_\mathrm{c}(x=0)$, the sudden jump in magnetization at zero field indicates the gas is ferromagnetic.\label{fig:yamafigure}}
\end{figure}

In Fig.\ \ref{fig:yamafigure}b, we sketch the magnetization of the Bose gas vs.\ applied field at several representative temperature.  At temperatures above $T_\mathrm{c}(x=0)$, the magnetization varies smoothly, with finite slope at $B=0$, indicating that the gas is paramagnetic.  For temperatures in the range $T_\mathrm{c}(x=0) < T < T_\mathrm{c}(x \rightarrow \pm \infty)$, the onset of Bose-Einstein condensation at non-zero magnetic field, occurring when the paramagnetic gas is sufficiently magnetized so that the density of one of the Zeeman sublevels rises above the critical density for condensation, causes a discontinuity only in the second derivative of the magnetization.

At temperatures below $T_\mathrm{c}(x=0)$, the magnetization is discontinuous at zero magnetic field.  The magnetic susceptibility for this gas (slope of magnetization vs.\ field at zero field) is infinite.  The spinor Bose-Einstein gas in this regime is thus ferromagnetic, even in the absence of any interaction among the atomic spins.  Magnetic ordering in this case can be thought of as a parasitic phenomenon, taking advantage of the fact that Bose-Einstein condensation creates a fraction of the gas with no entropy.

Our model also provides an approximation for the thermodynamic properties of the spinor Bose-Einstein gas.  We can use ideal-gas formulae to calculate the kinetic energy $U_{m_F}$ and entropy $S_{m_F}$ within each Zeeman sublevel of the gas:
\begin{eqnarray}
U_{m_F} & = & N_{\mathrm{th}, m_F} \frac{3}{2} \, k_B T \, \frac{g_{5/2}(z_{m_F})}{g_{3/2}(z_{m_F})} \label{eq:energy} \\
S_{m_F} & =& N_{\mathrm{th}, m_F} k_B \left( \frac{5}{2}\,  \frac{g_{5/2}(z_{m_F})}{g_{3/2}(z_{m_F})} - \frac{\mbox{\textmu}_{m_F}}{k_B T} \right) \label{eq:entropy}
\end{eqnarray}

Lines of constant energy and entropy determined by these formulae are shown on Fig.\ \ref{fig:yamafigure}a.  In the quantum-degenerate regime, we see that, at constant temperature, the energy and entropy of the gas decrease with the magnitude of the applied magnetic field.  As discussed above, there is one Zeeman sublevel for which the chemical potential is zero, and thus the normal fraction of atoms within this sublevel is saturated at an energy and entropy that is determined solely by the temperature.  Therefore, as the magnitude of the magnetic field is increased, the populations of the remaining magnetic sublevels decrease, decreasing the amount of thermal energy and entropy contained in these populations while increasing the number of condensed atoms.

This thermodynamic tendency is at the root of demagnetization cooling in a spinor Bose gas, which has been demonstrated both for alkali \cite{olf15cooling} and for high-spin spinor gases \cite{fatt06demag,nayl15demag}.  In the work with alkali gases, a fully longitudinally magnetized $F=1$ spinor gas was Bose-Einstein condensed in an optical trap.  Within the model discussed here, the full magnetization of the gas is equivalently described by the gas' coming to free equilibrium (without the magnetization constraint for alkali spinor gases) at an effective magnetic field $B_\mathrm{eff} = \infty$.  The magnetization of the gas was then deliberately lowered, simply by rotating the atomic spin slightly from the longitudinal direction, using rf fields so that the spin rotation was spatially uniform.  The gas was then allowed to come to thermal equilibrium under the constraint of constant magnetization.  Owing to the rotational symmetry of the s-wave contact interactions, the spin rotation pulse did not change the internal energy of the gas.  Of course, it did change the Zeeman energy of the gas greatly, but, as stated previously, the alkali gas is impervious to such energy changes.  Thereafter, the gas was allowed to equilibrate at constant magnetization.  The equilibration proceeds roughly at constant energy, assuming that the optically trapped gas is completely isolated from its surroundings.  As such, the gas undergoes isoenergetic demagnetization, following the yellow dashed line in Fig.\ \ref{fig:yamafigure}a to an equilibrium at a lower value of $B_\mathrm{eff}$.  As shown in the figure, following this line implies that demagnetization (of the coherent transverse magnetization produced by the spin-rotation pulse) leads to cooling.  Such cooling was indeed observed and used to drive down both the temperature and, upon ejecting the spin-flipped atoms, the entropy per particle of the spinor gas.

In the experiments on high-spin gases, demagnetization cooling is amplified by the fact that the gas can access the Zeeman energy.  A spin polarized gas was prepared with its magnetic moment oriented along an applied magnetic field.  Upon reducing the strength of the magnetic field, atoms were found to populate the magnetic sublevels with higher Zeeman energy, reaching those levels by converting thermal kinetic energy into Zeeman energy in a dipolar relaxation collision.  In the recent work of Ref.\ \cite{nayl15demag}, performed with a Bose-Einstein condensed chromium gas, both these demagnetization cooling effects contributed, i.e.\ both the conversion of kinetic energy to Zeeman energy and also the cooling effect of allowing magnetic excitations into the previously longitudinally polarized gas led to a reduction in temperature and entropy.

\subsection{Spin-dependent s-wave interactions in more recognizable form}

The above model emphasized the role of Bose-Einstein statistics in causing a spinor Bose-Einstein gas to become magnetically ordered.  Now we turn to the role of interactions in determining what type of magnetic order will emerge in conditions that are poorly described by our simple non-interacting-atom model, namely under the application of very weak magnetic fields, or, equivalently, in gases that are only weakly magnetized in the longitudinal direction.

We consider the effects of spin-dependent s-wave interactions, which we characterized in Eq.\ \ref{eq:swaveprojection} in terms of projection operators, $\hat{P}_{F_{\mathrm{tot}}}$, onto two-particle total spin states.  It is helpful to rewrite these projection operators onto more familiar forms of spin-spin interactions.  These interaction terms should retain the rotational symmetry of the interaction.

For the case of spin-1 gases, we find that two rotationally symmetric terms suffice to describe the s-wave interaction.  Considering operators that act on the states of two particles, labelled by indices $i$ and $j$, we note the following identities:
\begin{eqnarray}
\left[\hat{I}_i \otimes \hat{I}_j\right]_\mathrm{sym} & = & \hat{P}_0 + \hat{P}_2 \\
\left[\hat{\mathbf{F}}_i \cdot \hat{\mathbf{F}}_j\right]_\mathrm{sym} & = & \left[\frac{\hat{F}_\mathrm{tot}^2 - \hat{F}_1^2 - \hat{F}_2^2}{2}\right]_\mathrm{sym} = \hat{P}_2 - \hat{P}_0
\end{eqnarray}
Here, the subscript ``sym'' reminds us that we should include only states that are symmetric under the exchange of the two particles.  We can then replace the spin-dependent interaction of Eq.\ \ref{eq:swaveprojection} with the following expression:
\begin{equation}
\hat{V}_\mathrm{int} =  \frac{1}{2} \left(c_0^{(1)} + c_1^{(1)} \hat{\mathbf{F}}_i \cdot \hat{\mathbf{F}}_j \right) \delta^3(\mathbf{r}_i - \mathbf{r}_j)
\label{eq:vintwithc0c1}
\end{equation}
where the coefficients $c_0^{(1)}$ and $c_1^{(1)}$ are defined as
\begin{align}
c_0^{(1)} &=  \frac{4 \pi \hbar^2}{m} \frac{2 a_2 + a_0}{3} &
c_1^{(1)} &=  \frac{4 \pi \hbar^2}{m} \frac{a_2 - a_0}{3}. \label{eq:c0andc1}
\end{align}

The spin-dependent interactions are thus distilled to a form of isotropic Heisenberg interactions ($\hat{\mathbf{F}}_i \cdot \hat{\mathbf{F}}_j$).  The sign of the interaction divides spin-1 spinor gases into two types.  For $c_1^{(1)}<0$, the spin-dependent interaction is ferromagnetic, favoring a state where atomic spins are all maximally magnetized, i.e.\ the maximum spin-projection eigenstate along some axis $\mathbf{n}$.  For $c_1^{(1)}>0$, the spin-dependent interaction is anti-ferromagnetic.  We recall that in solid-state magnets, anti-ferromagnetic interactions between neighbors in a crystal lead to a variety of interesting magnetic states, such as N\'{e}el-ordered anti-ferromagnets, valence-bond solids, and spin liquids, the selection among which requires careful examination of the specific geometry and connectivity of the lattice.  Similarly, in anti-ferromagnetic spin-1 spinor gases, identifying the specific nature of magnetic ordering requires some careful thinking.

It may be useful here to remind the reader that the spin state of a spin-1 particle is characterized, at the one-body level, by a 3 $\times$ 3 density matrix.  Such a Hermitian matrix is defined by nine real quantities: three real diagonal matrix elements plus six real numbers that define the three independent off-diagonal complex matrix elements.  Corresponding to each of these real numbers, there is a Hermitian matrix that defines an observable quantity.  Thus, the space of one-body observables for a spin-1 particle is spanned by nine basis operators.  We may select these operators to be the identity operator, the three vector spin operators (denoted by symbols $\hat{F}$), and the five spin quadrupole operators (denoted by symbols $\hat{Q}$).   In the basis of $\hat{F}_z$ eigenstates, these operators can be written in the following matrix form:
\begin{align}
F_x &= \frac{1}{\sqrt{2}} \begin{pmatrix} 0 & 1 & 0 \\ 1 & 0 & 1 \\ 0 & 1 & 0 \end{pmatrix} &
F_y &= \frac{1}{\sqrt{2}} \begin{pmatrix} 0 & -i & 0 \\ i & 0 & -i \\ 0 & i & 0 \end{pmatrix} &
F_z &= \begin{pmatrix}1 & 0 & 0 \\ 0 & 0 & 0 \\ 0 & 0 & -1 \end{pmatrix} \label{eq:firstspin1matrix}\\
Q_{yz} &= \frac{1}{\sqrt{2}} \begin{pmatrix} 0 & 1 & 0 \\ 1 & 0 & - 1 \\ 0 & - 1 & 0 \end{pmatrix} &
Q_{xz} &= \frac{1}{\sqrt{2}} \begin{pmatrix} 0 & -i & 0 \\ i & 0 & i \\ 0 & i & 0 \end{pmatrix} &
Q_{xy} &= \begin{pmatrix} 0 & 0 & -i \\ 0 & 0 & 0 \\ i & 0 & 0 \end{pmatrix} \\
Q_{x^2-y^2} &= \begin{pmatrix} 0 & 0 & 1 \\ 0 & 0 & 0 \\ 1 & 0 & 0 \end{pmatrix} &
Q_{zz} &= \frac{1}{3} \begin{pmatrix} -1 & 0 & 0 \\ 0 & 2 & 0 \\ 0 & 0 & -1 \end{pmatrix} &
I &=  \begin{pmatrix} 1 & 0 & 0 \\ 0 & 1 & 0 \\ 0 & 0 & 1 \end{pmatrix}  \label{eq:lastspin1matrix}
\end{align}
All nine operators (except the identity operator) are traceless.  Except for $\hat{Q}_{zz}$, the traceless operators also have zero determinant and the same eigenvalues: $\{1, 0, -1\}$.

With these operators in mind, let us express the operator $\hat{V}_\mathrm{int}$, written in Eq.\ \ref{eq:vintwithc0c1}, in terms of Bose field creation $\hat{\psi}_{m_F}^\dagger(\mathbf{r})$ and annihilation operators $\hat{\psi}_{m_F}(\mathbf{r})$ which are written in the basis of the projection of the atomic spin along some axis (owing to the symmetry of the spin-dependent interaction, it does not matter which axis we choose).  We obtain the following,
%\begin{eqnarray}
%\hat{V}_\mathrm{int} &=& \frac{1}{2} \int d^3 \mathbf{r} \, \left(c_0^{(1)} + c_1^{(1)}\right) \left( \hat{\psi}^\dagger_1 \hat{\psi}_1 + \hat{\psi}^\dagger_0 %\hat{\psi}_0 + \hat{\psi}^\dagger_{-1} \hat{\psi}_{-1} \right)^2 \nonumber \\
%&  + & c_1^{(1)} \left( -4 \hat{\psi}^\dagger_1 \hat{\psi}^\dagger_{-1} \hat{\psi}_1 \hat{\psi}^\dagger_{-1}   - \left(\hat{\psi}^\dagger_0 \hat{\psi}_0\right)^2 + 2 %\left( \left.\hat{\psi}^\dagger_0\right.^2 \hat{\psi}_{1} \hat{\psi}_{-1} + \hat{\psi}_{1}^\dagger \hat{\psi}_{-1}^\dagger \hat{\psi}^2_0  \right) \right)
%\end{eqnarray}
\begin{eqnarray}
\hat{V}_\mathrm{int} &=& \frac{1}{2} \int d^3 \mathbf{r} \, \left[ \left(c_0^{(1)} + c_1^{(1)}\right) \left(\psipd \psip\right)^2 + c_0^{(1)} \left(\psizd \psiz \right)^2 +  \left(c_0^{(1)} + c_1^{(1)}\right) \left(\psimd \psim\right)^2 \nonumber \right. \\
& + & 2 \left(c_0^{(1)} + c_1^{(1)}\right) \psizd \psiz \left( \psipd \psip + \psimd \psim \right) + 2 \left(c_0^{(1)} - c_1^{(1)}\right) \psipd \psip \psimd \psim \nonumber \\
 & + & \left. 2 c_1^{(1)} \left( \left. \hat{\psi}^\dagger_0\right.^2 \psip \psim  + \psipd \psimd \psiz^2  \right) \right]
\end{eqnarray}
where, noting the delta-function spatial dependence, we have dropped the explicit dependence on position $\mathbf{r}$ which is the same for all field operators in the integrand.

\subsection{Ground states in the mean-field and single-mode approximations}

One common approach to identifying the state of the degenerate gas is to assume that a Bose-Einstein condensate forms in a coherent state, i.e.\ a many-body state that is a product state of identical single-particle states.  Let us simplify even further by stipulating that the single-particle state is a product state of a spatial wavefunction $\psi(\mathbf{r})$ (normalized to unity) and a spin wavefunction $|\zeta\rangle$, and consider only the lowest-energy choice for the spin wavefunction.  This is a variational, mean-field, and single-spatial-mode approximation.

We are left to minimize the following energy functional:
\begin{equation}
E = \langle \Psi | \hat{H}_\mathrm{motion} + \hat{V}_\mathrm{int} | \Psi \rangle
\end{equation}
where $|\Psi\rangle$ is our variational wavefunction, $\hat{H}_\mathrm{motion}$ includes the kinetic and potential energy terms.  Let us assume that $\hat{H}_\mathrm{motion}$  is spin independent.  Next, we assume that the spatial wavefunction $\psi(\mathbf{r})$ is fixed, and that only the spin wavefunction $|\zeta\rangle$ is to be varied.  With these assumptions, the only term remaining to consider is the spin-dependent energy, which we conveniently normalize by the number of atoms in the condensate:
\begin{eqnarray}
E_\mathrm{spin} &=& \frac{1}{N} \langle \Psi | \sum_{i,j} \frac{c_1^{(1)}}{2} \hat{\mathbf{F}}_i \cdot \hat{\mathbf{F}}_j  \, \delta^3(\mathbf{r}_i - \mathbf{r}_j) | \Psi \rangle \\
& = &  \frac{c_1^{(1)} n}{2} \langle \mathbf{F} \rangle^2
\end{eqnarray}
Here, $N$ is the particle number, $n = N \int |\psi(\mathbf{r})|^4 d^3\mathbf{r}$ is the average density, and $\langle \mathbf{F} \rangle = \langle \zeta | \hat{\mathbf{F}} | \zeta \rangle$.

This mean-field energy functional has two extremal values.  One extremum is obtained for the state that maximizes the length of the average spin vector, i.e.\ the state $|m_n = +1 \rangle$ with the maximum projection of its spin along some direction $\mathbf{n}$, giving $|\langle \mathbf{F} \rangle| = 1$.  This state minimizes the energy for gases with ferromagnetic interactions, with $c_1^{(1)}<0$.  The state is maximally magnetized, with $\mathbf{n}$ defining the direction of the magnetization (or the opposite of that direction, depending on the sign of the gyromagnetic ratio).  In our single-mode approximation, the spinor part of the wavefunction is constant in space.  So a gas in the $|m_n = +1\rangle$ state is ``ferromagnetic'' in the sense that, like in a solid-state ferromagnet, it has a macroscopic magnetization that has a constant direction across the entire volume of the material.

A second extremum in the mean-field energy function occurs for the state that minimizes the length of the average spin vector.  This is the state $|m_n = 0\rangle$, which is the eigenstate for the spin projection along the $\mathbf{n}$ axis with eigenvalue zero.  The average spin in directions orthogonal to $\mathbf{n}$ is also zero, so one obtains $|\langle \mathbf{F} \rangle| = 0$.  This state then minimizes the energy for gases with antiferromagnetic interactions, with $c_1^{(1)}>0$.  We can regard this state as being nematic, in that it is aligned with an axis $\mathbf{n}$ (which we may call the \emph{director}), but that it is oriented neither in the $+\mathbf{n}$ or $-\mathbf{n}$ directions.  A condensate with a spatially uniform nematic spin wavefunction is ``nematically ordered'' in that the director $\mathbf{n}$ is constant in space.  The $|m_n = 0\rangle$ is also termed \emph{polar} because of the fact that rotating the state by $\pi$ radians about an axis transverse to $\mathbf{n}$ produces a final state that still is a zero-valued eigenstate of the operator $\hat{F}_n = \hat{\mathbf{F}} \cdot \mathbf{n}$; however, the rotation results in the final state differing from the initial state by a minus sign.  This property is exhibited also by the p$_z$ electronic orbital of atomic and chemical physics, in which the electron density is symmetric under a $\pi$ rotation about, say, the $\mathbf{x}$ axis, but where the phase of the electronic wavefunction is antisymmetric under such rotation.

\subsection{Mean-field ground states under applied magnetic fields}
\label{sec:meanfieldgs}

In Sec.\ \ref{sec:nonint}, we discussed the thermodynamics of a non-interacting spinor gas in an applied magnetic field, or, equivalently, under the constraint of constant longitudinal magnetization.  Let us extend this consideration also to the interacting spinor gas in order to understand how the application of symmetry breaking fields tears at the rotationally symmetric interaction Hamiltonian and its solutions.

We consider two symmetry-breaking external influences.  The first is a linear Zeeman shift, or equivalently, as discussed before, a constraint of constant longitudinal magnetization.  Keeping with nomenclature in early papers on spinor Bose-Einstein condensates, we account for this energy through the term $p \hat{F}_z$.  In early experiments on spinor Bose-Einstein condensates, a magnetic field gradient was applied deliberately to the gas to provide clear experimental signatures of the spin-dependent interactions \cite{sten98spin}.  Such a gradient can be accommodated by making $p$ spatially varying; we revisit this point in Sec.\ \ref{sec:experimentalevidence}.

Secondly, we include also a quadratic Zeeman energy.  As mentioned earlier, the spinor gas can evolve by the spin-mixing collision.  This process is insensitive to the linear Zeeman energy, but depends on the quadratic Zeeman shift, which shifts the average energy of the $|m= \pm 1 \rangle$ states (Zeeman energy of the right side of Eq.\ \ref{eq:spinmixingreaction}) by an energy $q$ with respect to that of the $|m_F = 0\rangle$ state (Zeeman energy of the left side of Eq.\ \ref{eq:spinmixingreaction}). The quadratic Zeeman shift comes about from the fact than an applied magnetic field mixes hyperfine levels, causing energy repulsion between states in, say, the upper and lower hyperfine states with the same value of $m_F$ in the electronic ground state of alkali gases.  A quadratic Zeeman energy can also be imposed by driving the atoms with microwave-frequency magnetic fields that are near resonant with hyperfine transitions \cite{gerb06rescontrol}.  The latter approach allows one to reverse the sign of the quadratic Zeeman energy (microwave-dressed hyperfine levels are made to move closer, rather than farther, in energy).  For simplicity, and to match with most experiments, we consider quadratic Zeeman energy shifts in the same basis as that determined by the linear Zeeman energy, i.e.\ through the term $q \hat{F}_z^2$.  As such, the Zeeman energies still preserve the symmetry of rotations about the longitudinal direction (the $\mathbf{z}$ axis).  More generally, the quadratic and linear Zeeman energies could, in principle, be applied along different axes, breaking rotational symmetry altogether.

We now obtain a spin-dependent energy functional of the form
\begin{equation}
E_\mathrm{spin} = p \langle \hat{F}_z \rangle + q \langle \hat{F}_z^2 \rangle + \frac{ c_1^{(1)} n}{2} \langle \mathbf{F} \rangle^2
\end{equation}
Expressing the spin wavefunction $|\zeta\rangle$ in the $\hat{F}_z$ eigenbasis, the energy functional has the form
\begin{equation}
\begin{split}
E_\mathrm{spin} = p \left( \left|\zeta_1\right|^2 - \left|\zeta_{-1}\right|^2 \right) + q \left( \left|\zeta_1\right|^2 + \left|\zeta_{-1}\right|^2 \right) \\
+ \frac{c_1^{(1)} n}{2} \left[ \left( \left|\zeta_1\right|^2 - \left|\zeta_{-1}\right|^2 \right)^2 + 2 \left| \zeta_1^* \zeta_0 + \zeta_0^* \zeta_{-1} \right|^2\right]
\label{eq:espinversion1}
\end{split}
\end{equation}

This expression, while correct, hides from us the fact that some of the information in the spin wavefunction $|\zeta\rangle$ is irrelevant for calculating the energy.  For one, the spinor is normalized to unity, so we need not keep track of all the populations $|\zeta_i|^2$.  We'll keep track just of two, choosing the quantities
\begin{eqnarray}
M &=& \left|\zeta_1\right|^2 - \left|\zeta_{-1}\right|^2 \\
\rho_0 &=& \left|\zeta_0\right|^2 = 1 - \left|\zeta_1\right|^2 - \left|\zeta_{-1}\right|^2
\end{eqnarray}
Here, $M$ is the longitudinal magnetization of the gas, normalized to lie in the range $-1 \leq M \leq 1$, and $\rho_0$ is the fractional population in the $|m_F = 0\rangle$ state.  Under a constraint of constant longitudinal magnetization, $\rho_0$ is constrained to vary in the range $0 \leq \rho_0 \leq 1 - |M|$.

In addition, the wavefunction $|\zeta\rangle$ contains three complex phases, for example, the complex arguments of the three probability amplitudes $\phi_i = \arg(\zeta_{i})$.  However, two linear combinations of these phases do not affect the energy.  Clearly, the energy is unaffected by multiplying the wavefunction by a global phase.  Thus, the combination $\bar{\phi} = \left( \phi_1 + \phi_0 + \phi_{-1} \right)/3$ is unimportant to us.  Also, the mean-field energy functional $E_\mathrm{spin}$ is invariant under rotations of the spin about the $\mathbf{z}$ axis.  Such a rotation would change the value of the phase difference $\phi_z = \phi_1 - \phi_{-1}$, but would not change the energy.  The only linear combination of phases that is still potentially important is the combination
\begin{equation}
\theta = \phi_1 + \phi_{-1} - 2 \phi_0
\end{equation}
In our later discussion of spin-mixing dynamics, we will see that it is this phase combination that controls the direction in which populations will flow in a spin-mixing collision (Eq.\ \ref{eq:spinmixingreaction}).

We now rewrite the spin-dependent energy as
\begin{equation}
E_\mathrm{spin} = p M + q (1 - \rho_0) + \frac{c_1^{(1)} n M^2}{2} + c_1^{(1)} n \rho_0 \left[(1-\rho_0) + \sqrt{(1-\rho_0)^2 - M^2} \cos\theta \right]
\label{eq:meanfieldenergy}
\end{equation}
The task of minimizing this energy is now much clearer.  The phase $\theta$ controls the degree of magnetization present when the spin wavefunction contains a superposition of states with $\Delta m_F = 1$.  We choose $\cos\theta = 1$ for ferromagnetic interactions ($c_1^{(1)} <0$) to maximize that magnetization, and $\cos\theta = -1$ for antiferromagnetic interactions ($c_1^{(1)} >0$) in order to minimize that magnetization.  All that is left is to minimize the energy with respect to the populations in the spin wavefunction, expressed through $M$ and $\rho_0$.

\begin{figure}[t]
\centering
\includegraphics[width=\textwidth]{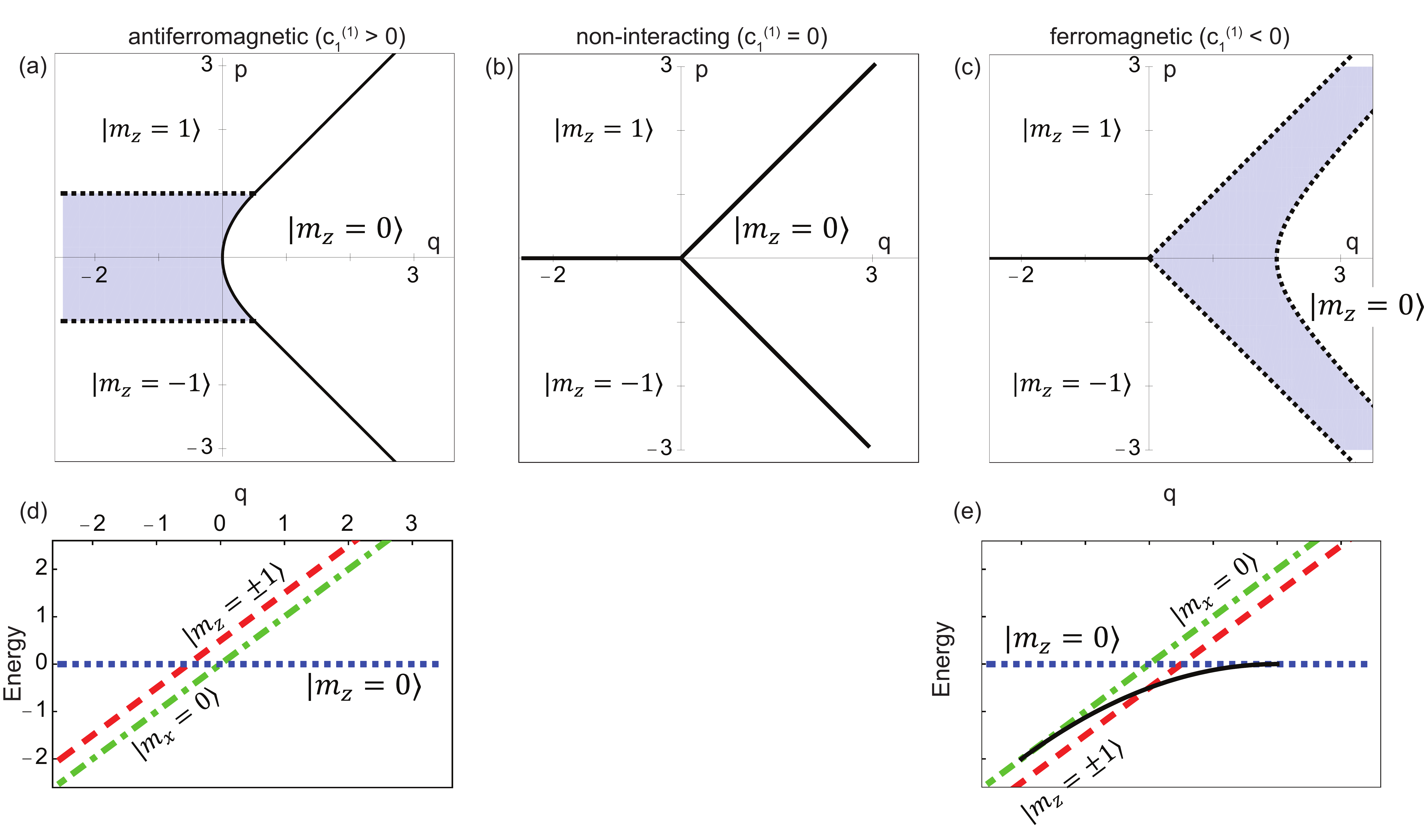}
\caption{Mean-field ground states predicted for an $F=1$ spinor gas with spin-dependent s-wave interactions.  The energy functional of Eqs.\ \ref{eq:espinversion1} and \ref{eq:meanfieldenergy} is considered.  Here $p$ (linear Zeeman energy, or Lagrange multiplier for magnetization constraint) and $q$ (quadratic Zeeman energy) are measured in units of the interaction energy $|c_1^{(1)} n|$.  Diagrams are shown for (a) antiferromagnetic interactions, (b) neglecting the effects of interaction, and (c) ferromagnetic interactions.   Gray regions in (a) and (c) indicate the ground state is a mixture of Zeeman sublevels.  Solid lines indicate discontinuous phase changes, while dashed lines indicate continuous changes.  Also shown are the mean-field energies of three extremal states -- the longitudinally magnetized state, the longitudinal polar state ($|m_z =0\rangle$, for which the nematic director is the $\mathbf{z}$ axis) and the transverse polar state (for example $|m_x = 0\rangle$, for which the nematic director is the $\mathbf{x}$ axis) -- for either (d) antiferromagnetic or (e) ferromagnetic interactions, where we set $p=0$.  In the antiferromagnetic case, the polar state is always the lowest energy state, changing its alignment at $q=0$.  For the ferromagnetic case, the longitudinal polar state is the ground state for $q>2$, while the longitudinally magnetized state is the ground state for $q<0$.  Between these regions, a state of non-zero transverse magnetization is favored (energy given by black line), with spin wavefunction given in Eq.\ \ref{eq:transversespin}. \label{fig:pandq}}
\end{figure}

Such minimization was performed initially in Ref.\ \cite{sten98spin}.  The results are presented in Fig.\ \ref{fig:pandq}.  Focusing first on the non-interacting case, the phase diagram is simple: For $q \leq 0$, the lowest energy state is either of the longitudinally magnetized states, $|m_z = \pm 1\rangle$, depending on the sign of $p$.  The situation here matches that of the non-interacting $F=1$ gas considered in Sec. \ref{sec:nonint}.  For $q>0$, the quadratic Zeeman shift favors the $|m_z = 0\rangle$ state where $q > |p|$.

The spin-dependent s-wave interaction modifies this phase diagram.  We observe that, for example, the region in the $p$-$q$ plane occupied by the $|m_z = 0\rangle$ ground state (this is the longitudinal polar state, with the nematic vector aligned with the longitudinal $\mathbf{z}$ axis) is increased in the case of antiferromagnetic interactions, and decreased in the case of ferromagnetic interactions.

The diagram also shows evidence of the miscibility and immiscibility of different magnetic sublevels.  The miscibility of two Bose-Einstein condensed components $i$ and $j$ of a gas (with equal masses, as here) is governed by the relation between the s-wave scattering lengths, $a_{ii}$, $a_{jj}$ and $a_{ij}$ where the indices indicate the two components colliding.  If $a_{ij} < \sqrt{a_{ii} a_{jj}}$, then the components are miscible in that a superposition of the two components has less interaction energy than two phase-separated domains of the two components.  The s-wave scattering lengths for different collisions among $F=1$ atoms are all determined by the values of $a_{F_\mathrm{pair} = 0}$ and $a_{F_\mathrm{pair} = 2}$, as shown in Table \ref{tab:scatteringlengths}.

\begin{table}[tb]
\begin{center}
\begin{tabular}{r|cc|cc|cc}
 & \multicolumn{2}{c|}{$|m_F = +1\rangle$} & \multicolumn{2}{c|}{$|m_F = 0\rangle$} & \multicolumn{2}{c}{$|m_F = -1\rangle$} \\
 %$|m_F = 0\rangle$ & $|m_F = -1\rangle$ \\
\hline
$|m_F = +1\rangle$ & $c_0^{(1)} + c_1^{(1)} $ & $ a_2$ & $c_0^{(1)} + c_1^{(1)} $ & $ a_2$ & $c_0^{(1)} - c_1^{(1)} $ & $ \dfrac{a_2 + 2 a_0}{3}$ \\[0.2 cm]
$|m_F = 0\rangle$ &  $c_0^{(1)} + c_1^{(1)} $ & $ a_2$ & $c_0^{(1)} $ & $ \dfrac{2 a_2 + a_0}{3} $ & $c_0^{(1)} + c_1^{(1)} $ & $ a_2$ \\[0.2 cm]
$|m_F = -1\rangle$ & $c_0^{(1)} - c_1^{(1)} $ & $ \dfrac{a_2 + 2 a_0}{3}$ & $c_0^{(1)} + c_1^{(1)} $ & $ a_2$ & $c_0^{(1)} + c_1^{(1)} $ & $ a_2$ \\[0.2 cm]
%$|m_F = +1\rangle$ & $c_0^{(1)} + c_1^{(1)} \sim a_2$ & $c_0^{(1)} + c_1^{(1)} \sim a_2$ & $c_0^{(1)} - c_1^{(1)} \sim (a_2 + 2 a_0)/3$ \\
%$|m_F = 0\rangle$ &  $c_0^{(1)} + c_1^{(1)} \sim a_2$ & $c_0^{(1)} \sim (2 a_2 + a_0)/3 $ & $c_0^{(1)} + c_1^{(1)} \sim a_2$ \\
%$|m_F = -1\rangle$ & $c_0^{(1)} - c_1^{(1)} \sim (a_2 + 2 a_0)/3$ & $c_0^{(1)} + c_1^{(1)} \sim a_2$ & $c_0^{(1)} + c_1^{(1)} \sim a_2$\\
\hline
\end{tabular}
\caption{Interaction strengths that characterize binary s-wave collisions among atoms in the $F=1$ manifold.  The interaction strength is characterized both in terms of the scalar and Heisenberg-like interaction strengths, $c_0^{(1)}$ and $c_1^{(1)}$ respectively, that are defined in Eqs.\ \ref{eq:c0andc1}, and also in terms of the scattering lengths $a_{F_\mathrm{pair}}$ that describe s-wave collisions between atoms with total angular momentum $F_\mathrm{pair} = 0$ and $F_\mathrm{pair} = 2$.}
\label{tab:scatteringlengths}
\end{center}
\end{table}

We see that for antiferromagnetic interactions, the $|m_z = 1\rangle$ and $|m_z = 0\rangle$ components are immiscible.  This fact is reflected in the ground-state phase diagram (Fig.\ \ref{fig:pandq}a) by the sharp boundary between the regions in which either of these components is the ground state.  In contrast, in the case of ferromagnetic interactions, the transition between these two regions is gradual.  The gray region between them in Fig.\ \ref{fig:pandq}c indicates that the mean-field ground state is a superposition of states including the $|m_z = 1\rangle$ and $|m_z=0\rangle$ state; such a mixture allows the gas to be partly magnetized and thus take advantage of the spin-dependent interaction.  One also sees that, for antiferromagnetic interactions, the $|m_z = 1\rangle$ and $|m_z=-1\rangle$ states are miscible.  This miscibility is indicated in Fig.\ \ref{fig:pandq}a by the gray region that lies between the $|m_z=1\rangle$ and $|m_z=-1\rangle$ ground state regions at $q<0$, in which a superposition of these two states reduces the mean-field energy.  Indeed, for $q<0$ and $p=0$, the minimum energy state for antiferromagnetic interactions is the transversely aligned polar state, i.e.\ the state $|m_n = 0\rangle$ for which the nematic director $\mathbf{n}$ lies in the plane transverse to $\mathbf{z}$ (in the $x$-$y$ plane).  Such a state is a superposition of the $|m_z = \pm 1\rangle$ states with equal population in the two states.

\subsection{Experimental evidence for magnetic order of ferromagnetic and antiferromagnetic $F=1$ spinor condensates}
\label{sec:experimentalevidence}

Compelling experimental evidence for the realization of such mean-field ground states for $F=1$ spinor gases has been obtained.  The ferromagnetic case is exemplified by the $F=1$ spinor gas of $^{87}$Rb.  The first experiments on this system were performed by the groups of Chapman \cite{chan04} and Sengstock \cite{schm04}. They examined optically trapped, Bose-Einstein condensed gases that were prepared with zero longitudinal magnetization (i.e.\ with $p=0$), by preparing all atoms in the $|m_F = 0\rangle$ state.  Then, the gases were allowed to evolve under an applied, uniform magnetic field, which produced an adjustable, positive value of $q$.

At high $q$, the $|m_F = 0\rangle$ state being lower than energy than the average energy of the $|m_F =\pm 1\rangle$ states that would be produced by spin mixing, we expect the $|m_F = 0\rangle$ state to be the preferred ground state of the atoms.  At low $q$, one expects the gas to acquire magnetization.  Consistent with the initial magnetization of the gas, one expects the gas to become transversely magnetized, described by the single-atom spin wavefunction
\def\verticaldistance{5pt}
\begin{equation}
|\zeta\rangle = \begin{pmatrix}  e^{-i \phi} \frac{\sqrt{\left(1 - q/q_0 \right)}}{2} \\[\verticaldistance]  \frac{\sqrt{\left(1 + q/q_0 \right)}}{\sqrt{2}}\\[\verticaldistance] e^{i \phi}  \frac{\sqrt{\left(1 - q/q_0 \right)}}{2} \end{pmatrix},
\label{eq:transversespin}
\end{equation}
over the range $0 \leq q \leq 2 |c_1^{(1)} n| = q_0$.  Here, the transverse magnetization has the amplitude $\sqrt{1 - (q/q_0)^2}$, i.e.\ ranging from zero magnetization at the transition from the paramagnetic to the ferromagnetic phase ($q = q_0$) to full magnetization at $q=0$ where the spin-dependent interactions are the only spin-dependent energy term.  In the experiment, the Zeeman-state populations of the gas were measured, and indeed one found the expected ratio of populations in the $|m_F = \{1, 0, -1\}\rangle$ states.

The Zeeman state populations on their own do not confirm that the gas has adopted a state with nonzero magnetization; for this, one needs to establish that the Zeeman populations are coherent with one another, and that the phase relations among them lead to a non-zero magnetic moment.  For example, compare the two spin wavefunctions
\begin{align}
|\zeta_\mathrm{F}\rangle &= \begin{pmatrix} 1/2 \\[\verticaldistance] 1/\sqrt{2} \\[\verticaldistance] 1/2 \end{pmatrix} & |\zeta_\mathrm{P}\rangle &= \begin{pmatrix} 1/2 \\[\verticaldistance] i/\sqrt{2} \\[\verticaldistance] 1/2 \end{pmatrix}.
\end{align}
The first of these, $|\zeta_\mathrm{F}\rangle$, is the state magnetized along the $\mathbf{x}$ axis.  The second of these, $|\zeta_\mathrm{P}\rangle$, differing from the first only in a phase factor on one of the probability amplitudes, is a polar state with zero magnetization.  Without measurements of the relative phase between Zeeman-state probability amplitudes, one cannot discern these states.

Two separate experiments confirmed that these relative phases are indeed those that maximize the magnetization in the case of ferromagnetic interactions.  The Chapman group utilized spin-mixing dynamics, which we discuss in Sec.\ \ref{sec:dynamics}, to cause the magnetic state of the gas to reveal itself \cite{chan05nphys}.  After the gas had come presumably to equilibrium, the experimenters applied a briefly pulsed quadratic Zeeman energy shift.  This pulse knocked the state away from its equilibrium state and initiated a tell-tale oscillation of the Zeeman-state populations, the dynamics of which were in excellent agreement with numerical calculations and indicative of the initial state being ferromagnetic as expected.

\begin{figure}[t]
\centering
\includegraphics[width=0.8\textwidth]{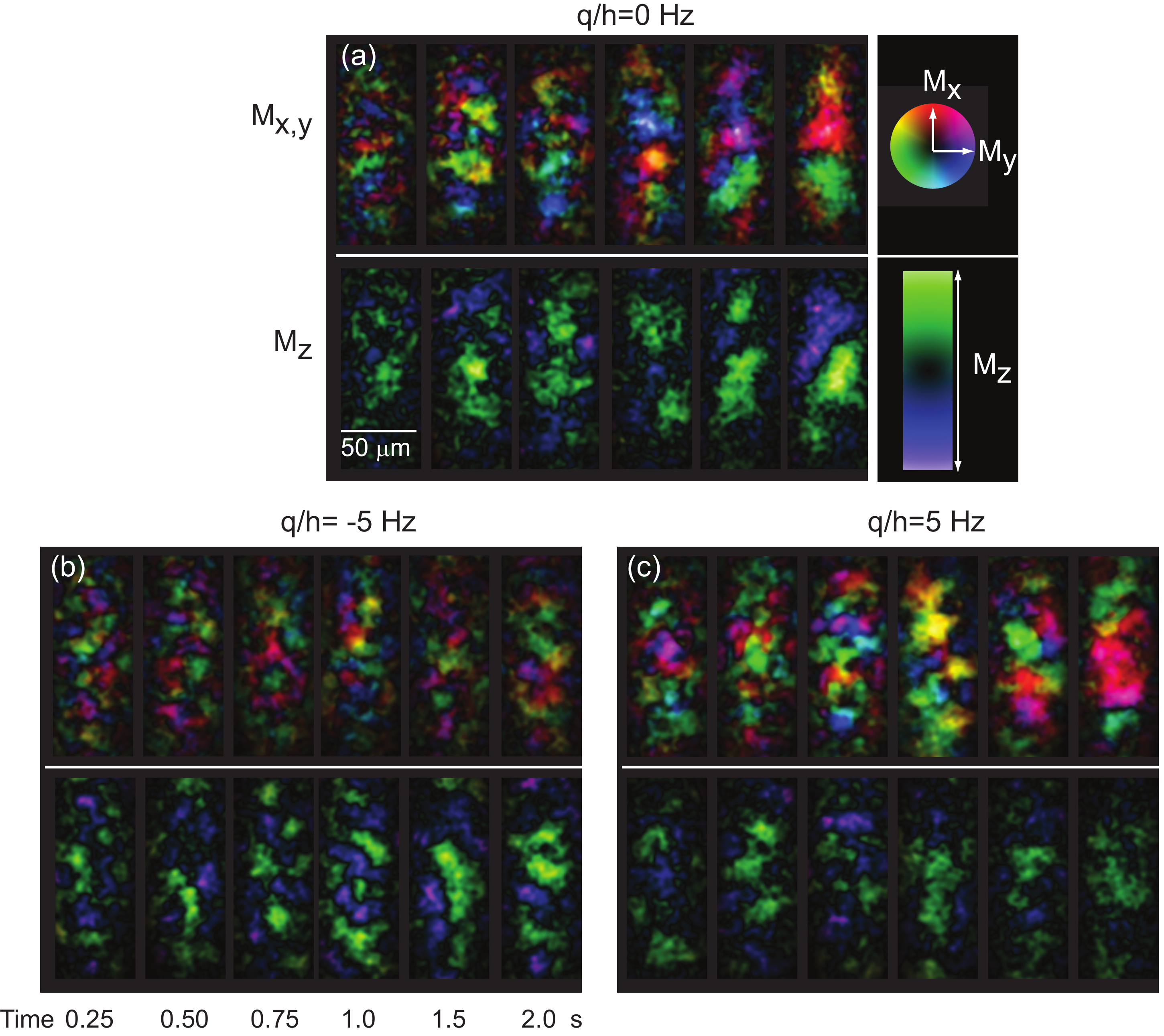}
\caption{Direct imaging of the formation of ferromagnetic order in a $F=1$ $^{87}$Rb spinor gas.  An unpolarized gas was prepared above the Bose-Einstein condensation transition temperature.  The gas was then cooled by evaporation, and allowed to equilibrate, for the time indicated on the figure, before its vector column magnetization was measured by magnetization-sensitive phase-contrast imaging.  The magnetization for each single sample probed is represented in two images stacked vertically; the top shows the transverse magnetization and the bottom the longitudinal magnetization.  The color scheme used for all these images is shown.  The gas is shown to develop ferromagnetically ordered regions of increasing size over time.  The magnetization realized with $q=0$ is isotropically distributed.  The magnetization realized for $q>0$ tends to lie in the transverse plane, while for $q<0$ it tends to point longitudinally.  Figure adapted from Ref.\ \cite{guzm11}.\label{fig:rbequilibration}}
\end{figure}

More direct evidence of the ground-state structure of the $F=1$ $^{87}$Rb gas comes from direct measurements of the gas magnetization.  The imaging methods used to measure such magnetization are discussed in Sec. \ref{sec:imaging}.  In the work of Ref.\ \cite{guzm11}, such measurements confirmed that a ferromagnetic spinor gas prepared with no initial spin coherence whatsoever will, upon being cooled to quantum degeneracy and allowed to evolve for long times, become spontaneously magnetized.  For a positive quadratic Zeeman energy ($q>0$), the magnetization was found to lie in the transverse plane, while for $q<0$ (realized by the application of microwave fields) the magnetization tended to orient along the magnetic field axis (Fig.\ \ref{fig:rbequilibration}).

The case of antiferromagnetic interactions is exemplified by the $F=1$ spinor condensate of sodium.  The first experiments on spinor condensates, performed with such a gas, confirmed the increased energetic stability of the polar state by placing the trapped gas in a magnetic field gradient \cite{sten98spin}.  Under such a gradient, a gas within a net zero longitudinal magnetization would tend to become polarized, lowering its energy by adopting the strong-field seeking state ($|m_F = +1\rangle$ in this case) at the high-field end of the trap, and the weak-field seeking state ($|m_F = -1\rangle$) at the low-field end.  However, such an arrangement maximizes the spin-dependent interaction of the antiferromagnetic spinor gas.  A lower energy configuration is achieved by reducing the magnetization of the gas at the cloud center.  Depending on the sign of the quadratic Zeeman shift, the favored non-magnetic state is a polar state ($|m_n = 0\rangle$) with the director $\mathbf{n}$ pointing either transversely (for $q<0$) or axially ($q>0$).  Measurements of the one-dimensional spin-state distribution of the gas (along the direction of the field gradient) showed the expected magnetization pattern, in quantitative agreement with expectations.

\subsection{Correlations in the exact many-body ground state of the $F=1$ spinor gas}
\label{sec:manybody}

It is clear, however, that these mean-field product states cannot be the true ground state of the interacting spinor gas.  Consider specifically the case of antiferromagnetic interactions, the ones we suspect, based on solid-state magnetism, are going to be the most interesting.  Mean-field theory predicts the state with all atoms in the $|m_F = 0\rangle$ Zeeman state to be a ground state; for a gas of $N$ atoms, we may write this state in the Fock-state basis as $|0,N,0\rangle$ where we enumerate the atoms in the $|m_F = \{+1, 0, -1\}\rangle$ states sequentially.  However, the many-body Hamiltonian contains a term describing spin-mixing collisions, meaning that this mean-field state is coupled, for example, to the orthogonal state $|1, N-2, 1\rangle$.  The true ground-state of the system must contain a superposition of (at least) these two many-body states.

A similar scenario applies in the interacting scalar Bose gas.  The mean-field ground state for a condensate in a uniform container is one with all atoms in the zero-momentum state.  However, the Hamiltonian includes terms describing elastic collisions that conserve the total momentum of the colliding atom pair.  One such term describes a collision in which two atoms are excited out of the zero-momentum state and promoted to states of non-zero counter-propagating momenta $\pm \mathbf{p}$.  Thus, the true many-body ground state must include also a population of atoms in non-zero momentum states.  Yet, the number of such momentum-excited atom pairs remains small owing to the energy penalty for creating such excitations.  The total number of excited pairs comprises the quantum depletion, which remains small in the case of weak interactions \cite{bogo47}.

Is there a similar energy penalty for spin-mixing collisions, which might suppress the population of atoms outside the $|m_F = 0\rangle$ spin state and constrain the quantum depletion atop the mean-field state?  Consider the state generated from our initial mean-field ansatz by a slight geometric rotation, i.e.\ all atoms in the polar state that is aligned with an axis just slightly tilted from the $\mathbf{z}$ axis.  Such a state also contains a superposition of the $|0,N,0\rangle$ and $|1, N-2, 1\rangle$ states, and yet the energy of this rotated state is no higher than the initial state.  Thus we suspect, without rigorous proof yet, that the spin-mixing interaction will mix new spin configurations into the mean-field initial state without an energy penalty, different from the situation with a weakly interacting scalar Bose gas. If this suspicion is correct, then the quantum depletion of the mean-field state will be massive, and mean-field state would no longer be a good approximation to the many-body ground state.

This vague argument was cast more rigorously and elegantly by Law, Pu and Bigelow \cite{law98spin2}.  One realizes that the exact many-body Hamiltonian, containing a sum of spin-dependent interactions between all pairs of atoms in the gas, treats all atoms symmetrically.  We might expect, therefore, that the Hamiltonian can be written not just in the basis of spin operators $\hat{\mathbf{F}}_i$ acting on each individual atom within the gas, but also in terms of collective spin operators that describe the joint spin state of all atoms in the gas.  This expectation turns out to be correct.  Defining the collective spin operator as $\hat{\mathbf{F}}_N = \sum_i \hat{\mathbf{F}}_i$, we find the spin-dependent many-body Hamiltonian for a spinor gas of atoms, each assumed to be in the same state of motion, can be written as
\begin{equation}
\hat{H} = \frac{c_1^{(1)} n}{2} \left( \frac{\hat{\mathbf{F}}_N^2}{N} - 2\right)
\end{equation}
where $n$ is the average density of the gas.

In the case of antiferromagnetic interactions ($c_1^{(1)}>0$), the energy is minimized by minimizing $\hat{\mathbf{F}}_N^2$.  Such minimization is achieved by the state of zero collective spin $|F_N = 0, m_{F_N} = 0\rangle$.  This state can be pictured as being composed of pairs of atoms (letting $N$ be even), each of which is in the spin-singlet state, written for a pair of particles as
\begin{equation}
|F_\mathrm{pair}=0, m_{F_\mathrm{pair}}=0\rangle = \sqrt{\frac{2}{3}} |1, 0, 1\rangle - \sqrt{\frac{1}{3}} |0, 2, 0\rangle
\end{equation}
using the Fock-state basis introduced above.  The many-body state is a product state of $N/2$ such pairs, fully symmetrized under particle exchange.

This many-body state has several distinct features.  First, it is a state that does not break the rotational symmetry of the Hamiltonian.  Indeed, one may write this state as the superposition of mean-field polar states, integrated over all alignment axes \cite{barn10rotor}.  Second, this ground state is unique up to an overall phase factor.  Third, the state is not a standard Bose-Einstein condensate, which would be defined, according to the criterion of Penrose and Onsager,  as a state for which there is just a single, single-particle state that is macroscopically occupied \cite{penr56}.  Finally, this state has measurable features that distinguish it from the mean-field ground state.  Consider that for an $N$-atom mean-field polar ground state, i.e.\ all atoms in the $|m_n = 0\rangle$ state, there is a defined axis $\mathbf{n}$ along which the spin projection of the atoms is strictly zero.  However, measured along any other axis, while the net atomic spin projection has zero average, it fluctuates by an amount on the order of $\sqrt{N}$.  In contrast, the many-body ground state is rotationally symmetric, and thus has strictly zero net spin projected on any quantization axis.  Zeeman-state populations can be measured experimentally with single-atom sensitivity, so this signature of the many-body ground state should be accessible in future experiments.

In the case of ferromagnetic interactions ($c_1^{(1)}<0$), the energy is minimized by maximizing $\hat{\mathbf{F}}_N^2$.  In this case, the many-body ground state is degenerate.  Its subspace is spanned by basis vectors of the form $|F_N = N, m_{F_N}\rangle$, with $m_{F_N}$ taking one of $2N + 1$ values.  This subspace includes the mean-field ground state, and our guess that a Bose-Einstein condensate would form that breaks rotational symmetry spontaneously is not invalidated.

%\DeclareGraphicsExtensions{.eps}
% !TEX root = ./Varenna_spinors_main.tex
\section{Imaging spinor condensates\label{sec:imaging}}
\dmskedit{\subsection{Introduction to imaging: absorptive and dispersive techniques}}{}

How do we experimentally probe a multicomponent condensate? A spinless condensate is characterized by a complex order parameter $\psi(\mathbf{r})$. The density $|\psi(\mathbf{r})|^2$ can be directly measured by conventional imaging techniques, for instance by measuring how much light is absorbed by the sample. Measuring the phase is more difficult. One approach is to take multiple images and observe the density change with time, since the velocity $\mathbf v = (-i \hbar/m) \psi^\ast \mathbf\nabla \psi$ depends on the phase. Alternatively, the phase can be measured with an interference experiment. Together, these yield all the information needed to reconstruct $\psi(\mathbf{r})$.

For a multicomponent condensate, the order parameter is more complicated. Ideally, we would like to know the spin density matrix at each point in the condensate. For a condensate near its ground state, at low temperature, we might expect to need to measure the $2F+1$ complex components of the wavefunction; that is, the population of each spin component and the coherences between those components. %In many situations, the required information is far less.

%First of all, in many multicomponent condensates, the spin-dependent interactions are much weaker than the spin-independent interactions. This leads to a separation of scales between the highly energetic excitations of the density and relatively low energy excitations of the spin. In many experiments that study quenches and spin dynamics, the density profile of the condensate is static while the spin may freely evolve.

%Second, spins may evolve within a restricted space. For instance, when $c^{(1)}_1<0$, spin-dependent s-wave interactions favor a fully magnetized ground state. Low energy excitations still satisfy $|\langle \mathbf F(\mathbf r) \rangle|=1$ at every point in the condensate, with changes in the orientation of $\langle \mathbf{F} \rangle$ throughout the sample. As we will see, it is then sufficient to measure changes in the orientation of $\langle \mathbf{F} \rangle$ to determine the spin wavefunction.

In this Section, we discuss three imaging techniques used to measure the spin evolution of the condensate. In brief, a light beam passes through a spinor condensate and is altered by its interaction with the atoms. The first imaging technique uses dispersive interactions, causing the light to acquire a phase shift, while the second and third techniques are absorptive, in which case atoms scatter photons out of the laser beam. Of particular interest are nondestructive techniques where many images of a single condensate show its temporal evolution.
How do we make optical imaging of an atomic gas sensitive to the spin state of that gas?  One can think about the degrees of freedom at hand: polarization, frequency, transverse position in an image, and time at which an image is taken.  Imaging techniques have been developed that translate each of these degrees of freedom into information on the spin composition of a gas.

\subsection{Stern-Gerlach imaging}

The first experiment on spinor gases \cite{sten98spin}, and many more recent studies, used spin-dependent forces to separate spin components spatially before imaging.  For instance, the spin-up atoms might be pushed to the right and the spin-down atoms pushed to the left before an image of all the atoms is taken.  The image can then be processed to surmise what was the spin distribution of the gas before the forces and imaging were applied.  A spin-dependent force is easily achieved by applying a magnetic field gradient. Most atoms have a magnetic moment on the order of a Bohr magneton, for which a moderate field gradient of just several G/cm will displace the different magnetic sublevels of a cloud of atoms by 100's of microns of one another within 10's of milliseconds.  Ultracold atoms have extremely narrow momentum distributions such that they barely expand over this time: a $100\;\mathrm{nK}$ thermal (non-condensed) gas of ${}^{87}$Rb will expand $30\;\mu\mathrm{m}$ over $10\;\mathrm{ms}$, while a $10\;\mathrm{G/cm}$ magnetic field gradient will separate the three $F=1$ spin states by ten times this distance in the same time. Thus, atoms in the different sublevels can be well separated from one another, just as the spots on a glass plate formed by silver atoms with different initial spin states were separated in the experiment of Stern and Gerlach \cite{gerl24}; hence the name ``Stern-Gerlach imaging.''  A conventional, destructive, spin-independent absorption image can then be taken, allowing the number of atoms in each of the magnetic sublevels to be determined.  Assuming that the inhomogeneous magnetic field that separates out the different spin components is ramped up and down gradually in time, as is typically the case, the measurement is performed in the basis of projections of the longitudinal field (the $\hat{F}_z$ eigenbasis).  But one can apply rf pulses before the separation/image is performed, rotating the spins so that the data can be interpreted as measuring the Zeeman populations along a rotated spin axis in the gas being analyzed.

Stern-Gerlach imaging has been a powerful probe of spinor Bose gases.  The spin distribution of spinor gases along one dimension of an elongated optical trap was determined by reversing (in software) the displacement between the different spin components, noting, for example that one component came predominantly from the top of the trap and the other predominantly from the bottom.  This analysis enabled the observation of between a few \cite{sten98spin} and very many \cite{mies99meta,kron10spont,book11} domains of different spin composition.  It enabled the observation of distinct spin-mixing resonances into different spin fluctuation modes of a trapped gas, as described in Sec.\ \ref{sec:spinmixing}.  Combined with the diffraction of atoms from an imposed optical lattice potential, it has allowed for the identification of states with coupled spin-orbital motion that are increasingly used for realizing effective magnetic fields \cite{lin09bfield} and magnetism in ``synthetic dimensions'' \cite{celi14synth,stuh15,manc15ribbon}.  Combined with absorption imaging with single-atom sensitivity, achieved by scattering very many photons off each single atom, Stern-Gerlach imaging has allowed for the detection of strong correlations between populations in different atomic levels, enforced by angular momentum conservation \cite{book11correlations} or parametric amplification \cite{haml12squeezing,gros11spinor,luck11squeeze}.

The limitation of Stern-Gerlach imaging is that it provides poor spatial resolution, and that it is a destructive imaging method.  Poor spatial resolution results from the separation of the different magnetic sublevels to resolvable locations in an image.  During that separation, the atoms expand out from their initial position and blur the image.  Once pulled apart, the gas cannot be put back together again, and so the image is destructive in that it can only be taken once.  Thus, if one wanted to measure several projections of the atomic spin, or to measure the evolution of an atomic gas in a single experiment by taking several images spaced in time, one would have to turn to different imaging techniques.  These limitations are not fundamental -- it may be possible to freeze the relative motion of atoms in a single spin state during the Stern-Gerlach separation into magnetic sublevels, and it is certainly possible to dribble out small fractions of a trapped atomic gas for Stern-Gerlach imaging while leaving the majority of atoms in the trap to continue their evolution and to be measured again later on.

\subsection{Dispersive birefringent imaging}

Rather than applying spin-dependent forces to the spinor gas, we can detect the spin state distribution of the trapped gas by discerning differences in how those spin states interact with light.  The topic of spin-dependent light-atom interactions is covered in other sources \cite{budk02rmp,sute05book,stam13rmp}, so we will be brief.

We recall that the effects of light-atom interactions are quantified by matrix elements of the interaction Hamiltonian between ground and excited states --  e.g.\ $- \mathbf{\mathcal{E}} \cdot \mathbf{d}_{e,g}$ for electric dipole transitions with $\mathbf{\mathcal{E}}$ being the electric field of the light and $\mathbf{d}_{e,g} = \langle e | \hat{\mathbf{d}} | g\rangle$ being the matrix element of the electric dipole moment operator between the ground ($|g\rangle$) and excited ($|e\rangle$) states -- and also by transition half-linewidths $\gamma$, and frequency differences $\delta_{e,g}$ between the light applied to the atoms and their resonance frequencies.  For example, the linear absorption cross section for absorbing light of polarization $\boldsymbol{\epsilon}$ on the $|g\rangle \rightarrow |e\rangle$ transition is $\sigma \propto \left|\boldsymbol{\epsilon} \cdot \mathbf{d}_{e,g}\right|^2 / (\delta_{e,g}^2 + \gamma^2)$ and the dispersive phase shift imparted on light passing the atom is proportional to $- \left|\boldsymbol{\epsilon} \cdot \mathbf{d}_{e,g}\right|^2 \delta_{e,g}/ (\delta_{e,g}^2 + \gamma^2)$.

In both of these expressions, coupling to the spin arises through the dot product $\boldsymbol{\epsilon} \cdot \mathbf{d}_{e,g}$ and through the detunings $\delta_{e,g}$.  In this subsection, we discuss imaging methods that make use of the former spin dependence -- a dependence on the relative orientation of the optical polarization and the atomic dipole moment that leads to dichroism and birefringence.  In the next subsection, we discuss an imaging method that takes advantage the latter spin dependent term, by driving resonant transitions at different frequencies in a spin-selective manner.  The frequency selectivity in that method is achieved on narrow microwave-frequency resonances for magnetic dipole transitions rather than broad optical-frequency resonances for electric dipole transitions.

\subsubsection{Circular Birefringent Imaging}

\begin{figure}[t]
\begin{center}
\includegraphics[width=5.5in]{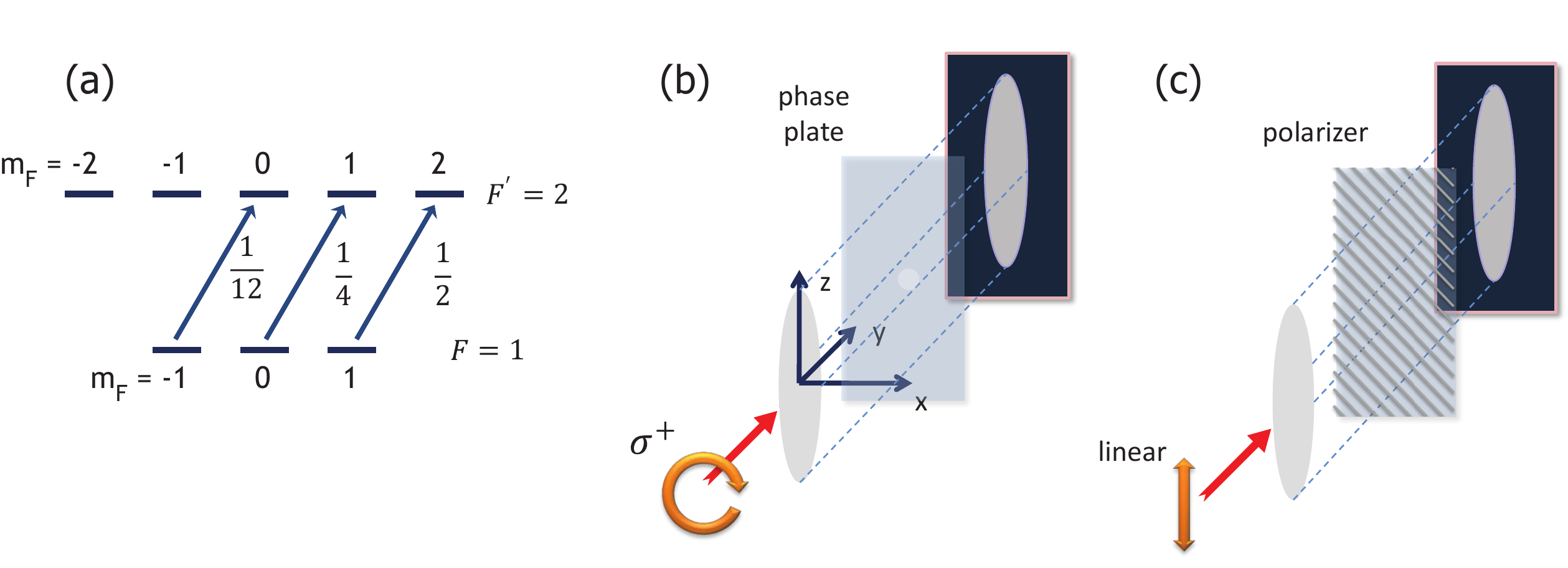}
\end{center}
\caption{Circular birefringence imaging of an $F=1$ spinor gas.  (a) The relative oscillator strengths for circular polarized light driving an optical transition from the $F=1$ ground state to an $F^\prime = 2$ excited state are shown.  Atoms polarized in the $|m_F=+1\rangle$ sublevel, quantized along the direction of the optical helicity, interact more strongly with the light.  A dispersive image taken with light near this transition therefore contains information on the magnetization of the gas along that direction.   The dispersive signal is turned into an image in one of two standard ways.  (b) One way is to convert the phase shift on a circularly polarized light field into an intensity image using a phase plate placed in the Fourier plane of the image, this being a form of phase contrast imaging \cite{hech89}.  (c) A second way is to illuminate the sample with linear polarized light.  The circular birefringence causes a rotation of the linear polarization, and this rotation is analyzed by passing the light through a linear polarizer before the camera.  Figure taken from Ref.\ \cite{stam15seeing}.}
\label{fig:birefimaging}
\end{figure}

A ``nondestructive'' direct imaging method has been developed that uses circular birefringence of an atomic gas to measure its magnetization \cite{higb05larmor}.  Circular birefringence is exemplified by an atom in a ground state with angular momentum $J\geq 1/2$ being driven near a transition to an excited state with angular momentum $J+1$ (here $J$ stands for some generic angular momentum operator).  Light with $\sigma^+$ circular polarization drives transitions from the $|g\rangle = |J, m_J\rangle$ ground state to the $|e\rangle = |J^\prime, m_J^\prime = m_J + 1\rangle$ excited state.  We assume that the energy of the states $|g\rangle$ and $|e\rangle$ are independent of the values of $m_J$ and $m_J^\prime$.  One finds that the dispersive phase shift produced by this transition, proportional to $|\boldsymbol{\epsilon} \cdot \mathbf{d}_{e,g}|^2$, increases monotonically with $m_J$ (the particular example of a transition from the $F=1$ ground state to the $F^\prime = 2$ excited state is illustrated in Fig.\ \ref{fig:birefimaging}a).  In other words, an atom with its spin oriented with the helicity of the light will experience stronger optical interactions than an atom with its spin oriented against the optical helicity.  The phase shift imprinted on circular polarized light that passes through the atomic gas   therefore carries information on the projection of the atomic spin along the direction of light propagation.

\begin{figure}[t]
\begin{center}
\includegraphics[width=4in]{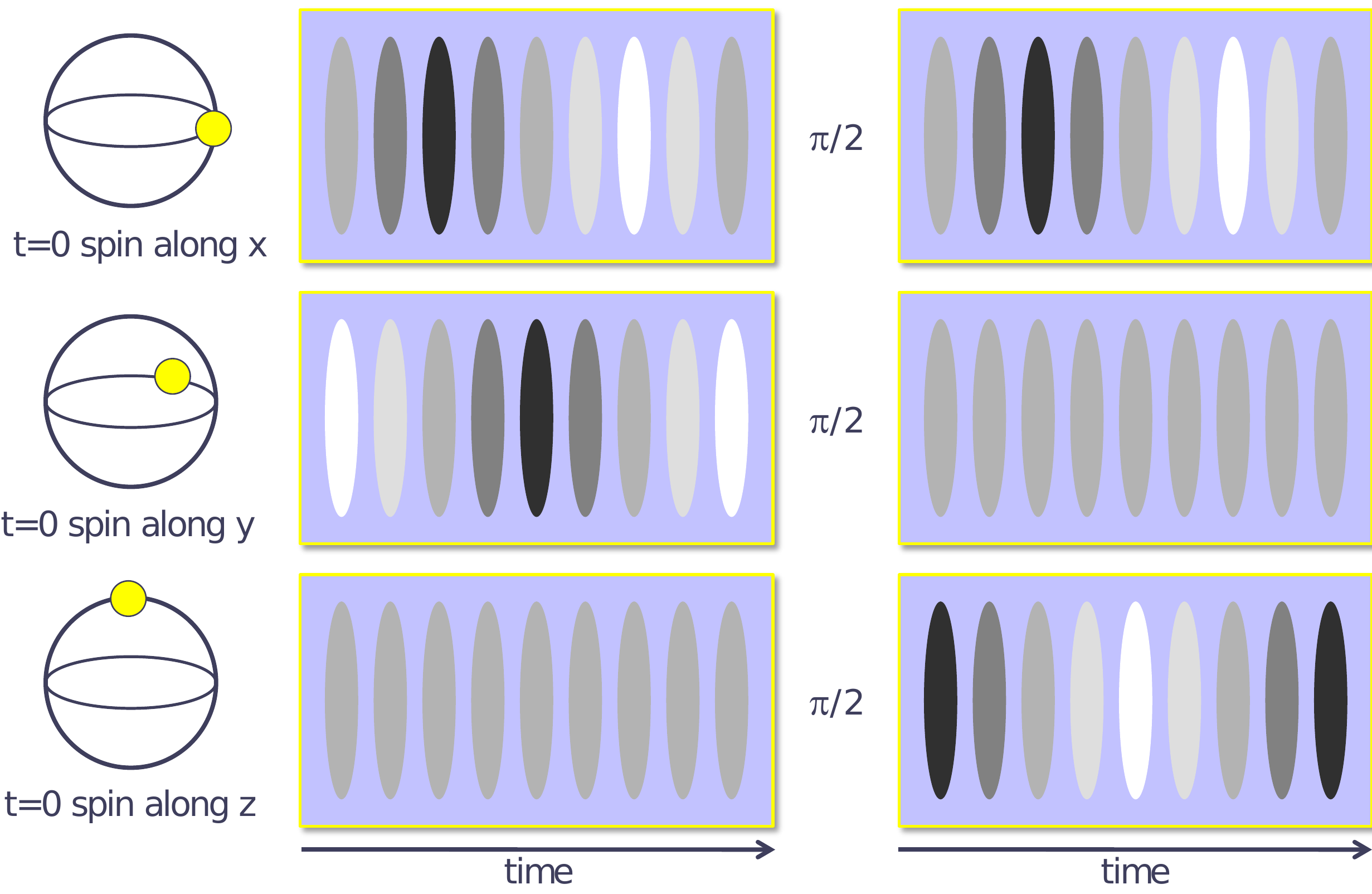}
\end{center}
\caption{All three components of the vector magnetization can be quantified by taking a sequence of imaging frames while the gas magnetization precesses in an applied magnetic field, and then is re-oriented using rf pulses.  Shown is the expected image frame sequence for a gas initially magnetized along each of the cardinal directions.  The imaging axis is $\mathbf{y}$, as indicated in Fig.\ \ref{fig:birefimaging}.  The first set of panels shows the expected signal during nine equally spaced image frames taken during Larmor precession.  The image signal from transversely magnetized sample shows a temporal oscillation, the phase of which determines the direction of the magnetization within the transverse plane.  Following a $\pi/2$ rf pulse, which causes the longitudinal magnetization to be rotated into the transverse plane, a second set of several image frames can be analyzed to quantify that transverse magnetization strength.  A two-dimensional image of the vector column magnetization is reconstructed from such an imaging sequence.  The entire sequence occurs within just a few ms, faster than the dynamical timescales for spin transport and mixing dynamics.  Figure taken from Ref.\ \cite{stam15seeing}}
\label{fig:lpimaging}
\end{figure}

Recall that the electric dipole operator directly modifies the spatial wavefunction of the electrons in an atom, but does not directly modify the electronic spin or nuclear spin.  Matrix elements $\mathbf{d}_{e,g}$, and thus the strength of optical interactions, can, therefore, be calculated by considering matrix elements of $\hat{\mathbf{d}}$ between the spatial parts of the atomic wavefunction.  For a high-spin atom, the ground state often has a non-zero orbital angular momentum $L$.  Thus, the matrix elements of the dipole operator connecting the ground state to an isolated excited state already clearly show the effects of circular birefringence.

In contrast, for an alkali atom, $L=0$ in the electronic ground state.  Circular birefringence in this case results only due to the spin-orbit interaction of the excited state.  If the fine-structure splitting of the excited state is very small compared to the detuning $\delta_{e,g}$ of probe light from the excited state, then the atom-light interactions can be considered as occurring simply between an $L=0$ ground state and an $L^\prime = 1$ excited state, and, therefore, there would be no circular birefringence.  If the probe is brought closer to atomic resonance -- detuned by less than the fine structure splitting \footnote{In principle one could apply a very large static field that would separate the excited state energies by an amount larger than the fine structure splitting, and then obtain circular birefringence at correspondingly larger detunings from resonance.} -- then the probe ``sees'' the excited state hyperfine structure.  The transition then effectively takes place between a $J=1/2$ ground state and a $J'=1/2$ or $J'=3/2$ excited state (whichever is closer to resonance), and one obtains circular birefringence. In ${}^{87}$Rb, the $J=1/2\to J'=1/2$ and $J=1/2\to J'=3/2$ optical transitions are separated in frequency by $7\;\mathrm{THz}$, much greater than the $6\;\mathrm{MHz}$ linewidths.  So it is possible to use the dispersive circular birefringence of $^{87}$Rb atoms with light that is still detuned enough from the atomic transitions to avoid absorption.

On tuning even closer to resonance, so that one can resolve also the excited state hyperfine structure, the optical interactions can be regarded as being those between a ground state with hyperfine spin $F$ and an excited state with specific hyperfine spin $F^\prime$.  In this case, with $F\geq 1$, the atom shows also linear birefringence, which reflects the alignment between the atomic spin quadrupole moment and the polarization vector of the optical field.  In practice, one typically uses probe light that is rather far detuned from the excited states so as to avoid undue spontaneous emission, suppressing the linear birefringence.  In contrast, for the high-spin atoms, such linear birefringence can be retained even at large detuning from resonance.  As such, dispersive imaging will likely be a very powerful tool for characterizing spinor Bose gases of high-spin atoms.

The application of circular birefringence imaging to an $F=1$ spinor gas is illustrated in Fig.\ \ref{fig:birefimaging}.  Light propagating along the $\mathbf{y}$ axis passes through a spinor gas.  Ignoring linear birefringence, this probe light acquires a phase shift $\phi_{\pm} \propto a \tilde{n} \pm b \tilde{M}_y$ that depends on whether the light has polarization $\sigma^+$ or $\sigma^-$.  Here $\tilde{n}$ is the column number density and $\tilde{M}_y$ the projection of the column magnetization along the probe axis ${\mathbf y}$.  The constants $a$ and $b$ derive from Clebsch-Gordan coefficients and the detuning between the probe light and nearby atomic resonances.  To convert this optical phase shift into a quantitative image, two different methods were used: phase-contrast imaging using probe light with definite circular polarization \cite{higb05larmor} (Fig.\ \ref{fig:birefimaging}b), or polarization-contrast imaging using linear probe light and a linear polarizer that quantified the optical rotation induced by the atomic birefringence \cite{guzm11} (Fig.\ \ref{fig:birefimaging}c).

A single image taken of the atomic gas by either of the above techniques provides information on one of the projections of the atomic magnetization.  What about the other two projections?  To gather information about them, a sequence of images was taken of the same atomic gas while the magnetization rotated dynamically owing to the application of uniform magnetic fields and brief rf pulses.  Taking advantage of the ``time at which the image is taken'' degree of freedom, these additional images were synchronized so that different projections of the spinor-gas column magnetization $\tilde{\mathbf{M}}(t_i)$, evaluated at the time $t_i$ at which the image sequence began, were brought into view.  For example, Fig.\ \ref{fig:lpimaging} illustrates how the combination of Larmor precession and the application of a single $\pi/2$ spin-rotation pulse allows one to obtain clear signatures of each of the magnetization vector components from a sequence of repeated dispersive images \cite{higb05larmor,sadl06symm,veng07mag}.  In later applications of the method, a sequence of $\pi/2$ spin-rotation and $\pi$ spin-echo pulses was used to make the imaging method more reliable and to reduce the number of images needed to record the vector magnetization \cite{guzm11}.

\dmskedit{\subsubsection{Limitations}
When imaging magnetization, the signal depends on the differences in polarizability between internal states (or, equivalently, the change of polarizability with optical polarization). A stricter practical limitation is placed by superradiant scattering. Special precautions must be taken to minimize the build up of superradiance, including (1) orienting the linear polarization of the laser along the longest condensate axis to suppress Rayleigh scattering, (2) detuning to the blue of resonance to suppress Rayleigh superradiance, and (3) applying a simultaneous pulse resonant to the D2, $F=2 \to F'=3$ transition to dephase atoms Raman scattered to $F=2$.}{}

\subsection{Projective imaging\label{sec:projective_imaging}}
While dispersive imaging is a powerful tool with broad applicability to characterizing the spin composition of Bose and Fermi gases of many different elements, it does have some limitations.  For one, while billed as ``nondestructive,'' dispersive imaging is necessarily somewhat destructive.  As we discuss below (Sec.~\ref{sec:noise_imaging}), there is always some residual optical absorption, no matter how far one detunes the probe light from atomic resonance (except, perhaps, at extremely low optical frequencies where probing is impractical), and this residual light scattering disturbs the gas being probed.

Second, dispersive birefringence, in the linear regime of light atom interactions, provides information only on the density, spin-vector and spin-quadrupole moments of the gas. This is because dispersive birefringence encodes information of the atomic spin onto the photon polarization, a spin-1 object. For atoms with spin $F>1$, it is not possible to use dispersive birefringence to characterize the spin state completely.  For alkali atoms, the linear birefringence that provides sensitivity to the spin-quadrupole moments is weak and cannot be generally measured with a good signal to noise ratio. For this reason, only the spin-vector moments are typically measured.

%\commentforDan{The reasons given here for improved jumping are rather technical and really jump ahead. ``Such imaging therefore cannot even characterize the spin order of $F=1$ alkali gases; for example, $F=1$ polar states with different alignment vectors, which emerge within the cores of polar-core vortices of ferromagnetic spinor gases \cite{sadl06symm} and within spin textures of antiferromagnetic gases \cite{choi12skyrmion} cannot be clearly distinguished by circular birefringence alone.''}

%The populations and coherences can be measured for different experimental realizations of identical ensembles with Stern-Gerlach separation, even to the single atom level. However, it can be important to simultaneously measure the populations and coherences, especially in chaotic systems (such as quenches) where each experimental realization produces substantially different final distributions. Furthermore, during the time needed to spatially separate the spin states, spatial information may be lost as diffraction blurs the signal. We solve these problems with a related technique: ASSISI, absorptive spin-sensitive \textit{in-situ} imaging. Discussion on measuring spin along multiple axes is in Sec.~\ref{sec:multiaxis}.

In regard to this second limitation, we see that Stern-Gerlach imaging has some capabilities that circular birefringence imaging lacks.  Specifically, consider three different polar states of an $F=1$ gas, aligned along either of the three cardinal directions $\mathbf{x}$, $\mathbf{y}$ or $\mathbf{z}$.  These are described by the spin wavefunctions
\begin{align}
|\zeta_x\rangle = |m_x\!=\!0\rangle =\!\begin{pmatrix}\nicefrac{1}{\sqrt{2}} \\ 0 \\ \nicefrac{-1}{\sqrt{2}} \end{pmatrix} &
\qquad |\zeta_y\rangle = |m_y\!=\!0\rangle =\!\begin{pmatrix}\nicefrac{i}{\sqrt{2}} \\ 0 \\ \nicefrac{-i}{\sqrt{2}} \end{pmatrix}  &
\qquad |\zeta_z\rangle = |m_z\!=\!0\rangle =\!\begin{pmatrix}0\\1\\0\end{pmatrix}
\label{eq:cartesian}
\end{align}
where we express the spinors as column vectors in the $\hat{F}_z$ eigenbasis.  All three states have zero magnetization ($\langle \mathbf{F} \rangle = \mathbf{0}$), and will thus appear identical in a circular birefringence image.  In contrast, Stern-Gerlach imaging performed along the $\mathbf{z}$ quantization axis will distinguish the longitudinal polar states $|\zeta_z\rangle$ from the transverse polar states $|\zeta_x\rangle$ and $|\zeta_y\rangle$: for the former, all the atomic population appears in the $|m_z = 0\rangle$ image, while for the latter, all the population appears in the $|m_z = \pm 1\rangle$ images.

Such Stern-Gerlach analysis is, at this point, unable to discern the two transversely aligned polar states, $|\zeta_x\rangle$ and $|\zeta_y\rangle$; the difference between these states reflects the phase of the $\Delta m_F = 2$ coherence between the $|m_z = \pm 1\rangle$ Zeeman sublevels. Detecting this phase coherence requires some sort of interference experiment. Such interference is achieved simply by rotating the atomic spin. For example, a $\pi/2$ spin rotation coherently transfers some of the $|m_z=+1\rangle$ and $|m_z=-1\rangle$ components to the $|m_z=0\rangle$ state, where they interfere.  A Stern-Gerlach analysis of the gas after this spin rotation measures the $|m_z=0\rangle$ population and thereby reads out the interference between the initially occupied states.  This combination of rotation and Stern-Gerlach analysis along $\mathbf{z}$ is equivalent to a Stern-Gerlach analysis along a different axis. If the equivalent projection axis is $\mathbf{x}$, then we obtain a unique signature of the $|\zeta_x\rangle$ state, and similarly for $\mathbf{y}$ and the $|\zeta_y\rangle$ state.

Can one utilize the power of Stern-Gerlach analysis to discern a greater variety of spin states than  possible with dispersive birefringence imaging while maintaining the power of dispersive imaging \emph{vis a vis} measuring spin distributions repeatedly, along several spin axes, with high spatial resolution and minimal destruction of the atomic gas under measurement?  The answer, at least for $F=1$ gases of $^{87}$Rb, is yes, as detailed in the rest of this section.

\subsubsection{Absorptive spin-sensitive \textit{in-situ} imaging (ASSISI)\label{sec:assisi}}

\begin{figure}[t]	
\centering
\includegraphics[scale=0.8]{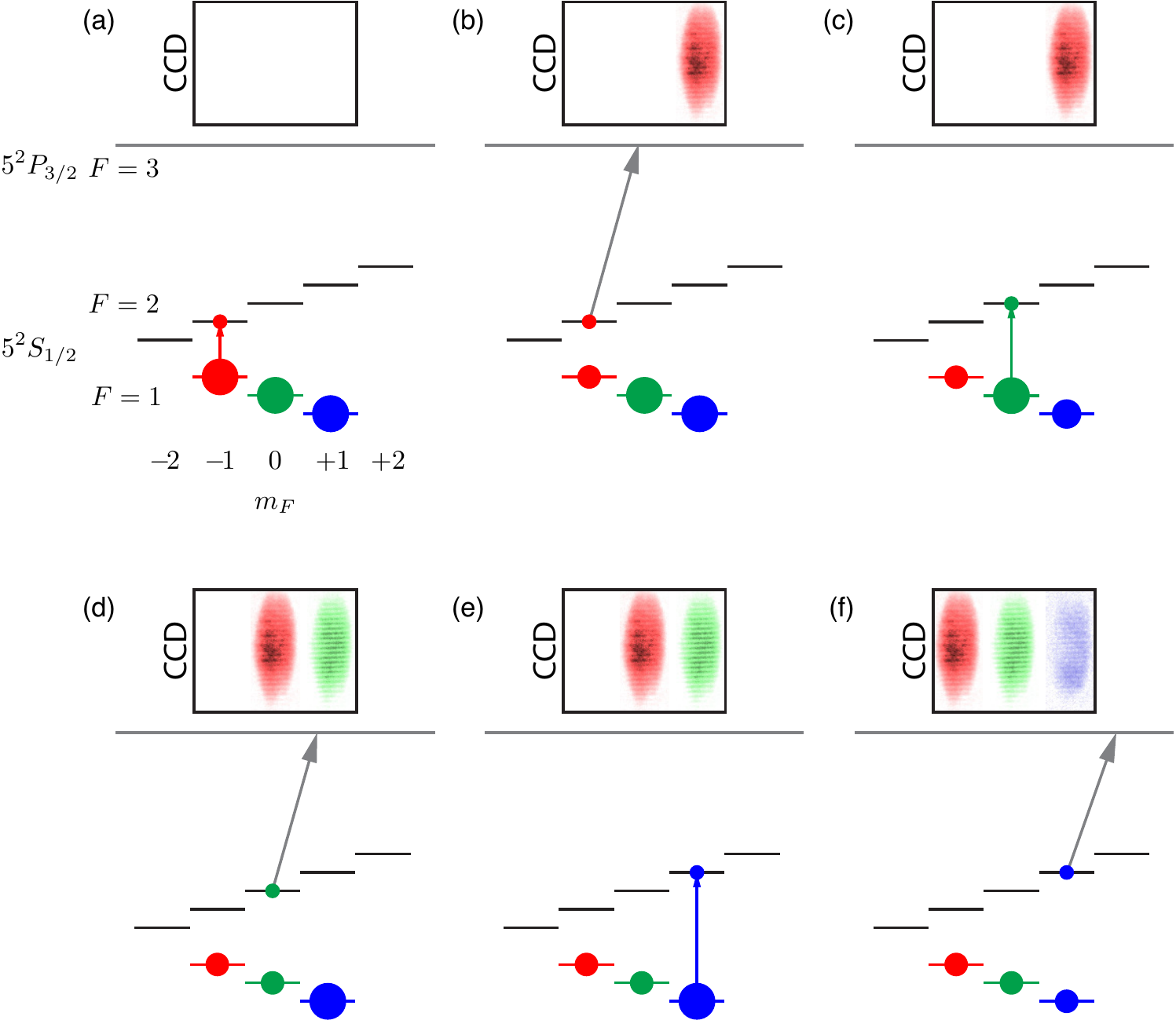}
\caption[ASSISI scheme]{ASSISI scheme (steps a through f) to image three projections of $F_z$ with interleaved microwave and optical pulses. First, a calibrated pulse resonant with the transition $|F=1,m_F=-1\rangle\to|F=2,m_F=-1\rangle$ (red circle) coherently transfers a small fraction of atoms to the $F=2$ level. Second, an imaging pulse destructively images the transferred atoms (red image, top) and pushes them out of the trap. The images are rapidly acquired using a CCD camera in frame transfer mode (top), where an image is stored as photoelectrons that are shifted across the CCD each time a new image is taken. The process is repeated for the $m_F=0$ (green) and $m_F=+1$ (blue) components of the same spinor condensate. Projections of the magnetization along other axes can be measured by following this sequence with an rf pulse and repeating the procedure.\label{fig:imaging-scheme}}
\end{figure}

ASSISI circumvents the \dmskedit{inherent}{} limitations of dispersive imaging by employing a three-level scheme to image the $F=1$ spin density (Fig.~\ref{fig:imaging-scheme}). In an $F=1$ ${}^{87}$Rb gas, this is accomplished by ``shelving'' atoms in the empty $F=2$ hyperfine manifold. First, a brief microwave pulse transfers a small number of atoms from one spin state in the $F=1$ manifold to the $F=2$ manifold. A weak magnetic field is sufficient to separate microwave transitions from different initial $m_F$ states spectroscopically. Then, the atoms in the $F=2$ states are imaged with a short, intense pulse of resonant imaging light on the D2, $F=2\to F'=3$ transition, to which the $F=1$ gas is dark owing to the large detuning of the probe light from the $F=1$ absorption lines.  In the Berkeley setup, each imaged atom typically scatters around $300$ photons, achieving a high signal-to-noise ratio by destructively imaging a small fraction of the sample (see Sec.~\ref{sec:noise_imaging} for a comparative discussion of imaging noise).

\begin{figure}[t]
\centering
\begin{subfigure}[t]{.2\textwidth}
	\centering
	\includegraphics[scale=0.667]{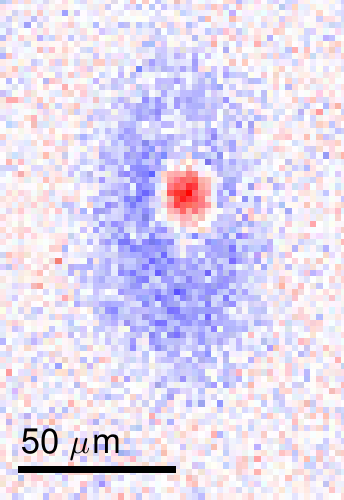}
	\captionsetup{skip=3pt}
	\caption{\label{fig:polz_contrast_skyrmion}}
\end{subfigure}
\hspace{5em}
\begin{subfigure}[t]{.433\textwidth}
	\centering
	\includegraphics[scale=0.667]{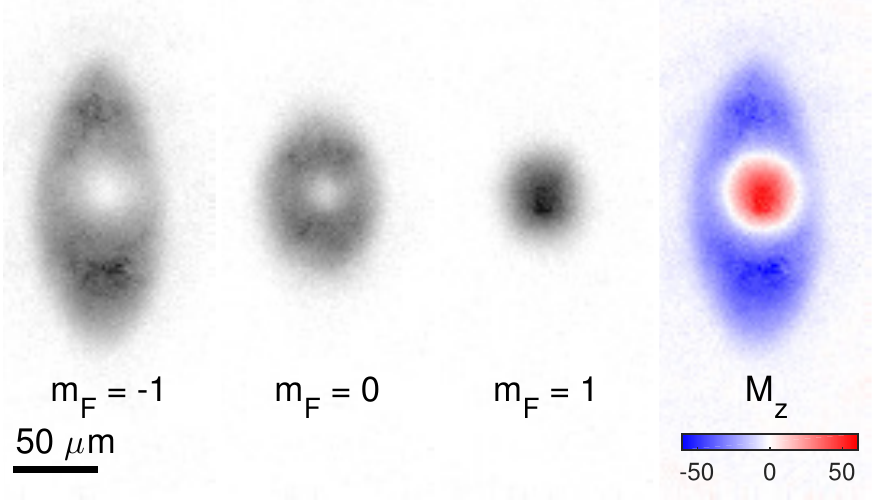}
	\captionsetup{skip=3pt}
	\caption{\label{fig:ferro_skyrmion_assisi}}
\end{subfigure}
\caption[Polarization-contrast and ASSISI]{(a) Polarization-contrast and (b) absorptive spin-selective \emph{in situ} images of a ferromagnetic skyrmion-like texture.  The dimensionless column magnetization image $\tilde{M}_z$ is the difference between the $m_F=1$ and $m_F=-1$ column density images. The color bar for the rightmost image in (b)  has units of $\mu\mbox{m}^{-2}$. \label{fig:ferro_skyrmion}}
\end{figure}

Spin distributions imaged with either circular birefringent (polarization contrast) imaging or ASSISI are shown in Fig.~\ref{fig:ferro_skyrmion}. The polarization contrast image (Fig.~\ref{fig:polz_contrast_skyrmion}) is taken with a light fluence  sufficient to destroy the sample. Using ASSISI (Fig.~\ref{fig:ferro_skyrmion_assisi}), we image only a small fraction (${<}10\%$) of the sample and yet have an enormous signal to noise ratio -- these images have no digital smoothing, binning, or filtering. The three image frames correspond to three repeated images taken of the same atomic gas after different microwave pulses on the $|F=1, m_F\rangle$ to $|F=2, m_F\rangle$ transition for the three difference $m_F$ sublevels, respectively.  From these three images, we reconstruct the Zeeman state population of the trapped gas with excellent spatial resolution.  The fact that three high quality image frames are obtained in succession also attests to the nondestructive nature of the imaging method.  That is, in spite of robust light scattering by the atoms sent by the microwave pulse into the $F=2$ hyperfine level, which causes those imaged atoms to be ejected from the optical trap, the remaining atoms in the $F=1$ levels remain trapped and relatively undisturbed.  The fraction of atoms selected for imaging by the microwave pulse is chosen so that the optical density of the $F=2$ atoms being imaged remains small, below unity. Imaging a larger fraction of the spins is detrimental as the signal-to-noise ratio decreases with high optical density. Absorption imaging is done with an imaging pulse at the saturation intensity, which gives optimal signal-to-noise ratios when the optical density of the transferred population is small.

\subsubsection{Noise in dispersive imaging and ASSISI\label{sec:noise_imaging}}

As we have stated, neither dispersive imaging nor ASSISI are truly nondestructive methods.  In both cases, some atoms in the gas are sacrificed for the sake of obtaining information about the gas.  Let us compare the noise limits of the two methods.  In particular, we ask what are the limits of each of the methods for determining the number of atoms $N_\mathrm{a}$ in the gas that occupy one specific Zeeman level and that occupy an area $A$ within the image.

One source of the destructiveness of dispersive imaging is the incoherent scattering of light by the atoms being imaged.  Suppose we image the gas with coherent probe light that is uniform in intensity, and that has a fluence of $N_\mathrm{p}$ photons with the imaged area $A$.  For simplicity, we treat the atoms as two-level atoms with a resonant cross section $\sigma_0$, a natural transition linewidth $\gamma$, and consider that the probe light is detuned by a frequency $\delta$ from the atomic resonance where $|\delta| \gg \gamma$.  In passing through the atomic gas, light acquires a dispersive phase shift of $\phi = - N_\mathrm{a} \sigma_0 \gamma / (2 A \delta)$ where we assume the atoms are distributed uniformly.  A measurement of this phase shift has an inherent imprecision stemming from optical shot noise, at the level $\Delta \phi = N_\mathrm{p}^{-1/2}$.  This shot-noise imprecision translates into an imprecision in the measurement of the atom number.

This measurement imprecision decreases with increasing photon fluence, suggesting that it is always advantageous to increase the probe fluence to obtain more precise measurements.  However, increasing the photon fluence increases also the destructiveness of the image.  In an ultracold atomic gas, a single photon-scattering event is sufficient to excite an atom to an energy that is very large compared to the inherent energy scales (notably the temperature) of the gas.  At best, an atom that scatters a photon will be removed from the trap without depositing its recoil energy within the gas.  The number of atoms that scatter photons and are then lost from the gas is given as $f_\mathrm{disp} N_\mathrm{a}$.  For dispersive imaging to have any semblance of ``nondestructiveness,'' this fraction of atoms that are lost in imaging, $f_\mathrm{disp} = N_\mathrm{p} \, \gamma^2 \sigma_0 / (4 \delta^2 A)$, should be kept small, and this requirement sets a limit on the number of photons $N_\mathrm{p}$ used for imaging.  We obtain
\begin{equation}
\left(\Delta N_\mathrm{a}\right)_\mathrm{disp} = \sqrt\frac{A}{\sigma_0 f_\mathrm{disp}} \label{eq:disp}
\end{equation}
for the atom-counting imprecision for dispersive imaging.

We recall that the imaged area $A$ will be at least several times the cross-section $\sigma_0$ owing to the optical diffraction limit.  Along with the nondestructiveness requirement $f_\mathrm{disp} \ll 1$, the atom-counting imprecision is much larger than one.  Nevertheless, at sufficiently high optical density $D = N_\mathrm{a} \sigma_0 / A$, it is possible to achieve dispersive imaging with imprecision below the atomic shot-noise limit, i.e.\
\begin{align}
1 &\ll \left(\Delta N_\mathrm{a}\right)_\mathrm{disp} = N_\mathrm{a}^{1/2} \sqrt{\frac{1}{D f_\mathrm{disp}}} < N_\mathrm{a}^{1/2} & \mbox{for\, } D > f_\mathrm{disp}^{-1}.
\end{align}
It is interesting to consider using such sub-atom-shot-noise imaging to squeeze the atomic spin distribution, and then to track the evolution of this conditionally squeezed distribution over time and space by performing repeated images on the gas.

In comparison, in the ASSISI method, one uses a microwave pulse to select a fraction of atoms $f_\mathrm{abs}$ that are transferred to a different internal state and then detected by absorption imaging.  During such imaging, we are willing to use a very high photon fluence, so that each of the selected atoms scatters very many imaging photons, enough so that the number of selected atoms $N_\mathrm{s}$ can, in principle and in practice, be determined at the single-atom level (i.e.\ with absolute precision).  The copious light scattering also ensures that the selected atoms are all expelled from the trap, rather than their remaining trapped and depositing excess energy into the unselected fraction of the gas that remains trapped.

However, our task was to determine not the number of selected atoms, but rather the total number of atoms $N_\mathrm{a}$.  The imprecision in this determination comes from the Poisson fluctuations in the number of atoms selected, with the rms value $\Delta N_\mathrm{s} = N_\mathrm{s}^{1/2}$ when $f_\mathrm{abs}$ is small, as a result of which our measurement of $N_\mathrm{a}$ has an uncertainty
\begin{equation}
\left(\Delta N_\mathrm{a}\right)_\mathrm{abs} = \sqrt{\frac{N_\mathrm{a}}{f_\mathrm{abs}}} = \sqrt{D} \sqrt{\frac{A}{\sigma_0 f_\mathrm{abs}}}.
\end{equation}

Comparing the two results with the same fractions of atoms lost to imaging, we observe that dispersive imaging is generally advantageous for imaging gases with high optical densities, while ASSISI is preferable at low optical densities.  In practice, the noise limits to dispersive birefringent imaging are higher than those implied by Eq.\ \ref{eq:disp}.  For one, the variation of $\phi$ with the spin state of the atoms is small compared with $\phi$ itself, so the sensitivity the magnetization of the gas, rather than the density or atom number, is smaller.  Second, light absorption in high optical-density samples can be amplified by collective effects \cite{inou99super2,veng07mag}.  Third, atoms that scatter just a single photon will sometimes remain within the trap where they can cause greater destruction to the trapped gas, for example by knocking several atoms out of a Bose-Einstein condensate rather than just removing a single atom.  On the other hand, for large numbers of atoms $N_\mathrm{s}$, it is hard to achieve absorption detection with single-atom precision.  In practice, ASSISI appears to be a superior method for measuring spin structures within cold gases that have optical depths (along the imaging axis) below about 30.

\subsection{Spin-spin correlations and magnetic susceptibility}
A powerful application of these imaging techniques is observing correlations and fluctuations.  A system at thermal equilibrium is characterized by average physical properties and also by thermal fluctuations away from those average values. According to the fluctuation-dissipation theorem, fluctuations of measurable parameters (energy, density, spin) are related to linear response coefficients that dictate how a perturbed system relaxes to equilibrium.

Spin-sensitive imaging techniques allow one to measure the local magnetic order of a spinor Bose-Einstein gas.  Even in the alkali spinor gases, where the global longitudinal magnetization may be a fixed quantity, the magnetization will still vary locally at thermal equilibrium provided that the system size is large compared to the length scale over which fluctuations are correlated.  Perhaps the length defined by the spin-dependent mean-field energy (typically a few microns) or the thermal deBroglie wavelength (also typically on the order of a micron) provides such a length scale; both scales can be small compared to condensed spinor gases with dimensions of 10's to 100's of microns.  We can then treat well-separated regions of a sample as independent realizations of a thermodynamic experiment. The fluctuation-dissipation theorem predicts that the spin susceptibility $\chi = \partial M/\partial B$ is related to the root-mean-square magnetizations as $\delta M^2 = k_B T \chi$. Thus, local measurements of magnetization in a system at thermal equilibrium can characterize the magnetic susceptibility, a measurement that becomes particularly interesting near a magnetic phase transition. This type of measurement and analysis has been carried out in experiments on spin mixtures in ultracold Bose \cite{guzm11} and Fermi gases \cite{mein12}.

In Refs.\ \cite{veng10periodic,guzm11}, the vector magnetization was measured in $F=1$ $^{87}$Rb gases that were initially unmagnetized and were cooled slowly across the Bose-Einstein condensation phase transition.  Results of the latter of these experiments are shown in Fig.\ \ref{fig:rbequilibration}.  The images indeed show strong fluctuations of the magnetization, varying all the way from one region with a gas fully magnetized in one direction to another region where the gas is fully magnetized in the opposite direction.  So, qualitatively at least, these images indicate a very large magnetic susceptibility which is indeed what we would expect for a $F=1$ spinor Bose-Einstein condensate in the ferromagnetic phase. What prevents this analysis from being made quantitative is that the rubidium spinor gases in these experiments is found not to be at thermal equilibrium, owing perhaps to the very small spin-dependent interaction in that gas or to very slow coarsening dynamics in an effectively two-dimensional magnetic system \cite{barn11pretherm}.  In other spinor Bose-Einstein gases with stronger spin-dependent interactions, or perhaps in more three-dimensional settings, the method of characterizing thermodynamic phases by measuring local fluctuations may be more successful.

\subsection{Multi-axis imaging and topological invariants\label{sec:multiaxis}}
The above discussion focused on local, small scale thermal perturbations of an ordered material and their relation to thermodynamic susceptibilities.  Another class of excitations in an ordered system are topological excitations, which describe large-scale persistent deviations away from a uniform equilibrium state.  Such topological excitations arise in systems that are characterized by an order parameter that can occupy a continuous order parameter manifold.  Depending on the nature of this order parameter manifold, one identifies a group of topological excitations as a mapping of order parameter manifold onto one-, two-, or three- dimensional spheres  (this classification gives rise to the first, second, or third homotopy groups, respectively).

%\commentforDan{It is probably better to call them topological excitation to topological defect. Otherwise it's confusing to discuss defectless topological excitations.}

%\commentforDan{We should be careful to distinguish the order parameter space (not necessarily a group), the group of symmetries that take the order parameter from one value to another, and the homotopy groups. For a scalar BEC the first two look similar, since they are $S^1$ and $U(1)$, respectively, while the first homotopy group is $\mathbb{Z}$. The first homotopy group is mapping a circle onto the order parameter space. We should cite someone; maybe Sethna's book or Mermin?}

For example, a scalar Bose-Einstein condensate is characterized by a complex order parameter.  The condensate energy does not change if we multiply the order parameter (the condensate wavefunction) by a complex phase $e^{i \phi}$.  Thus, the degenerate order parameter space is a circle $S^1$, corresponding to the circle in the complex plane upon which the complex order parameter may lie without changing the system's energy. One type of topological excitation involves varying the condensate phase by $2 \pi$ along a closed path. It is topological in the sense that it is a defect that cannot be undone by smooth local variations of the order parameter. Equivalently, this topological excitation has a quantized topological charge: the phase accumulated by the condensate wavefunction along the closed-loop path (divided by $2 \pi$ so that charge is  an integer).

%\commentforDan{Charge is a little misleading, since the actual description can be much more complicated. Charge only makes sense for homotopy groups like $\mathbb{Z}$ or maybe $\mathbb{Z}_n$ where there is a sense of `how many times we went around' rather than the more accurate `how did we go around.' (Crystal defects in a 2D lattice have a first homotopy group $\mathbb{Z} \times \mathbb{Z}$, corresponding to defects along one of two axes.) I do see the validity of discussing charge since it's what we'll use later.}

How is such a topological excitation identified? In two or three dimensions, topological excitations given by the aforementioned procedure of mapping of a circle onto the order parameter space give a visible local ``glitch'' in the ordered system.  For example, the topological excitation of a scalar Bose-Einstein condensate generates a localized defect in the superfluid -- the vortex line, within which the superfluid order parameter is forced to zero.  This vortex line has been experimentally measured in a variety of systems. In cold atom experiments, the superfluid can be allowed to expand by releasing it from its trap.  During this expansion, the vortex line can grow in thickness to where it is resolvable by conventional optical imaging.  In experiments on superfluid helium, vortex cores were identified by decorating the cores with detectable impurities, such as electrons \cite{sand69,will74} and hydrogen clusters \cite{bewl06}.

Alternately, one can attempt to map the order parameter spatially and then integrate up the relevant variation in the order parameter along a surface (of variable dimension) in order to quantify, directly, the possible non-zero topological charge.  In the case of scalar Bose-Einstein condensates, such measurements have been performed by using interference between two Bose-Einstein condensates to detect the $2\pi$ phase winding around a closed path that might exist owing the presence of a vortex in one of the condensates \cite{matt99vort,inou01singularity}. This is particularly valuable for ``defectless'' topological excitations that do not have a local ``glitch''.

Here, we discuss how spin-sensitive imaging allows one to identify topological properties of magnetic spin textures in spinor Bose-Einstein condensates.  We focus on the task of measuring the differential curvature in a two-dimensional gas with spatially varying magnetization vector $\langle \mathbf{F} \rangle$, describing first how the local value of $\langle \mathbf{F} \rangle$ is measured using the ASSISI technique, and, second, how a spatial map of the magnetization is analyzed to determine the local differential curvature and also its integral over the entire spinor gas.  We apply these techniques to a spin texture created within a ferromagnetic $F=1$ $^{87}$Rb spinor condensate, a structure studied earlier within a sodium spinor gas in Ref.\ \cite{choi12njp}.

%\commentforDan{This is not why the spin texture is topologically nontrivial. My understanding is that it's because the boundary (at infinity) has circulation (though I'm still confused as to how this argument compares to a scalar vortex). See more comments below. I would put off the discussion until later. ``As we point out at the end of our discussion, this spin texture is \emph{not} a topological excitation of the ferromagnetic $F=1$ spinor gas, because the integrated magnetization curvature, which is detected by our imaging method, turns out to be offset by an equal valued phase winding of the condensate phase, which is not detected in our images.  Nevertheless, the spin texture provides a nice test target for our analysis method.''}

\subsubsection{Multi-axis imaging of ferromagnetic structures}
To image the column magnetization of an $F=1$ $^{87}$Rb spinor gas, we apply the ASSISI method described in Sec.\ \ref{sec:assisi} three times to the same gas.  We begin as before by taking a set of three sequential image frames which measure the number column densities $\tilde{n}_{m_z}$ in each of the three magnetic sublevels $|m_z = \{1, 0, -1\}\rangle$, where, as before, the subscript indicates that the quantization axis for these magnetic sublevels is the $\mathbf{z}$ axis.  We then repeat the image sequence two more times.  Before each repetition, we apply an rf pulsed magnetic field to the gas, which effects a spatially uniform $\pi/2$ rotation of the magnetization.  If the $\pi/2$ rotations are perfectly controlled, then the second set of three image frames measures the column densities in the states $|m_x = \{1, 0, -1\}\rangle$, while the third set of image frames measures the column densities in the states $|m_y = \{1, 0, -1\} \rangle$.  Altogether, we obtain nine image frames, as shown in Figs.\ \ref{fig:full_magnetization_vortex12} and \ref{fig:full_magnetization_helix2} for the characterization of two different spin textures.

The challenge in controlling the two $\pi/2$ pulses required in this imaging system is that the background magnetic field varies slightly between repetitions of our experiment.  While the first rf pulse will quite accurately rotate the spin vector by $\pi/2$, bringing \emph{some} component of the initial transverse magnetization onto the longitudinal axis along which we select populations with microwave pulses in the ASSISI technique, we find that it is difficult to control which axis in the initial $y$-$z$ spin plane is brought into view by the second $\pi/2$ pulse.  A similar challenge arises in the application of circular birefringence imaging, and our solution to this challenge to both imaging methods has been similar.  Namely, we either perform a statistical decorrelation to extract the $\mathbf{y}$-axis populations from images that contain some information on both the $\mathbf{y}$-axis and $\mathbf{z}$-axis populations, or else we intersperse the imaging sequence with spin-echo pulses to reduce sensitivity to magnetic field variations.

%\commentforDan{In this section, I decided to calculate the normalized column magnetization $\langle F_i \rangle = M_i/N$, with values ranging from $-1$ to $1$, rather than the column magnetization. Maybe $\tilde M_i$ is confusing, since it looks like the column integrated magnetization.}

The nine image frames obtained by this multi-axis imaging approach can be processed to obtain the column magnetization vector, with components
\begin{equation}
\tilde{M}_i(\mathbf{r}) = \tilde n_{m_i=+1}(\mathbf r) - \tilde n_{m_i=-1}(\mathbf r),
\end{equation}
or, relevant to the discussion below, the column-averaged atomic spin vector, $\langle \mathbf{F} \rangle$, by normalizing to the total atomic density as follows
\begin{equation}
\langle F_i (\mathbf r)\rangle = \frac{\tilde n_{m_i=+1}(\mathbf r) - \tilde n_{m_i=-1}(\mathbf r)}{\tilde n_{m_i=+1}(\mathbf r) + \tilde n_{m_i=0}(\mathbf r) + \tilde n_{m_i=-1}(\mathbf r)}.
\end{equation}

For cases where we believe the gas being imaged is an $F=1$ spinor Bose-Einstein condensate whose order parameter resides locally within the ferromagnetic superfluid order parameter space, i.e.\ a condensate in which the atoms at each location are fully magnetized along some spatially varying axis, the local condensate spin wavefunction can be taken as
\begin{equation}
|\zeta(\mathbf r) \rangle = R(\phi(\mathbf r), \theta(\mathbf r), \gamma(\mathbf r)) \left(
\begin{array}{c} 1 \\[5 pt] 0 \\[5 pt] 0 \end{array} \right)
=  e^{-i \gamma(\mathbf r)} \left(
\begin{array}{c}
e^{-i \phi(\mathbf r)} \frac{1}{2}(1 + \cos \theta(\mathbf r)) \\[5 pt]
\frac{1}{\sqrt{2}} \sin \theta(\mathbf r) \\[5 pt]
e^{i \phi(\mathbf r)} \frac{1}{2}(1 - \cos \theta(\mathbf r))
\end{array} \right) \label{eqn:wignerD}
\end{equation}
where $\theta(\mathbf r)$, $\phi(\mathbf r)$, and $\gamma(\mathbf r)$ are locally defined Euler angles, and the wavefunction is written in the $\hat{F}_z$ eigenbasis.  We note that the probability densities in the $\hat{F}_z$ eigenbasis are entirely determined by $\theta$; thus, $\theta$ can be surmised from just the $\mathbf{z}$-axis ASSISI image frames (under the assumption that the gas is fully magnetized).  The $\Delta m_z = \pm 1$ coherences are determined by $\phi$; determining $\phi$ thus requires imaging data along the $\mathbf{x}$ and $\mathbf{y}$ axes as well.  Variations in the third angle $\gamma$ contribute excitations to the superfluid that are not seen by in ASSISI imaging. Now, the components of $\langle \mathbf F(\mathbf r) \rangle$ have the familiar form of a unit vector with polar angle $\theta$ and azimuthal angle $\phi$.
\begin{equation}
\langle F_x(\mathbf r) \rangle = \sin \theta \cos \phi \qquad \langle F_y(\mathbf r) \rangle = \sin \theta \sin \phi \qquad \langle F_z(\mathbf r) \rangle = \cos \theta
\end{equation}

%\dmskedit{The populations of the three components of $\psi_{m_F}$ are entirely quantified by $\theta$, while the coherences are quantified by $\phi$. Fig.~\ref{fig:skyrmion_transverse} demonstrates a sequence of six images determining $F_x$ and $F_z$. An example of measure spin projections along three axis is shown in Figs.~\ref{fig:full_magnetization_vortex12} and \ref{fig:full_magnetization_helix2}.}{}

%\dmskedit{\begin{figure}[t]
%	\centering
%	\includegraphics[scale=0.8]{figures/skyrmion_transverse}
%	\caption[Multi-axis magnetization imaging of a topological structure]{Multi-axis magnetization imaging of a topological structure. The first three images measure the longitudinal magnetization $M_z$, i.e. the populations $|\psi_{m_F}(\mathbf r)|^2$ of $m_F$ states along $\hat {\mathbf z}$, as in Fig.~\ref{fig:ferro_skyrmion_assisi}. We then apply an rf $\pi/2$ pulse to rotate the magnetization by $90^\circ$ and repeat the procedure. These images constitute the transverse magnetization $M_x$,  which are sensitive to coherences between $\psi_{m_F}$ states. The color bar represents the column magnetization in units of atoms per square micron. All images are from a single experimental realization of a spinor condensate.
%\label{fig:skyrmion_transverse}}
%\end{figure}
%}{}

\subsubsection{Magnetization curvature}
Images of the magnetization along three spatial axes allow us to extract topological parameters from the data, in this case the solid angle swept out by the magnetization. Topological invariants give us a means to solve yes or no questions. For a two-dimensional Heisenberg magnet (where the only degrees of freedom are the orientation of the magnetization), if the system has fixed uniform magnetization along a boundary, can that magnetization be continuously unwrapped to a uniform magnetization inside the region? The answer can be determined by integrating the solid angle swept out by the local magnetization. That is, around each infinitesimal square $dx\,dy$, the vector $\mathbf F$ sweeps out a solid angle $d \Omega = K \, dx\, dy$ (for the reminder of this section, we will only consider the mean value of $\mathbf F$). Integrating over the whole region gives the total curvature of the magnetization texture $\Omega$.
\begin{equation}
\Omega = \int K\, dx\,dy \qquad K = \mathbf F \cdot \left( \frac{\partial \mathbf F}{\partial x} \times \frac{\partial \mathbf F}{\partial y} \right)\label{eqn:curvature}
\end{equation}
Given the boundary condition specified above, the curvature must be a multiple of $4\pi$, the solid angle of a sphere. Continuous deformations cannot let $\Omega$ jump from one value to the next. If $\Omega=0$, then the magnetization inside the region can be continuously unwrapped to a constant orientation (a trivial structure) unless a defect is introduced. Otherwise, the region is topologically robust.

In a solid-state ferromagnet, the two-dimensional topological structure over which $\Omega$ covers a non-zero integer multiple of $ 4 \pi$ is a ferromagnetic skyrmion.  Unlike for a ferromagnetic solid, in a magnetically ordered superfluid, the local curvature $K$ must be accompanied by superfluid flow (the $\gamma(\mathbf r)$ term in Eq.~\ref{eqn:wignerD}).  This skyrmion-like structure in a ferromagnetic superfluid is not classified at a topological excitation because of this flow. The integrand $K$ is often called the topological density, Pontryagin density \cite[sec. 1.19]{girv00qhlecture}, or Berry curvature.

\begin{figure}[t]
	\centering
	\includegraphics[scale=0.7]{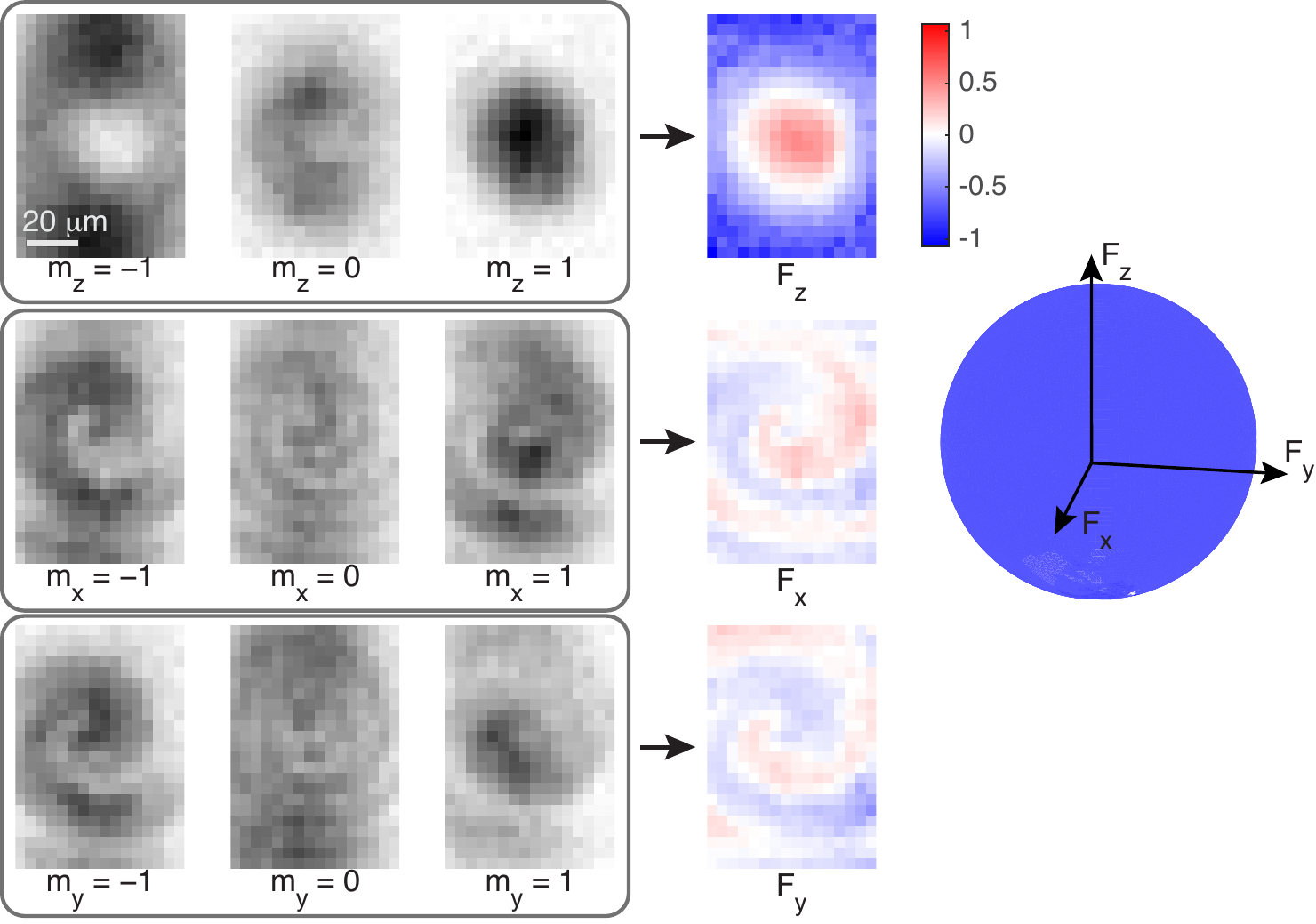}
	\caption[Full magnetization and topological imaging of a skyrmion-like texture.]{Vector magnetization of a skyrmion-like texture. Nine image frames of the $m_i = -1, 0, 1$ projections for $i = x, y, z$ are used to calculate the three estimates of the spin projection $F_i$.  (Right) The magnetization sphere. Each triplet $(F_x, F_y, F_z)$ is projected onto a sphere. Blue and red denote regions of the magnetization sphere covered by the images in clockwise and counter-clockwise directions, respectively. For this texture, nearly the entire sphere is covered in a clockwise fashion. \label{fig:full_magnetization_vortex12}}
\end{figure}

Experimentally, Eq.~\ref{eqn:curvature} is calculated on a square grid of imaged pixels with spacing $a$, a discretized form of $\mathbf r$. We split one square patch in the image defined by four neighboring pixels at locations $(x,y)$, $(x+a,y)$, $(x+a, y+a)$ and $(x, y+a)$ into a lower triangle with points $\mathbf r_1 = (x,y)$, $\mathbf r_2 = (x+a, y)$, and $\mathbf r_3 = (x, y+a)$, and an upper triangle with points $\mathbf r_1' = (x+a,y+a)$, $\mathbf r_2' = (x, y+a)$, and $\mathbf r_3'= (x+a, y)$. For each triangle, we calculate the solid angle swept out by the three measured values of $\mathbf F(\mathbf r_1)$, $\mathbf F(\mathbf r_2)$, and $\mathbf F(\mathbf r_3)$. It is important to use the exact formula for solid angle to avoid building up numerical errors \cite{oost83}:
\begin{equation}
\tan \frac{d\Omega}{2} = \frac{ \left| \mathbf F_1 \mathbf F_2 \mathbf F_3 \right| }
{ |\mathbf F_1| |\mathbf F_2| |\mathbf F_3| + \mathbf F_1 \cdot \mathbf F_2 \, |\mathbf F_3| + \mathbf F_2 \cdot \mathbf F_3 \, |\mathbf F_1| + \mathbf F_3 \cdot \mathbf F_1\, |\mathbf F_2| } \label{eq:domegaformula}
\end{equation}

Fig.~\ref{fig:full_magnetization_vortex12} shows the magnetization for a skyrmion-like structure. The nine image frames from the multi-axis ASSISI sequence are shown in black and white, along with the three estimates of the spin projection. For the structure imaged, the core is small and its finer structures are poorly resolved. Regardless, the measured $\Omega = -12.5$ is close to the expected value of $-4\pi$. $\Omega$ is not exactly a multiple of $4\pi$ because the boundary is not uniform.

% Great stuff: 20130311

\begin{figure}[t]
	\centering
	\includegraphics[scale=0.7]{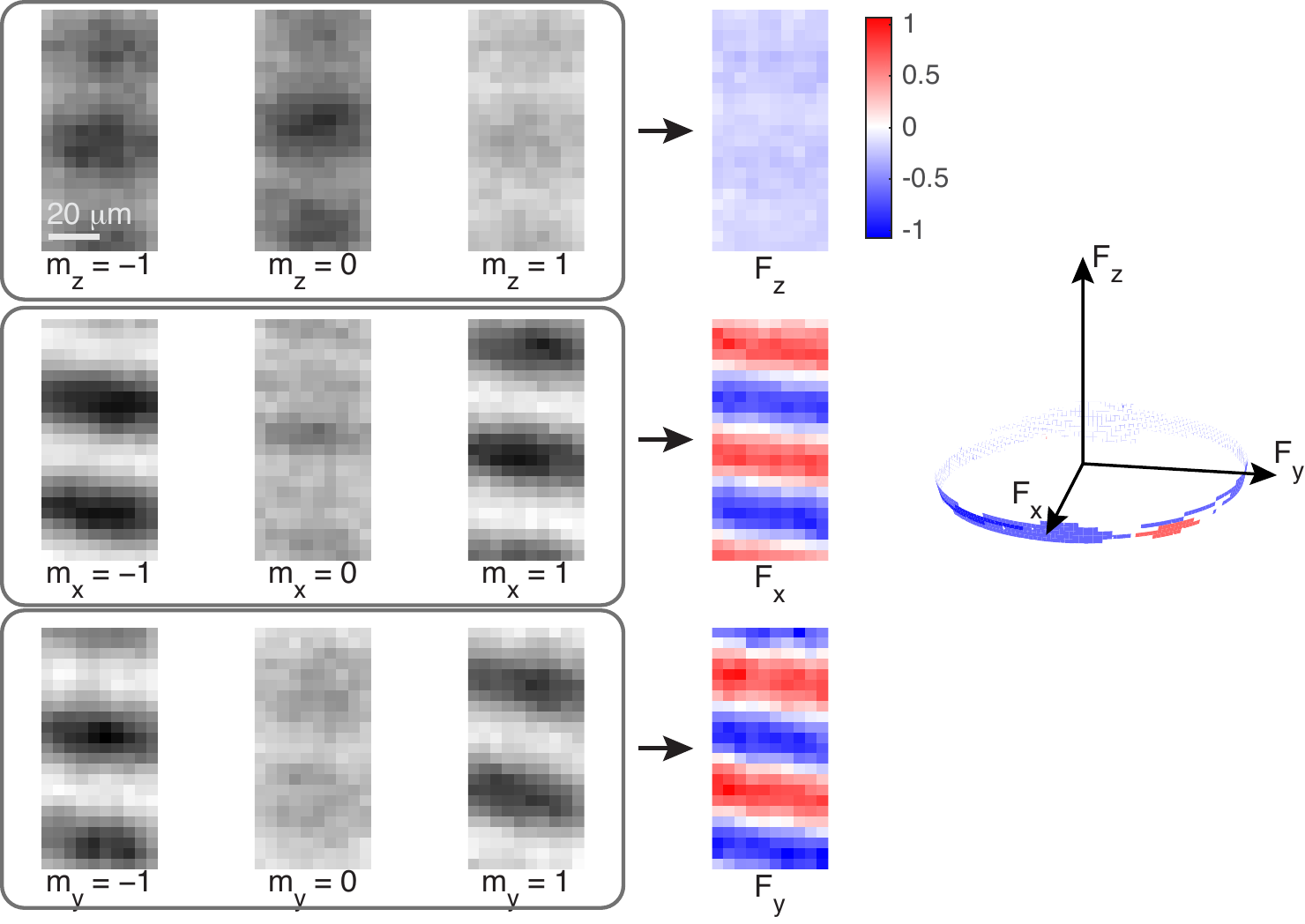}
	\caption[Full magnetization and topological imaging of a spin helix.]{Vector magnetization of a spin helix, reconstructed as in Fig.\ \ref{fig:full_magnetization_vortex12}.  Here, the mapping of the image onto the magnetization sphere shows the spin helix texture spans only a band at fixed latitude, and with no net encompassed solid angle $\Omega$. \label{fig:full_magnetization_helix2}}
\end{figure}

By comparison, a spin helix -- a spin texture in which the spin lies in the transverse plane with an azimuthal angle $\phi$ that advances linearly in position -- has a small integrated solid angle and occupies a small fraction of the magnetization sphere (Fig.~\ref{fig:full_magnetization_helix2}). %This analysis works despite the non-uniform boundary of the chosen region.

%Another way to observe the geometric difference between the skyrmion-like and spin-helix structures is to plot the area swept out on a magnetization sphere, with axes $F_x$, $F_y$, and $F_z$. The value of $\mathbf F$ at one pixel within the image maps to a point on this sphere. For each triangle of pixels $\mathbf r_1$, $\mathbf r_2$, and $\mathbf r_3$, we fill in the surface area on the sphere between the three vectors $\mathbf F(\mathbf r_1)$, $\mathbf F(\mathbf r_2)$, and $\mathbf F(\mathbf r_3)$. The region is filled red if the vectors circle the region in a counterclockwise fashion, and blue for clockwise (reflecting the fact that $d\Omega$ in Eq.\ \ref{eq:domegaformula} is a signed quantity). For the skyrmion-like structure, the entire sphere is covered blue (Fig.~\ref{fig:full_magnetization_vortex12}). For the spin helix, only the region near the equator is covered with equal likelihood for red and blue regions (Fig.~\ref{fig:full_magnetization_helix2}).

%\commentforDan{I can try to generate a new figure rotated. On my computer, I can see through the white sphere and see the red/blue smattering in the back.} 
%\DeclareGraphicsExtensions{.pdf}
\section{Spin dynamics}
\label{sec:dynamics}

One of the features of spinor Bose gases that makes their behavior so rich is that their spin order is dynamic, evolving even from simple, uniform, uncorrelated initial patterns into a richly landscaped quantum field.  Fortunately, these dynamics take place on rather slow timescales (milliseconds) and also on rather long length scales (micrometers) so that the detection methods described in previous Sections can provide a detailed view of their intricate evolution.

In this Section, we discuss dynamics brought about by spin-mixing collisions within $F=1$ spinor Bose-Einstein condensates.  Starting from a tractable treatment of the exact spin mixing oscillations of just two atoms, we will consider dynamics that involve also a macroscopic number of atoms within a spinor gas.  Approximate treatments of such many-body dynamics will reveal spin mixing dynamics to be source of significant quantum spin correlations, and some early views of such correlations in recent experiments will be presented.  This material is meant to serve as a primer for understanding a far broader range of dynamical phenomena that have been, or will be, explored in spinor Bose gases.

\subsection{Microscopic spin dynamics}

As we argued above based on rotational symmetry, the interactions between two atoms within a spinor gas conserves their total angular momentum.  In the case of s-wave interactions, this total angular momentum corresponds to the total spin $F_\mathrm{pair}$ of the colliding pair.  Therefore, in the absence of any external influences that break rotational symmetry, the energy eigenstates of two interacting atoms are also eigenstates of the pair angular momentum operator.

We consider such a symmetry preserving configuration: two spinor-gas atoms in the $F=1$ manifold, trapped in a state-independent, spherical trap, in the absence of any applied fields.  We focus on the lowest energy states of motion.  These low energy eigenstates include ones in which the spins of the particles are in the $|F_\mathrm{pair}=2, m_{F_\mathrm{pair}}=\{-2, -1, 0, 1, 2\}\rangle$ states or the $|F_\mathrm{pair}=0, m_{F_\mathrm{pair}}= 0\rangle$ state.  The state of motion of the two particles in these different states is, in general, different; for example, for a ferromagnetic $F=1$ spinor gas, the atoms in the $|F_\mathrm{pair}=0, m_{F_\mathrm{pair}} = 0\rangle$ state repel one another more strongly, so that the motional ground state might extend out to a larger volume or be more strongly correlated.  To simplify our considerations, so that we can focus just on spin dynamics and not on motional dynamics, let's assume that the spin state of the atoms does not strongly influence their state of motion.  For example, the atoms might be held in a tight trapping container where the ground and excited states of motion are separated by large energy gaps that the interaction energy cannot span.  We would then conclude that the motional wavefunction of each atom in both states is the same ($\psi(\mathbf{r})$).

We are interested in spin-mixing dynamics, and so let us consider that the two particles are prepared so that the initial value of $m_{F_\mathrm{pair}}$ is zero.  The ensuing dynamics, which preserve $m_{F_\mathrm{pair}}$, take place in the subspace spanned by the following two energy eigenstates:
\begin{eqnarray}
\psi(\mathbf{r}_1) \psi(\mathbf{r}_2) \otimes |F_\mathrm{pair}=2, 0\rangle & = & \psi(\mathbf{r}_1) \psi(\mathbf{r}_2) \otimes \left( \sqrt{\frac{2}{3}} |0,2,0\rangle + \sqrt{\frac{1}{3}} |1, 0, 1\rangle \right) \\
\psi(\mathbf{r}_1) \psi(\mathbf{r}_2) \otimes |F_\mathrm{pair}=0, 0\rangle & = & \psi(\mathbf{r}_1) \psi(\mathbf{r}_2) \otimes \left(  -\sqrt{\frac{1}{3}} |0,2,0\rangle + \sqrt{\frac{2}{3}} |1, 0, 1\rangle  \right)
\end{eqnarray}
The energies of the two states are
\begin{eqnarray}
\hbar \omega_2 & = & E + \frac{4 \pi \hbar^2 a_2}{m} \langle n \rangle \\
\hbar \omega_0 & = & E + \frac{4 \pi \hbar^2 a_0}{m} \langle n \rangle,
\end{eqnarray}
respectively, with $E$ being the kinetic and potential energy, and the remainders giving the interaction energy with $\langle n \rangle = \int |\psi^4(\mathbf{r})| d^3\mathbf{r}$ being the average density of the states.

When these two particles are placed in a superposition of states of different total spin, we expect them to undergo dynamics.  Let's specifically consider the case where the particles are both prepared in the $|m_F = 0\rangle$ single particle state.  The temporal evolution of this two particle state, $\Psi(t)$, is given simply as
\begin{eqnarray}
|\Psi(t)\rangle & = & \psi(\mathbf{r}_1) \psi(\mathbf{r}_2) \otimes \left[c_{0,2,0} |0,2,0\rangle + c_{1,0,1} |1, 0, 1\rangle \right] \\
& = & \psi(\mathbf{r}_1) \psi(\mathbf{r}_2) \otimes \left[\left(\frac{2}{3} e^{-i \omega_0 t} + \frac{1}{3} e^{-i \omega_2 t} \right) |0,2,0\rangle \nonumber \right.\\
& & \left. + \left( \frac{\sqrt{2}}{3} e^{-i \omega_0 t} - \frac{\sqrt{2}}{3} e^{-i \omega_2 t}\right) |1, 0, 1\rangle \right]
\label{eq:2bodyspinmixing}
\end{eqnarray}
We observe an oscillation brought about by spin-mixing collisions, coherently converting two atoms from the $|m_F = 0\rangle$ state to the state with one particle in the $|m_F = +1\rangle$ state and another in the $|m_F = -1\rangle$ state.  The dynamics proceed at a frequency $\omega_2 - \omega_0$ that is proportional to the difference in s-wave scattering lengths $a_2$ and $a_0$, and also to the density $\langle n \rangle$.

We note also that these spin-mixing oscillations generate entanglement.  The initial state $|\Psi(0)\rangle = |0,2,0\rangle$ is a product state.  The state produced by spin mixing, $|1,0,1\rangle$, is not a product state; rather, it is a two-particle Bell state:
\begin{equation}
|1,0,1\rangle = \frac{|m_F = 1, m_F = -1\rangle + |m_F = -1, m_F = 1\rangle}{\sqrt{2}}
\end{equation}
where now we use a separable basis for the two-particle wavefunction.  During the spin mixing oscillation, the two particles are in a superposition of pure Fock states.  The relative phase in that superposition, $\theta = \arg(c_{0,2,0}^* c_{1,0,1})$ controls the direction of the spin-mixing reaction, i.e.\ determining whether the population in the $|1, 0, 1\rangle$ state should increase or decrease in time (as is standard in Rabi oscillations between two states).  If the phases of the single particle Zeeman sublevels are changed by the amounts $\phi_{m_F}$, say by imposing single-particle energy shifts for a short time, the relative phase $\theta$ increases by the amount $\Delta \theta = \phi_1 + \phi_{-1} - 2 \phi_0$.  Thus we recognize $\theta$ as the quantum phase that controls the spin-mixing interaction, as we surmised earlier in our discussion of the mean-field spin-dependent energy functional (Sec.\ \ref{sec:meanfieldgs}).

\subsection{Mean-field picture of collective spin dynamics}

The spin-mixing dynamics in a gas with larger particle number is generally far more complex.  Not only does the Hilbert space of spin states become far larger, but also we should allow for motional dynamics that accompany the spin-mixing dynamics, producing quantum fluids with both spatial and spin structure.  While such complexity makes it difficult to understand spin-mixing dynamics in complete detail, it also bestows such dynamics with intriguingly rich phenomenology that is being slowly uncovered experimentally and theoretically.

Let us simplify this problem greatly by adopting the mean-field approximation, that all particles in the spinor gas share the same single particle wavefunction.  This is generally a spatially varying spin wavefunction, with probability amplitudes of $\psi_{m_F}(\mathbf{r})$ for the particle being in the Zeeman sublevel with quantum number $m_F$ and at location $\mathbf{r}$.  With this assumption, the procedure for generating equations of motion is formally equivalent to the construction of the Gross-Pitaevskii equation:  One starts with the many-body Hamiltonian, which may be written in terms of spatial integrals of Bose field operators $\hat{\psi}_{m_F}(\mathbf{r})$ and their adjoints $\hat{\psi}^\dagger_{m_F}(\mathbf{r})$.  One then makes the Bogoliubov approximation of replacing all those field operators with complex numbers reflecting the macroscopically occupied single-particle state, i.e.\
\begin{eqnarray}
\hat{\psi}_{m_F}(\mathbf{r}) & \rightarrow & \sqrt{N} \psi_{m_F}(\mathbf{r}) \\
\hat{\psi}^\dagger_{m_F}(\mathbf{r}) & \rightarrow & \sqrt{N} \psi_{m_F}^*(\mathbf{r})
\end{eqnarray}
The factors $\sqrt{N}$ come about here because we chose to normalize the single-particle wavefunction to unity.  We then obtain dynamical equations by the standard variational approach, i.e.\
\begin{equation}
i \hbar \frac{d}{dt} \psi_{m_F}(\mathbf{r}) = \frac{\partial H_\mathrm{mf}}{\partial \psi_{m_F}^*(\mathbf{r})}
\end{equation}
where $H_\mathrm{mf}$ is the mean-field Hamiltonian we obtained after making those substitutions above.

The spinor Gross-Pitaevskii equation derived in this manner is reproduced in several references \cite[and others later]{ho98,ohmi98}.  We will make use of this equation later on to describe magnetic excitations in spatially extended spinor Bose-Einstein condensates (Sec.\ \ref{sec:excitations}).

Here, we wish to focus just on internal-state dynamics, and so we will make a second approximation: the single-mode approximation, where we express the macroscopically occupied single-particle wavefunction as $\psi(\mathbf{r}) |\zeta(t)\rangle$, i.e.\ where the spatial and spin degrees are separable, and where only the spin wavefunction varies in time.  We obtain equations of motion for $|\zeta(t)\rangle$ from the spinor Gross-Pitaevskii equation after integrating out the spatial dependence.  The evolution from the spin-independent energies can be removed from these equations, obtaining finally \cite{zhan05cohspin}
\begin{eqnarray}
i \hbar \frac{d \zeta_{\pm 1}}{d t} & = & (\pm p + q) \zeta_{\pm 1} + c_1^{(1)} n \left[ \left(\rho_{\pm 1}+\rho_0 - \rho_{\mp 1}\right)\zeta_{\pm 1} + \zeta_0^2 \zeta^*_{\mp 1} \right] \label{eq:GPE1} \\
i \hbar \frac{d \zeta_0}{d t} & = & c_1^{(1)} n \left[\left(\rho_{+1} + \rho_{-1} \right) \zeta_0 + 2 \zeta_{+1} \zeta_{-1} \zeta_0^* \right] \label{eq:GPE2}
\end{eqnarray}
Here, $\rho_{m_F} = |\zeta_{m_F}|^2$.

We see in these equations three effects.  First is the evolution of the spin wavefunction owing to single-particle energy shifts -- the linear Zeeman energy, quantified by $p$, and the quadratic Zeeman energy, quantified by $q$ (see Sec.\ \ref{sec:meanfieldgs}).  Second is the evolution owing to a mean-field interaction energy shift of the Zeeman sublevels.  We see, for example, that for an antiferromagnetic spinor gas ($c_1^{(1)}>0$), the energy of atoms in the $|m_F = +1\rangle$ state is increased by the population of atoms in the $|m_F = 1\rangle$ and $|m_F = 0\rangle$ state, and decreased by the population of atoms in the $|m_F = -1\rangle$ state.  These relations reflect the fact that the fully magnetized state $|m_F = 1\rangle$ is penalized energetically by antiferromagnetic interactions, and the fact that populations in the $|m_F = 1\rangle$ state tend to phase separate from atoms in the $|m_F = 0\rangle$ state, while they tend to mix with populations in the $|m_F = -1\rangle$ state.  These phase separation/mixing tendencies are reflected in the mean-field ground-state diagram for the antiferromagnetic spinor gas (Fig.\ \ref{fig:pandq}), as we have discussed above.  Third, we see terms (the last entries in the square brackets) that lead to spin mixing dynamics through which the populations in the various magnetic sublevels will change.  The signs of these terms (i.e.\ of $\zeta_0^2 \zeta^*_{\mp 1}$ and  $\zeta_{+1} \zeta_{-1} \zeta_0^*$) and  are clearly sensitive to the phases of the spin wavefunction in the various magnetic sublevels.

A more intuitive form of these equations of motion was provided in Ref.\ \cite{zhan05cohspin}.  The authors derive the relations
\begin{align}
\frac{d \rho_0}{d t} = -\frac{2}{\hbar} \frac{d E_\mathrm{spin}}{d \theta} && \frac{d \theta}{d t} = \frac{2}{\hbar} \frac{d E_\mathrm{spin}}{d \rho_0}
\end{align}
In other words, the quantities $\rho_0$ and $\theta$, which span a conical surface, act like canonical coordinates, while the mean-field spin-dependent energy $E_\mathrm{spin}$, derived in Eq.\ \ref{eq:meanfieldenergy}, acts like a Hamiltonian governing their dynamics.  Such dynamics proceed along lines of constant $E_\mathrm{spin}$.

\begin{figure}[t]
\centering
\includegraphics[width=0.7\textwidth]{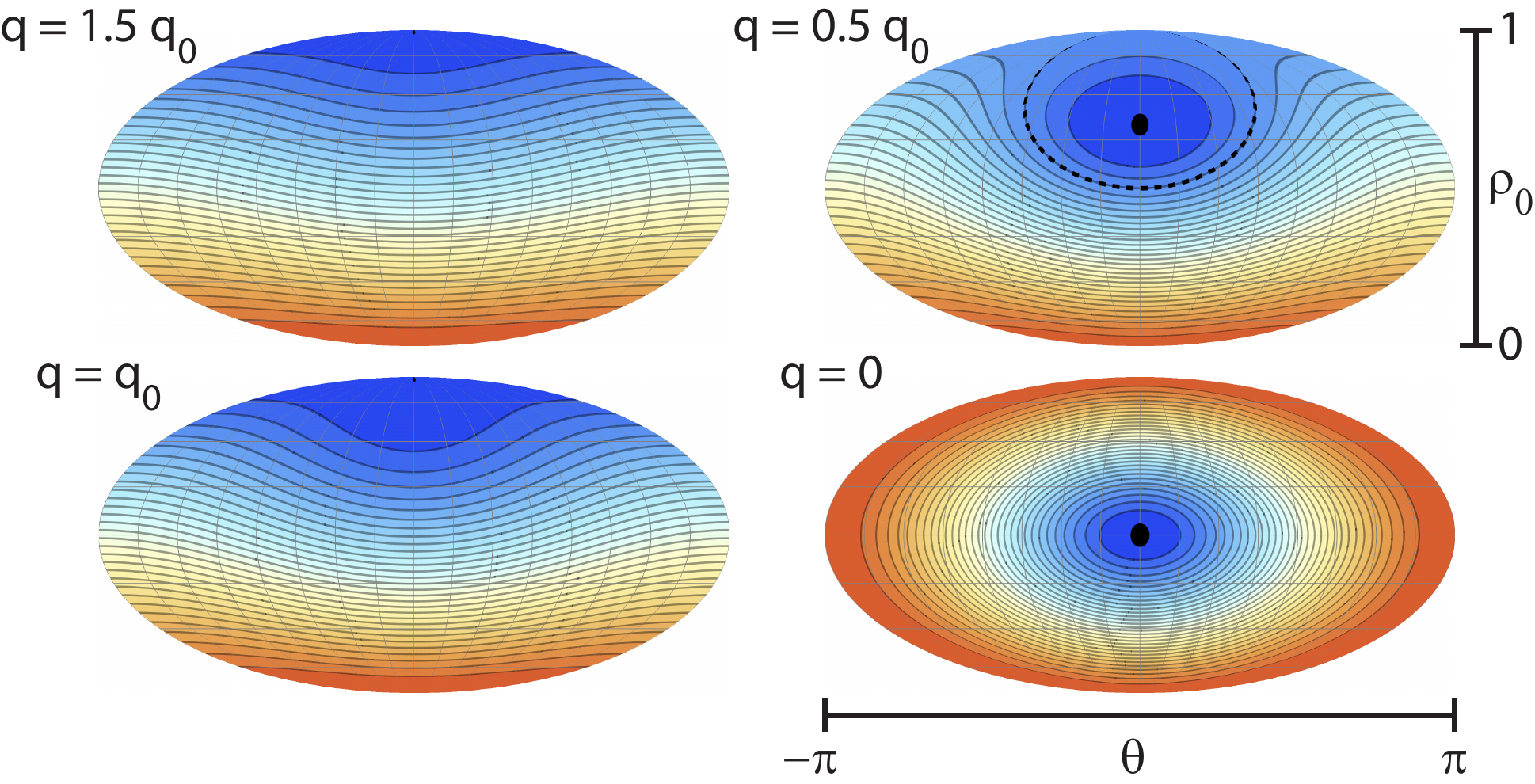}
\caption{Contour plots of the spin-dependent mean-field energy $E_\mathrm{spin}$ for a ferromagnetic gas $c_1^{(1)}<0$.  For the case shown, with $M=0$, the phase space spanned by the relevant parameters in the mean-field state, the relative population $\rho_0$ in the $|m_z = 0\rangle$ state and the phase $\theta$ that governs spin-mixing dynamics, has the geometry of a sphere.  Hence, the energy contours are shown here as Mollweide projections of the sphere, with the north pole representing the longitudinally aligned polar state $|m_z = 0\rangle$ ($\rho_0=1$), and the south pole representing the transversely aligned polar state, such as $|m_x = 0\rangle$ ($\rho_0 = 0$).  For $q<q_0 = 2|c_1^{(1)}n|$, the minimum energy state, shown by the black dot, has non-zero transverse magnetization.  The black dashed line marks the separatrix between two types of coherent spin-mixing dynamics.  This gray lines show lines of constant $\theta$ and $\rho_0$.  Red shading indicates high energy, and blue shading indicates low energy.\label{fig:contourplots}}
\end{figure}

Now that we understand how $E_\mathrm{spin}$ determines not only the ground state mean-field spin state, but also the dynamics of excited energy states, let us examine its form more closely.  Contour plots of $E_\mathrm{spin}$, calculated for the fixed value $M = 0$, are presented in Fig.\ \ref{fig:contourplots}.  The phase space spanned by $\rho_0$ and $\theta$ in this case has the geometry of a sphere (a cone pinched closed at both its base and its top), as the states at the north and south pole, with $\rho_0 = 1$ and $\rho_0=0$, respectively, are independent of the phase $\theta$.  Thus, we plot the energy contours using a standard cartographic projection of a spherical surface (the Mollweide projection).

The equal energy contours shown on such plots are of two types.  For $q>q_0$, all energy contours are circumpolar.  In this case, spin-mixing dynamics lead a state to cycle through the full range of $\theta$.  In the limit $q \gg q_0$, these dynamics are just the single-particle dynamics of orientation-to-alignment conversion that occurs in the presence of quadratic Zeeman or Stark energy shifts.  As $q$ approaches $q_0$ from above, the circumpolar orbits also show increasing variation in $\rho_0$.  The dynamics thus become increasingly visible to measurements of the Zeeman-state distribution of the spinor gas, as reported in Refs.\ \cite{chan05nphys,kron06tune}. Throughout this region, the minimum energy is attained for the polar state $|m_z=0\rangle$ (on the north pole in Fig.\ \ref{fig:contourplots}).

For $q<q_0$, the minimum energy state moves away from the north pole, describing a state of non-zero transverse magnetization, of the form given in Eq.\ \ref{eq:transversespin}, and shown by a black dot on the contour plots of Fig.\ \ref{fig:contourplots}.  A family of contours emerges that describes orbits in phase space that circle the minimum energy state and span only a limited range of $\theta$, as observed in Ref.\ \cite{chan05nphys}.  These contours are separated from the remaining circumpolar orbits by a separatrix, indicated by the black dashed line in the figures.  Spin mixing dynamics generally proceed along equal energy contours, but on this separatrix the velocity with which the state evolves along the contour goes to zero.  Evidence of this arrested evolution has been obtained in experiments \cite{blac07naspinor,liu09qpt}.

\subsection{Spin-mixing instability}
\label{sec:spinmixing}

While the mean-field description of spin-mixing dynamics presented above certainly makes for a nice story, with an intuitive dynamical picture that was confirmed by elegant experiments, it certainly cannot be the whole story.  Besides mean-field dynamics, we should expect also something more, the evolution of some quantum fluctuations atop the mean-field state.  But what is the nature of these fluctuations?  How can we detect them?  Under what conditions might their effect be particularly glaring?

One scenario that might bring forth answers to the questions above is a starting point from which mean-field physics predicts that \emph{nothing} should happen, but where beyond-mean-field physics shows that something definitely does happen.  We are led thereby to consider the evolution of a spinor gas at a dynamically unstable fixed point.  A good analogy for this scenario is a rigid pendulum that is initiated at the apex of its motion.  Placed exactly at the apex, the pendulum remains stationary.  However, small fluctuations away from the apex, perhaps necessitated by quantum mechanics, cause the pendulum to veer exponentially away from its fixed point.  During the early stage of motion, before it swings back up toward the apex, if the pendulum is found rotating clockwise, then it is also displaced clockwise from the apex.  As such, the position and momentum of the pendulum become highly correlated.  The evolution of such a pendulum is Hamiltonian, and thus phase space preserving.  The high correlations produced dynamically along one axis in phase space (positive correlations between position and momentum) must then also greatly narrow down the phase space distribution along an orthogonal axis.

Spin-mixing dynamics lead to several such unstable fixed points.  Let us focus on one in particular: the instability under ferromagnetic interactions of the axial polar mean-field state of the $F=1$ spinor gas.  In such a state, all atoms begin in $|m_F = 0\rangle$ Zeeman sublevel.  The mean-field energy functional, expressed in Eq.\ \ref{eq:meanfieldenergy} and shown in Fig.\ \ref{fig:contourplots}, shows this state to be a (local) maximum energy state when the quadratic Zeeman energy is sufficiently small ($q \leq 2 |c_1^{(1)} n|$).  Mean-field dynamics proceed on equal energy contours, but since this contour is just a single point, the initial state should undergo no mean-field dynamics.

However, we know that this state does undergo dynamics.  Return to the example given earlier of spin-mixing oscillations of two atoms prepared in the $|m_F = 0\rangle$ state.  The state certainly undergoes dramatic dynamics, as described by Eq.\ \ref{eq:2bodyspinmixing}.  How are these dynamics of just two atoms translated to a macroscopic gas of interacting atoms?

We need a theoretical description that goes beyond mean-field theory.  We turn to the equivalent of Bogoliubov theory, i.e.\ a theory of linear perturbations atop a mean-field state, accounting not only for the classical evolution of these perturbations but also for their quantum fluctuations.%  The interaction we are interested in is the spin-mixing interaction, which exchanges pairs of particles between the $|m_F = 0\rangle$ state and the $|m_F = +1\rangle$ and $|m_F = -1\rangle$ states.

We express the perturbations of the condensate in the Zeeman-state basis as follows:
\begin{equation}
\zeta = \left( 1+\zeta_z \right) \begin{pmatrix} 0 \\ 1 \\ 0\end{pmatrix} + \zeta_x \begin{pmatrix} \nicefrac{1}{\sqrt{2}} \\ 0 \\ \nicefrac{- 1}{\sqrt{2}} \end{pmatrix} + \zeta_y  \begin{pmatrix} \nicefrac{i}{\sqrt{2}} \\ 0 \\ \nicefrac{i}{\sqrt{2}} \end{pmatrix}
\end{equation}
Here, we choose to represent the fluctuations of the mean-field state in a Cartesian basis (recall the spin wavefunctions listed in Eq.\ \ref{eq:cartesian}).  Also, we've inserted these perturbations in a way that the spinor $\zeta$ is no longer normalized to unity; constraints of normalization would come in at higher order and can be disregarded anyhow if we're not too strict on particle number conservation.

Now we apply second quantization to these fluctuations. We'll leave aside the quantization of the fluctuation $\zeta_z$, and focus instead on the spin fluctuations.  In second quantization, we turn the quantities ${\zeta}_x$ and ${\zeta}_y$ into spin-mode Bose operators, $\hat{\zeta}_x$ and $\hat{\zeta}_y$, which annihilate a particle in the indicated spin state and in a spatial mode that is for now unspecified; we will adapt this mode function to describe various experimental situations.  The Bose operators have the standard commutation relations, $\left[\hat{\zeta}_x, \hat{\zeta}_x^\dagger\right] =  1$ and $\left[\hat{\zeta}_y, \hat{\zeta}_y^\dagger\right] =  1$, while operators for different polarizations of spin fluctuations mutually commute.

Working through the second-quantized Hamiltonian of the $F=1$ spinor condensate, focusing on the spin dynamics part, we obtain the following:
\begin{equation}
H_\mathrm{fluc} = \sum_{\beta \in \{x,y\}} \left(\epsilon + q + c_1^{(1)} n \right) \hat{\zeta}_\beta^\dagger \hat{\zeta}_\beta - \frac{c_1^{(1)} n}{2} \left( \hat{\zeta}_\beta^\dagger  \hat{\zeta}_\beta^\dagger +  \hat{\zeta}_\beta  \hat{\zeta}_\beta \right)
\label{eq:spinfluchami}
\end{equation}
We've swept a fair bit under the rug to obtain this expression.  To summarize quickly, we've made the Bogoliubov approximation, replacing operators that create and annihilate particles in the mean-field condensate state with complex numbers.  The quantity $\epsilon$ represents the energy of the spin mode whose fluctuations we are considering.  This spin mode can be, for example, a spatially uniform spin excitation, or a spin excitation with a specific momentum, or a spin excitation in some excited motional state within a trap.  We assume that whatever mode we are considering, we have already taken care to define the mode properly and obtain its energy.  The quantity $q$ is again the quadratic Zeeman shift, which is inserted here because it turns out to be a valuable experimental control parameter.  The quantity $n$ represents, of course, the density of the Bose-Einstein condensate whose fluctuations we are considering.  For a non-uniform condensate, $n$ may be a spatially averaged density.

This Hamiltonian can be put in more familiar form with the following substitutions:
\begin{align}
\hat{Z}_\beta = \frac{\hat{\zeta}_\beta + \hat{\zeta}_\beta^\dagger}{2} && \hat{P}_\beta = \frac{\hat{\zeta}_\beta - \hat{\zeta}_\beta^\dagger}{2 i}
\end{align}
With these operators, the spin-fluctuation Hamiltonian takes the following form:
\begin{equation}
H_\mathrm{fluc} = \sum_{\beta \in \{x,y\}} \left(\epsilon + q \right) \hat{Z}_\beta^2 + \left( \epsilon + q + 2 c_1^{(1)} n \right) \hat{P}_\beta^2
\label{eq:zandpdynamics}
\end{equation}

The operators $\hat{Z}_\beta$ and $\hat{P}_\beta$ are like the position and momentum of two different harmonic oscillators, one for each polarization of spin fluctuations.  But what do these operators represent physically?  Let us return to Eq.\ \ref{eq:spinfluchami}, and focus on the $x$ polarization.  If we create such a fluctuation, we take atoms out of the $|m_F = 0\rangle$ state and transfer them $\pi$ out of phase to the $|m_F = 1\rangle$ and $|m_F = -1\rangle$ states.  As a matrix operation, represented in the $F_z$ eigenbasis, such an operation has the following form:
\begin{equation}
\hat{\psi}_x^\dagger \sim \left( \begin{array}{c c c} 0 & \nicefrac{1}{\sqrt{2}} & 0 \\ 0 & 0 & 0 \\ 0 & \nicefrac{-1}{\sqrt{2}} & 0 \end{array}\right)
\end{equation}
This matrix is not Hermitian, and neither is the operator $\hat{\psi}_x^\dagger$ it represents.  The operators $\hat{Z}_x$ and $\hat{P}_x$ are Hermitian, and we see, by examining the array of single-particle spin observables (Eqs.\ \ref{eq:firstspin1matrix} - \ref{eq:lastspin1matrix}), that these operators represent the observables $\hat{Q}_{xz}$ and $\hat{F}_{y}$ (within some proportionality factors).  Similarly, $\hat{Z}_y$ and $\hat{P}_y$ correspond to the operator pair $\hat{Q}_{yz}$ and $\hat{F}_{x}$.

In the initial state we are considering, a condensate of $N$ atoms in the $|m_F = 0\rangle$ state, the expectation values of all four of these spin observables is zero.  Position and momentum operators do not commute, and so while their expectation values may both be zero, their fluctuations are perforce non-zero and have a minimum uncertainty relation.  Similarly, the corresponding pairs of spin observables do not commute.  We obtain thereby an uncertainty relation between the variances of these operators:
\begin{equation}
(\Delta F_x)^2 (\Delta Q_{yz})^2 \geq \left( \frac{1}{2} \langle \left[\hat{F}_x, \hat{Q}_{yz}\right] \rangle \right)^2 = N^2
\end{equation}
and similar for the pair $\hat{F}_y$ and $\hat{Q}_{xz}$.  In the last equality, we evaluate the commutator on the initial (mean-field) state of the condensate.  The variances themselves initially have the values $\left( \Delta F_x\right)^2 = N$ and $\left( \Delta Q_{yz}\right)^2 = N$.  Thus, the initial state is a minimum uncertainty state for such spin fluctuation operators, with the uncertainty being evenly distributed between the two quadratures.

As for the Hamiltonian, we identify two distinct behaviors.  When $\epsilon+q$ has a value such that the prefactors before the $\hat{Z}^2$ and $\hat{P}^2$ terms have the same sign, the Hamiltonian is that of a normal harmonic oscillator (or perhaps one with negative mass).  The evolution of fluctuations under this Hamiltonian is the familiar stable precession in phase space at a frequency that is determined, just as it is for the harmonic oscillator, by the product of the two prefactors, to wit
\begin{equation}
\omega^2 = \hbar^{-2} \left(\epsilon + q \right) \left(\epsilon + q + 2 c_1^{(1)} n \right)
\label{eq:spinfreq}
\end{equation}
A familiar regime of such dynamics occurs at very high (positive or negative) values of $q$.  We observe stable harmonic precession in the phase space spanned by $\hat{Z}_\beta$ and $\hat{P}_\beta$ (one plane for each polarization).    A state that is initially displaced along, say, the $\hat{P}_x \sim \hat{F}_{y}$ axis rotates, within a quarter cycle of the oscillation, to be displaced along the $\hat{Z}_x \sim \hat{Q}_{xz}$ axis.  This is the physics of orientation-alignment conversion familiar from molecular and atomic physics, driven by quadratic Stark or Zeeman shifts.

In a different regime, the spin fluctuations undergo parametric amplification.  This occurs when the prefactors before the $\hat{Z}^2$ and $\hat{P}^2$ terms have opposite sign.  The Hamiltonian under these conditions is equivalent to that of a (positive mass) inverted harmonic oscillator, for which the potential energy decreases quadratically with distance.  The ensuing dynamics will cause an initial phase space distribution to be stretched out, with strong correlations between position and momentum.  The factor $r$ by which the distribution is stretched increases exponentially in time, $r \propto e^{|\omega| t}$, where the temporal gain $|\omega|$ is determined again from Eq.\ \ref{eq:spinfreq}.

In such dynamics, we identify the mechanism of spontaneous symmetry breaking.  If we observe the initial state at a macroscopic level, i.e.\ with a measurement instrument of not-exceptional resolution, it will appear to us that the values of all these spin fluctuation operators are consistent with zero.  In other words, our initial measurements will confirm the fact that our initial state has not broken rotational symmetry about the $\mathbf{z}$ spin axis.  In contrast, at long times after the onset of the dynamical instability, a single measurement of the transverse spin observables will likely show a value that is well distinguished from zero, and it will appear to us that the system has spontaneously broken its initial rotational symmetry, acquiring, for example, a large transverse spin vector.  Repeating the experiment many times, we would find that the transverse spin is typically distinctly different from zero, but that it points in random directions from shot to shot.  This observation would support our view that the rotational symmetry is indeed broken spontaneously and not by some explicit residual symmetry breaking in our experimental setup.

Of course, before we perform our measurement, rotational symmetry has not, in a strict sense, been broken: an initially rotationally symmetric state evolving under a rotationally symmetric Hamiltonian retains its symmetry.  The stretched-out state of quantum fluctuations generated by the spin-fluctuation Hamiltonian (Eq.\ \ref{eq:spinfluchami}) is consistent with all our expectations: it is a rotationally symmetric superposition of broken-symmetry states that are easily distinguishable by measurement.

But recall that the area of a phase-space distribution is conserved under Hamiltonian dynamics.  For the simple quadratic Hamiltonian treated here, the stretching out in phase space along one direction (say, positive correlations between $\hat{Z}$ and $\hat{P}$) is accompanied by squeezing in phase space along an orthogonal direction (negative correlations between $\hat{Z}$ and $\hat{P}$).  The squeezing generated by this dynamical instability, this parametric amplification of the initial spin fluctuations of the condensate, is observable, and it appears likely that this type of spin-squeezing (spin-nematic squeezing) can be manipulated in a way to reduce the noise in atomic sensors \cite{sau10njp}.

\subsubsection{Experiments in the single-mode regime}

All of these predicted phenomena -- the tuning between stable and unstable dynamics by varying the spin-mode energy, spontaneous symmetry breaking, and the generation of spin-nematic squeezing through spin-mixing interactions -- have been observed in experiments focusing on the single-mode regime.  Here, we highlight just a couple of early experimental results on the spin-mixing instability, but note that not only are there more early results than summarized here, but also that more advanced and impressive results are still being obtained at present.

\begin{figure}[t]
\centering
\includegraphics[width=\textwidth]{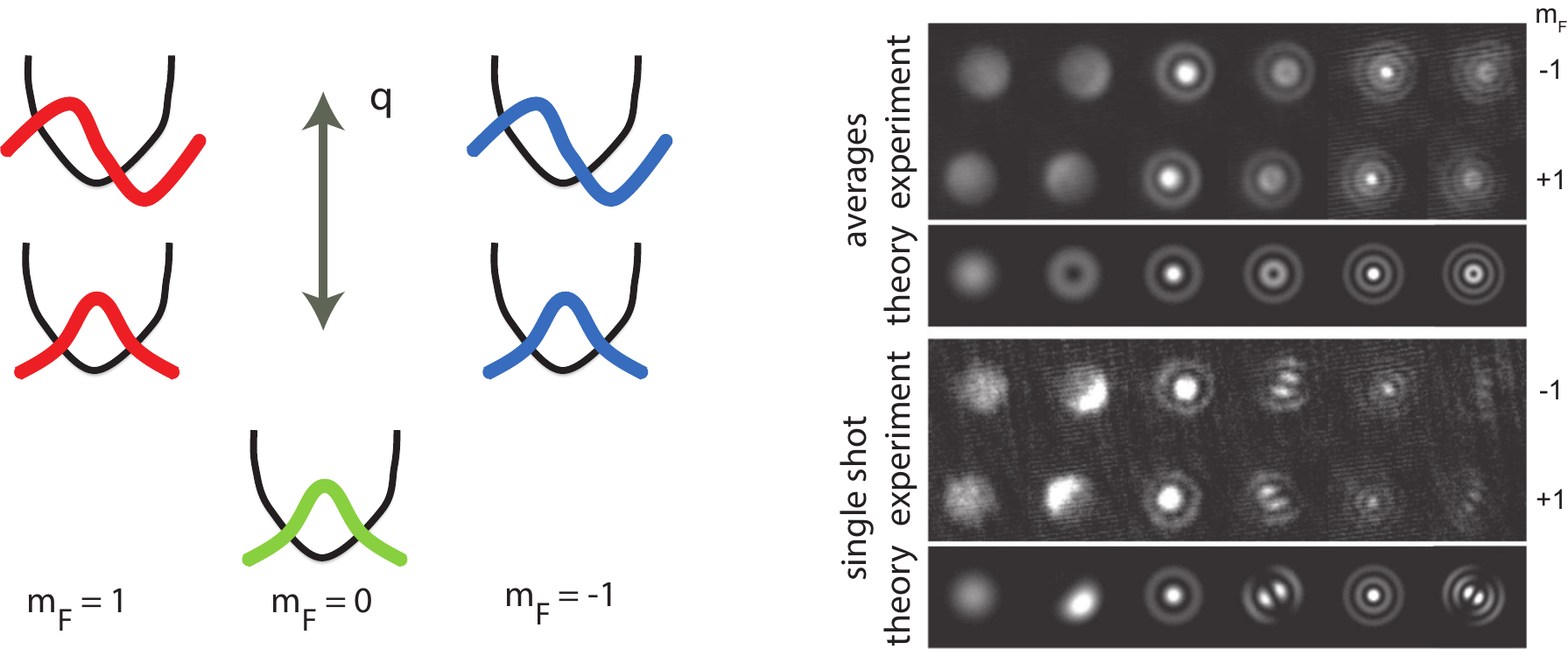}
\caption{Selective amplification of spin fluctuation modes.  A spinor condensate was prepared in the $|m_F = 0\rangle$ state and trapped tightly in an optical trap.  The tight trap confinement ensured that different spin excitation modes, sketched schematically in the left half of the figure, were at discrete energies.  A quadratic Zeeman shift tuned the energy of the $|m_F = 0\rangle$ state, bringing it into spin-mixing resonance with particular spin fluctuation modes.  The output of spin mixing was analyzed by releasing the atoms from the trap and imaging the spatial distributions of atoms in the different Zeeman sublevels.  The Bessel-mode-like patterns of the spin fluctuation modes matched well with theoretical predictions.  Right figure taken from Ref.\ \cite{sche10}.\label{fig:klemptfigure}}
\end{figure}

Experiments performed at Hannover demonstrated the tuning of the spin-mixing instability with energy \cite{klem09multi,klem10vacuum}.  Experiments were performed with atoms in a trap that was very tightly confining in two dimensions.  Spin excitations within these clouds with different transverse mode structure were separated by a large energy, corresponding to discrete and well-separated values of $\epsilon$, which appears in Eqs.\ \ref{eq:spinfluchami} and \ref{eq:zandpdynamics}.  As such, it was possible to define specific spin modes, by tuning the $q$ appropriately, so that they were dynamically unstable with the highest possible gain (the largest negative value of $\omega^2$ in Eq.\ \ref{eq:spinfreq}).  By imaging the distribution of atoms produced by spin mixing in the $|m_F = \pm 1\rangle$ states, it was possible to identify which spatial mode was selected for amplification (Fig.\ \ref{fig:klemptfigure}), and the results agreed nicely with theoretical predictions.

For evidence of parametric amplification in the phase space of spin fluctuations, one should look to experiments by the Chapman group \cite{haml12squeezing}.  There, a single-mode spinor Bose-Einstein condensate was prepared in the $|m_F = 0\rangle$ state, and then $q$ was set near zero to initiate the spin mixing instability.  After a set time, the gas was examined by releasing the atoms and then counting the number of atoms in each of the Zeeman sublevels along the field direction.  From such a measurement one obtains a precise, indeed atom-shot-noise limited \cite{book11correlations}, measurement of $\hat{F}_z$.  However, what one wants to measure to detect the spin-mixing dynamics is a transverse spin moment.  To achieve this, one simply applies a $\pi/2$ rotation to the atomic spins before separating the Zeeman sublevels.  Now, a measurement of $\hat{F}_z$ corresponds to a measurement of a projection of the collective atomic spin onto an axis in the transverse plane.  To go even further, one wants to measure some linear combination of spin and quadrupole moment in order to detect not only the amplification but also the squeezing of spin fluctuations.  For this, one inserts an additional step before measurement: One applies a large quadratic Zeeman shift, rotating the spin fluctuations around in the $\hat{F}_x$-$\hat{Q}_{yz}$ plane (also in the $\hat{F}_y$-$\hat{Q}_{xz}$ plane), so that the desired quadrature now lies along the $\hat{F}_x$ (also $\hat{F}_y$) axis.

\begin{figure}[t]
\centering
\includegraphics[width=\textwidth]{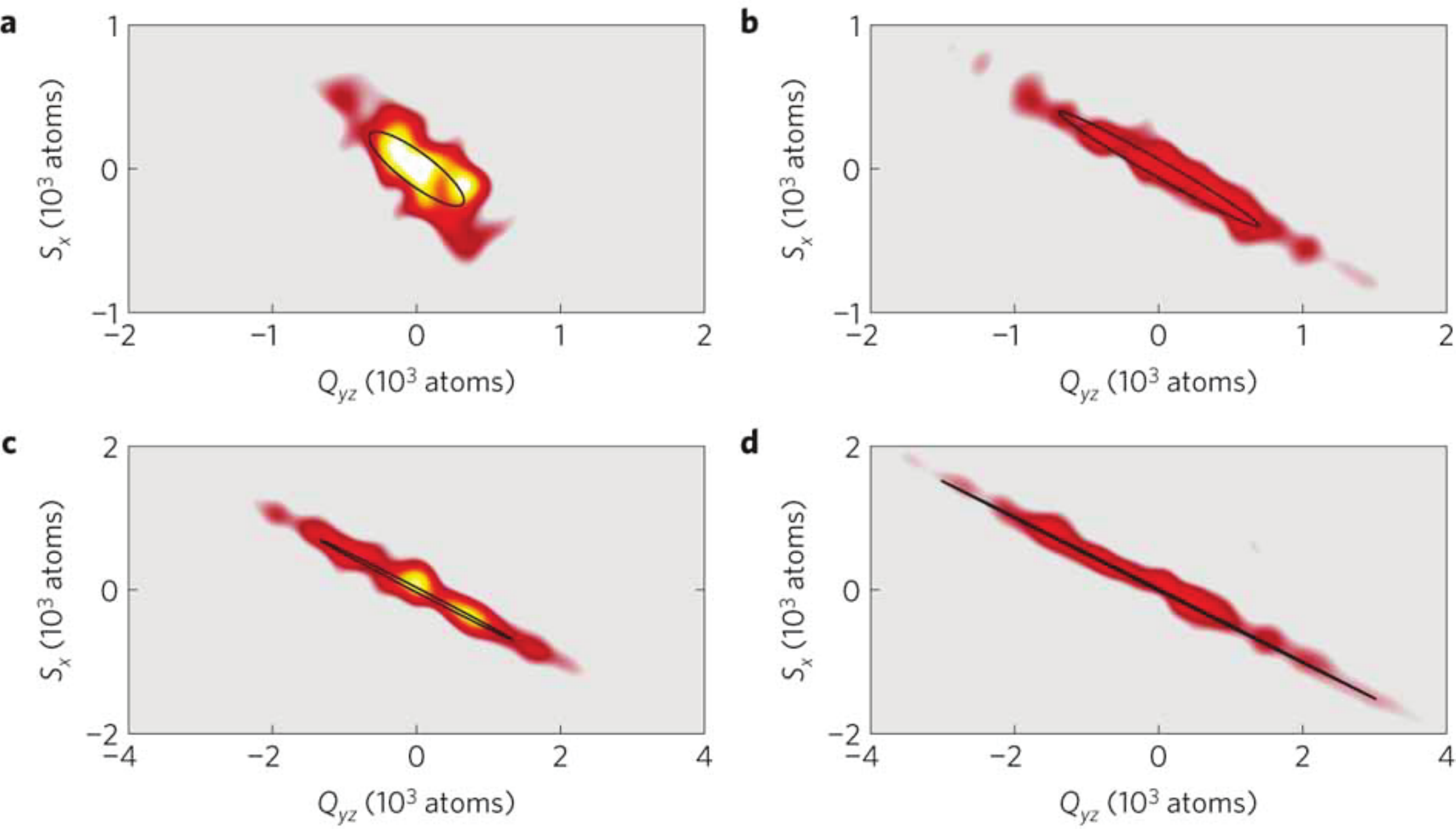}
\caption{Observation of spin-nematicity squeezing.  An $F=1$ $^{87}$Rb condensate in the single-mode regime was prepared in the $|m_z=0\rangle$ initial state and allowed to evolve for variable time under a spin-mixing dynamical instability.  The gas was then probed by (1) pulsing on a quadratic Zeeman energy to rotate spin fluctuations in the spin-nematicity plane, (2) applying a $\pi/2$ spin rotation pulse to rotate one component of the transverse spin onto the longitudinal axis, and (3) measuring the longitudinal spin with sub-atom-shot-noise sensitivity.  The phase-space distribution in the spin-nematicity plane was reconstructed from repeated measurements at evolution times of (a) 15, (b) 30, (c) 45, and (d) 65 ms, at different spin-nematicity rotation angles (measurement step (1)).  The squeezed and anti-squeezed quadratures are clearly observed.  Black circles indicate the $1/\sqrt{e}$ uncertainty ellipse calculated from theory. Figure taken from Ref.\ \cite{haml12squeezing}.\label{fig:chapmanfigure}}
\end{figure}

The outcome of these measurements brought spin-nematic squeezing directly into view (Fig.\ \ref{fig:chapmanfigure}).  The stretched-out axis shows the signature of spontaneous symmetry breaking discussed above: nearly all the measurements of the transverse spin-vector moment give results significantly different from zero, preferring neither positive nor negative values.  At the same time, the parametric amplification along one quadrature of fluctuations causes dramatic squeezing-in of spin fluctuations along an orthogonal quadrature, reducing the variance of measurements by a factor of at least 6 (8 dB).  Related work by other groups focused on other features of the two-mode squeezed state (atoms in both $|m_F = \pm 1\rangle$ states), confirming the emergence of quantum correlations \cite{gros11spinor,luck11squeeze}.  Work by these same groups in the ensuing years has revealed further details of quantum dynamics beyond spin-nematic squeezing, quantitative analyses of the influence of such dynamics on parameter estimation, and more.

\subsubsection{Quantum quenches in spatially extended spinor Bose-Einstein condensates}

In a spatially extended sample, the same treatment given above for the spin-mixing instability can be applied, except that we must now accommodate a continuum of spin excitation modes.  Again, we consider an $F=1$ spinor Bose-Einstein condensate prepared initially in the unmagnetized and axially symmetric $|m_F = 0\rangle$ single particle spin state.  The condensate is held in a large box and is at constant density $n$.

Consider first that a large positive quadratic Zeeman energy is applied to the gas.  Under these conditions, this initial state is indeed the mean-field zero-temperature ground state of the gas.  The gas should have three branches of excitations, corresponding to excitations at momentum $\hbar \mathbf{k}$ in each of the three Zeeman components of the gas.  Excitations in the $|m_F = 0\rangle$ internal state are the gapless Bogoliubov excitations of a scalar Bose-Einstein condensate.  Excitations in the $|m_F = \pm 1 \rangle$ state are the spin excitations described in Eq.\ \ref{eq:spinfluchami} where $\epsilon = \hbar^2 k^2 / 2 m$ is the kinetic energy.

\begin{figure}[t]
\centering
\includegraphics[width=\textwidth]{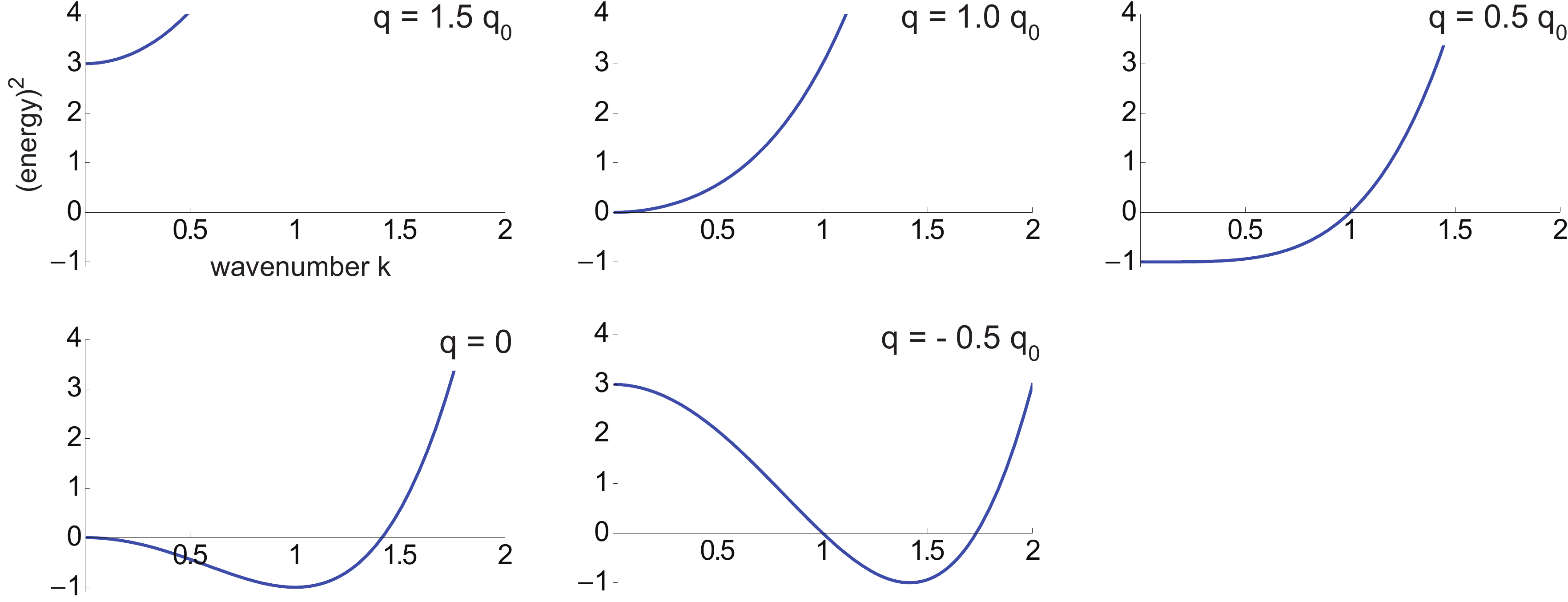}
\caption{Spectrum of spin excitations within the polar $|m_z = 0\rangle$ initial state for ferromagnetic interactions.  The squared energy $\hbar^2 \omega^2$ is shown vs.\ wavenumber $k$ of the excitation.  Here, energy is scaled in units of the spin-dependent energy $|c_1^{(1)} n|$ and the wavevector is scaled in units $k_s = (2 m |c_1^{(1)} n|/\hbar^2)^{1/2}$.  For $q \geq q_0 = 2 |c_1^{(1)} n|$, the excitations are stable, indicated by the squared energy's being positive.  For $q<q_0$, the spectrum shows dynamical instabilities at wavenumbers where $\hbar^2 \omega^2 < 0$. The wavenumber at which instability has maximum gain (maximum negative value of squared energy) grows as $q_0 - q$ grows.\label{fig:amplifierspectrum}}
\end{figure}

Under these conditions, these spin excitations are all stable, with the prefactors in Eq.\ \ref{eq:zandpdynamics} both being positive.  The stability is seen in the expression of Eq.\ \ref{eq:spinfreq} for the squared frequency of spin fluctuations, $\omega^2$, plotted in Fig.\ \ref{fig:amplifierspectrum} as function of wavevector.  At zero temperature, these excitations are occupied only by zero-point fluctuations.  These fluctuations quantify the projection noise we would see if we were, say, to measure the populations of atoms in the $\hat{F}_x$ eigenbasis in order to determine the transverse projection of the atomic spin vector.  Beyond mean-field theory, we expect these excitations to be slightly changed; to lowest order, the corrections resemble the Lee-Huang-Yang correction to the equation of state and speed of sound in an interacting scalar Bose-Einstein condensate.

If we now reduce the value of $q$, some portion of the branch of spin excitations becomes dynamically unstable.  Specifically, those excitations with kinetic energies $\epsilon$ that lie in the range between $-q$ and $-q - 2 c_1^{(1)} n$ acquire an imaginary frequency $\omega$, seen from the value of $\omega^2$ dropping below zero in Fig.\ \ref{fig:amplifierspectrum}.  If we restrict ourselves now to ferromagnetic $F=1$ spinor gases, then the instability is seen to set in with $q < 2 |c_1^{(1)}| n$, precisely where the mean-field ground state is ferromagnetic.  The unstable spin-mixing dynamics then describe the initial growth of magnetization, starting from initial zero-point spin fluctuations, in a gas that is quenched across a symmetry-breaking phase transition from a non-magnetic to a ferromagnetic phase.

The emergence of magnetization under such dynamics has been examined experimentally \cite{sadl06symm,lesl09amp}.  These experiments really brought to view the prediction of spontaneous symmetry breaking through spin-mixing dynamics.  In a large spinor Bose-Einstein condensate that is suddenly quenched across the symmetry-breaking transition, different portions of the gas break symmetry independently.  That is, every small patch of the gas independently undergoes the dynamics such as those revealed in the single-mode experiments described above.  Rather than repeating single-mode experiments many times to measure fluctuations, in a large spinor gas one can just image the entire gas at once and look at the fluctuations between one small region and another to characterize the distribution of spin states that is produced.

\begin{figure}[t]
\centering
\includegraphics[width=\textwidth]{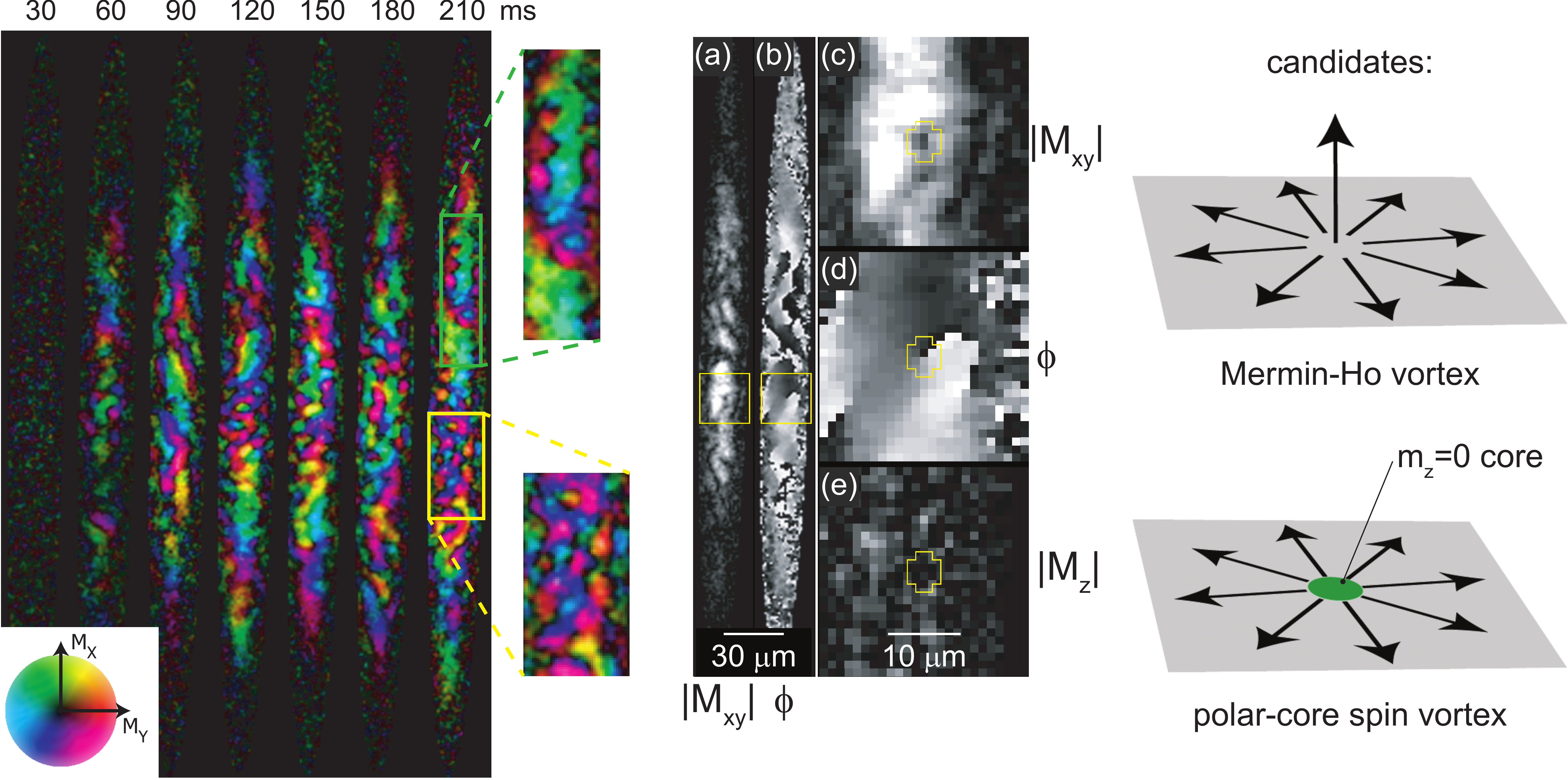}
\caption{Emergence of magnetization in a spinor condensate that is suddenly quenched across a symmetry-breaking phase transition. A $^{87}$Rb spinor condensate was prepared in the $|m_z = 0\rangle$ state and then allowed to evolve at low $q < 2 |c_1^{(1)} n|$ ($n$ being the central condensate density), before its vector magnetization was imaged.  An inhomogeneous spin texture emerged spontaneously, showing some regions with extended spin domains and others with small domains divided by sharp domain walls.  Right: Some images (example shown: transverse magnetization (a,c) amplitude and (b,d) orientation, with (b,d) showing zoomed-in region) showed spontaneously formed spin vortices, detected by identifying regions around which the transverse magnetization wound around by $2 \pi$.  The center of the spin vortex (within thin line in (c-e)) was found to be unmagnetized (longitudinal spin shown in (e)), identifying the vortex as a polar-core spin vortex, which is the only topological defect of a ferromagnetic superfluid.  Figure adapted from Ref.\ \cite{sadl06symm}.\label{fig:sadlerquench}}
\end{figure}

The spatial distribution of the magnetization produced through such a quench was richly variegated.  As shown in Fig. \ref{fig:sadlerquench}, the quench produces a spinor gas that is strewn with regions of large transverse magnetization.  The orientation of the transverse magnetization varies spatially across the gas.  Between these magnetized regions, one observes dark walls that lack transverse magnetization, at least when viewed at limited spatial resolution.  One observes also vortex structures, where the transverse magnetization rotates by a full $2 \pi$ along a closed path.  Within the core of these vortices, the gas appears unmagnetized, consistent with the predicted property of a polar-core spin vortex, which is the only stable topological defect in a two-dimensional, $F=1$ ferromagnetic quantum field \cite{make03defects}.  Experiments performed on quantum quenches to different values of $q$ revealed the post-quench spin texture to be characterized by a variable length scale, roughly consistent with the idea that the dominant structure of the texture is generated by the spin-fluctuation mode of greatest temporal gain.

Experiments on these spatially extended spinor condensates have yet to reveal the subtle details found in the single-mode experiments.  It is possible that the spin-mixing instability in the large samples produces a highly squeezed, multimode quantum field.  A theoretical description of the dynamics in terms of squeezing in spatial spin modes defined dynamically was derived in Ref.\ \cite{sau10njp}.  The Berkeley group attempted to quantify the magnitude of spin fluctuations to confirm that the dynamics are indeed described as the parametric amplification of zero-point spin fluctuations, but could not resolve discrepancies between the data and a quantum amplifier model because of systematic uncertainties in the system properties \cite{lesl09amp}.

The experiments in spatially extended samples did raise hopes of performing quantitative tests of the dynamics of quenched quantum systems using ultracold gases.  One of the foci of attention is on the Kibble-Zurek picture of quenches across symmetry-breaking phase transition \cite{kibb76,zure85cosmo}.  This picture describes the dynamics as occurring in three stages.  First, as the system crosses the phase transition, at some point the system dynamics depart from equilibrium and the existing fluctuations in the system at that point are frozen into the system.  These frozen-in fluctuations have a characteristic size $d$ that scales with the rate at which the system is quenched across the transition point.  Within each region of size $d$, these fluctuations seed the growth of the symmetry-breaking order parameter.  Second, these regions of discordant symmetry-breaking heal to form topological defects that remain within the system at long times.  Third, at very long times, the topological defects move and merge, allowing the system to coarsen and develop larger regions of common broken symmetry.  The Kibble-Zurek picture has been applied to thermal phase transitions in several systems, including, recently, in scalar Bose gases that are cooled across the Bose-Einstein condensation transition \cite{chom15coherence,navo15kz}.  Experiments on spinor Bose-Einstein condensates quenched into a ferromagnetic state raised the prospect of studying the Kibble-Zurek picture also in systems that transit a quantum phase transition, where it would be quantum, rather than thermal, fluctuations that freeze in and then grow to become macroscopic \cite{dzia05,polk05univ,zure05qpt}.

The predictions of the second and third stages of the Kibble-Zurek picture, the development of topological defects and their subsequent coarsening \cite{guzm11}, were seen in the large-sample spinor gas experiments.  However, clean evidence of the first stage, i.e.\ of fluctuations frozen in with a length scale related quantitatively to the quench rate, has not been obtained.  Further experiments are warranted.

%For this, one should measure linear combinations of the conjugate spin observables $F_x$, a component of the transverse spin, and $Q_{yz}$, a component of the spin quadrupole tensor.  Exactly such a measurement has been performed by the Chapman group \cite{haml12squeezing}, as we discuss below.
%The two-body spin-mixing dynamics are coherent, but their coherence is not captured by examining one-body spin observables, which are the limit of what is described by mean-field theory.  Here it will be useful to remind the reader of these observables.  The space of such spin observables is spanned by nine operators (corresponding to the nine real numbers that make up the $F=1$ density matrix).  These basis operators include the identity operator, the three vector spin operators (denoted by symbols $F$), and the five spin quadrupole operators (denoted by symbols $Q$). 
% !TEX root = ./Varenna_spinors_main.tex
\section{Magnetic excitations\label{sec:excitations}}

The low-energy excitations and low-temperature thermodynamic properties of a many-body system are intimately tied to its symmetry, and in particular to how the ground state breaks symmetries of the Hamiltonian. A scalar Bose gas breaks a global gauge (phase) symmetry when it becomes a condensate, and phonons arise as a consequence of this symmetry breaking \cite{ande97book}. A Heisenberg ferromagnet breaks rotational symmetry because its ground state chooses an orientation in space, and this results in excitations known as magnons \cite{ande97book}. A spinor condensate breaks both phase and rotational symmetry, and thus has both phonons and magnons \cite{wata12unified}. These emergent particles are known as Nambu-Goldstone bosons and are a general feature of spontaneous symmetry breaking.

It has been observed that many systems at low temperatures behave like a weakly interacting gas of elementary excitations. For example, the elementary excitations of a scalar Bose gas are phonons. At low temperatures, the thermodynamics of the scalar Bose gas can be determined by treating the Bose gas as a ``vacuum'' filled with weakly interacting phonons. For this reason, the thermodynamics of a low temperature Bose gas are remarkably similar to that of blackbody radiation, only for a Bose gas there is only one polarization of phonons and the speed of light is replaced by the speed of sound.

We are then motivated to measure properties of elementary excitations (i.e., their dispersion relation, particle statistics, and interactions), since from these we will be able to predict thermodynamic and transport properties of the system. Phonons are well studied in scalar Bose gases \cite{dalf99rmp}, and so in this section particular interest is paid to understanding the properties of magnons.

\subsection{Quasiparticles of a spin-1 spinor condensate}
Both the ferromagnetic and the polar states of $F=1$ spinor condensates break three continuous symmetries.  In both cases, the state of the condensate selects an axis in spin space, which we may define to be the $\mathbf{z}$ axis, along which the ferromagnetic state is in the $|m_z = +1\rangle$ eigenstate and the polar state is in the $|m_z = 0\rangle$ state.  In selecting this axis, the states breaks two symmetries of the Hamiltonian, the symmetries under rotations about either of the transverse axes ($\mathbf{x}$ and $\mathbf{y}$).  The third broken symmetry is the $U(1)$ symmetry of the energy of the system upon multiplying the condensate wavefunction by a complex phase.  In the case of the ferromagnetic state, there is a subtlety that it is possible to rotate the spin state about the $\mathbf{z}$ axis by an angle $\phi$ and then to multiply the condensate wavefunction by an overall phase $e^{i \phi}$ so that the net effect is to leave the state unchanged.  The state and the Hamiltonian remain invariant under this combined spin-gauge rotation.  In this case, the ``orthogonal'' spin-gauge rotation (rotate by $\phi$ and multiply by $e^{-i \phi}$) is more properly identified as belonging to the $U(1)$ symmetry group that is broken by the condensate state \footnote{At high temperatures, an $F=1$ Bose gas is symmetric under rotations and the addition of a quantum mechanical phase, summarized by the symmetry group $G=SO(3)\times U(1)$. At low temperatures, a ferromagnetic condensate adopts a state that still has a $H=U(1)$ spin-gauge symmetry. The broken symmetries are characterized by the order parameter manifold $M=G/H=SO(3)$ \cite{muke06topo,stam13rmp}.}

While both states break three continuous symmetries, they have different numbers of Nambu-Goldstone bosons.  The polar state has three such bosonic excitations: one phonon, associated with the broken gauge symmetry, and two independent magnons, associated with each of the broken rotational symmetries.  All three excitations are gapless and linearly dispersing with wavevector at low energy, as we derive in Sec.\ \ref{sec:polarmagnons}.  The ferromagnetic condensate has only two Nambu-Goldstone bosons (derived in Sec.\ \ref{sec:ferromagnetic_magnons}).  The broken spin-gauge symmetry results in a gapless and linearly dispersing phonon excitation as in the case of the polar condensate.  However, the two broken rotational symmetries result only in one magnon excitation and this excitation disperses quadratically with wavevector.  The difference in the case of the ferromagnetic state is that excitations produced by rotating the condensate magnetization about the $\mathbf{x}$ and $\mathbf{y}$ state become dynamically coupled.  The origin of this coupling is the non-commutativity of the generators of these two rotations.  As a result of this coupling, the number of Nambu-Goldstone bosons is reduced while the power of the dispersion relation is promoted ($E(\mathbf k) \propto |{\mathbf k}|^2$ instead of $E(\mathbf k) \propto |\mathbf k|$) \cite{wata12unified}.

\subsection{Linearized Schr\"odinger equation\label{sec:linearized_schrodinger}}
At low temperatures, spinor condensates can be described well by a nonlinear Schr\"odinger equation. Many of the low-energy excitations can be captured by linearizing a perturbation expansion of this equation around its ground state. This is particularly true for Nambu-Goldstone bosons, where we may know \textit{a priori} what a perturbation should look like. Moreover, expansion of the perturbation to higher order can show interactions between excitations. In this subsection, we use this approach to derive the dispersion relation for magnons and phonons in ferromagnetic and polar spinor gases.

Certain classes of excitations cannot be understood by merely linearizing the nonlinear Schr\"odinger equation. Topological excitations, such as vortices, may have defects that cannot be described as a small excursion from the ground state. Solitons, or non-dispersing waves, only exist because of nonlinearities, with properties that depend on their amplitude in a way that cannot be captured in a linearized analysis. These types of excitations are relatively energetic compared to the Nambu-Goldstone bosons, but are important in many systems because they are stable and can persist for experimentally relevant time.

\subsubsection{Ferromagnetic $F=1$ condensate\label{sec:ferromagnetic_magnons}}

The elementary excitations of a ferromagnetic condensate are phonons, fluctuations of the overall density and phase of the condensate, and magnons, fluctuations of the orientation of the spin direction. In addition, there is a gapped mode from fluctuations in the length of $\mathbf F$. Here we derive these three modes by linearizing perturbations of the wavefunction.

First, we will use a perturbation expansion to derive the phonon and magnon dispersion relations. A good ansatz is to perturb the ground state along directions of broken symmetry. The ground state of a ferromagnetic condensate ($c^{(1)}_1 < 0$) is one where $|\mathbf F| = 1$, and we orient our coordinate system so that $\mathbf F$ points along $\mathbf z$, with a ground state $\psi_\mathrm{ground} = \sqrt{n}\,(1, 0, 0)$ for a uniform condensate with density $n$\footnote{Be mindful that here, in keeping with notation in the literature, we are choosing to include a factor of $\sqrt{n}$ in condensate wavefunction, so that the wavefunction is now normalized to the number of particles.}. The broken symmetries are the phase of the condensate and orientation of the spin, so we perturb the wavefunction by multiplying the density by a complex number ($1+\chi$) and changing spin orientation ($\theta$, $\phi$):
\begin{equation}\label{eqn:excitation_ansatz}
\psi_\mathrm{ansatz} = \sqrt{n} e^{-i \mu t/\hbar} ( 1 + \chi)\,R(\theta, \phi)
\left( \begin{array}{c} 1 \\[0.2cm] 0 \\[0.2cm] 0 \end{array} \right)
\xrightarrow{\theta \ll 1}
\sqrt{n} e^{-i \mu t/\hbar} (1 + \chi) \left( \begin{array}{c}
1-\frac{1}{4} \theta^2 \\[0.2cm] \frac{1}{\sqrt{2}} \theta e^{- i \phi} \\[0.2cm] \frac{1}{4} \theta^2 e^{-2 i \phi}\end{array} \right)
\end{equation}

The ansatz can be simplified by rewriting the two (real) angles $\theta$ and $\phi$ as a single complex number $\xi = \frac{1}{\sqrt 2} \theta e^{-i \phi}$, which will turn out to be the wavefunction for $m_F=0$ component.
\begin{equation}\label{eqn:excitation_ansatz2}
\psi_\mathrm{ansatz} \approx
\sqrt{n} e^{-i \mu t/\hbar} \left( \begin{array}{c}
1 + \chi \\ \xi \\ 0 \end{array} \right)
\end{equation}
The $m=-1$ component is neglected because it is of order $\xi^2$. Contact interactions do not depend on $\theta$ or $\phi$ and so the interaction term is $\mu |1+\chi|^2 \psi$ with $\mu = (c_0^{(1)} + c_1^{(1)}) n$. We can now do a perturbation expansion in $\chi$ and $\xi$.
\begin{align}
i \hbar \dot \chi =&{} - \frac{\hbar^2}{2m} \nabla^2 \chi + 2\mu (\chi + \chi^\ast )\label{eqn:phonons_linear}\\
i \hbar \dot \xi =&{} - \frac{\hbar^2}{2m} \nabla^2 \xi \label{eqn:magnons_linear}
\end{align}

These two equations represent the two Nambu-Goldstone bosons of our system: phonons ($\chi$) and magnons ($\xi$). The eigenspectrum of Eq.~\ref{eqn:phonons_linear} is the well-known Bogoliubov phonon with $E_\chi = \sqrt{(\hbar^2 k^2/2m)(\hbar^2 k^2/2m + 2 \mu)}$ (see Ref.~\cite{dalf99rmp} for more on phonons). Magnons, as described by Eq.~\ref{eqn:magnons_linear}, have free-particle solutions $E_\xi = \hbar^2 k^2/2m$ with no gap and an effective mass identical to the bare mass of rubidium. As expected, both of these modes are gapless.

What does the wavefunction of a magnon look like? When $\chi=0$, Eq.~\ref{eqn:magnons_linear} is solved by $\xi = \xi_0\,e^{i \left( \mathbf k \cdot \mathbf x - \omega(k) t \right)}$. For small values of $|\xi_0|$, this solution represents a magnetization with a fixed polar angle $\theta = \sqrt{2} |\xi_0|$ and a periodic azimuthal angle $\phi=\mathbf k\cdot \mathbf x-\omega(k)t$.
\begin{equation}\label{eqn:magnon_solution}
\psi_\mathrm{magnon} = \sqrt{n} e^{-i \mu t/\hbar} \left( \begin{array}{c}
1 \\ \frac{1}{\sqrt{2}} \theta e^{i ( \mathbf k \cdot \mathbf x - \omega(k) t)} \\ 0
\end{array} \right)
\simeq \sqrt{n} e^{-i \mu t/\hbar}\,R(\theta, \mathbf k \cdot \mathbf x - \omega(k)t) \left( \begin{array}{c}
1 \\ 0 \\ 0 \end{array} \right)
\end{equation}

While phonons and magnons are gapless, there is in fact a gapped magnetic mode corresponding to fluctuations in the length of $\mathbf F$. This mode does not correspond to excitations along a broken symmetry direction. An ansatz for this mode is $\psi \approx \sqrt{n} e^{-i\mu t/\hbar}(\sqrt{1-\gamma^\dagger \gamma}, 0, \gamma)$, which has constant density but varying $|\mathbf F|$. We now do a perturbation expansion in $\gamma$.
\begin{equation}
i \hbar \dot \gamma = - \frac{\hbar^2}{2 m} \nabla^2 \gamma + 2 |c^{(1)}_1| n \gamma
\end{equation}
The energy of this mode is $E_\gamma = 2 |c_1^{(1)}| n + \hbar^2 k^2/2m$.

%How do these excitations propagate in a non-uniform condensate, as might occur in a real experiment? The density of most condensates varies smoothly from its peak in the center to zero on the edge. The dispersion relation of phonons depends on the local density, and so the propagation of phonons is strongly affected as it moves through a nonuniform medium \cite{Dalfovo1999}.

%In contrast, the magnons dispersion relation appears to be unrelated to the condensate density. For this reason, magnons can propagate through a nonuniform condensate unperturbed by changes in density, until the reach the edge. Magnons behave as an ideal gas within the condensate, even in the presence of a nonuniform trap.

\subsubsection{Polar $F=1$ condensate}
\label{sec:polarmagnons}
For a polar condensate, we instead find three Nambu-Goldstone bosons: phonons and two polarizations of magnons. Unlike the ferromagnetic case, these magnons have a linear dispersion relation. The ground state of the polar wavefunction oriented along the $\mathbf z$ is $\psi_\mathrm{ground} = \sqrt{n} (0, 1, 0)$. Before we write down our ansatz, it is helpful to rotate the wavefunction to guess the correct form of perturbation.
\begin{equation}
\sqrt{n} e^{-i \mu t/\hbar} ( 1 + \chi)\,R(\theta, \phi)
\left( \begin{array}{c} 0 \\[0.1cm] 1 \\[0.1cm] 0 \end{array} \right)
\xrightarrow{\theta \ll 1}
\sqrt{n} e^{-i \mu t/\hbar} (1 + \chi) \left( \begin{array}{c}
-\frac{1}{\sqrt 2}\theta e^{i \phi} \\[0.1cm] 1 \\[0.1cm] \frac{1}{\sqrt 2}\theta e^{-i \phi}\end{array} \right)
\end{equation}

To capture the two polarizations of magnons, we will need two complex variables $\xi_\pm = \frac{1}{\sqrt 2}\theta_\pm e^{-i \phi_\pm}$ for our perturbation expansion. Expressing fluctuations about the mean-field ground state in the form $\psi = \sqrt{n} e^{-i \mu t/\hbar} (\xi_+, 1, \xi_-^\dagger)$, we obtain the following equations of motion:
\begin{eqnarray}
i \hbar \dot \xi_+ &= - \frac{\hbar^2}{2m} \nabla^2 \xi_+ + c^{(1)}_1 n (\xi_+  + \xi_-^\dagger) \\
i \hbar \dot \xi_- &= - \frac{\hbar^2}{2m} \nabla^2 \xi_- + c^{(1)}_1 n (\xi_+^\dagger  + \xi_-)
\end{eqnarray}
These equations are coupled, reflecting the fact that a polar-state magnon is a superposition of excitations in the $|m_F = \pm 1\rangle$ magnetic sublevels; indeed, the formalism used in Sec.\ \ref{sec:spinmixing} to describe spin fluctuations in a Cartesian basis already captures this coupled form of the polar-state magnon.  From these equations, we obtain finally a magnon dispersion relation similar in form to that of Bogoliubov excitations of a scalar condensate (Eq.~\ref{eqn:phonons_linear}), namely
\begin{equation}
E_\mathrm{\xi_\pm} = \sqrt{(\hbar^2 k^2/2m)(\hbar^2 k^2/2m + 2 c_1^{(1)} n)}\;.
\end{equation}
We note that the dispersion relation is linear at long wavelengths.

\subsection{Making and detecting magnons\label{sec:making_detecting_magnons}}
A weak magnetic field perturbs an atom by adding a term $H_B = - \hat{\boldsymbol{\mu}} \cdot \mathbf{B} = - \frac{\mu}{F} \mathbf{B} \cdot \hat{\mathbf{F}}$ to the Hamiltonian. Light tuned to the right wavelength can create the same Hamiltonian, called the ac vector Stark shift or optical Zeeman effect, except the magnetic field $\mathbf{B}$ is replaced by a function of the intensity, detuning, and polarization of the light. Using light has a key advantage: the Helmholtz equation that governs optics allows for much more varied structures (e.g., local maxima) than Laplace's equations allow for magneto-statics. We use this method to create effective magnetic fields with either Gaussian or sinusoidal profiles.

\begin{figure}[t]
\begin{center}
\begin{subfigure}[t]{.48\textwidth}
	\centering\includegraphics[scale=1]{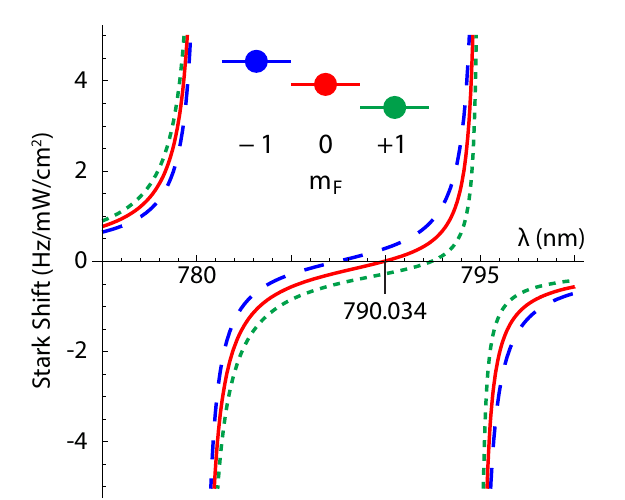}
	\captionsetup{skip=3pt}
	\caption{\label{fig:Stark-Shift}}
\end{subfigure}
\begin{subfigure}[t]{.48\textwidth}
	\centering\includegraphics[scale=1]{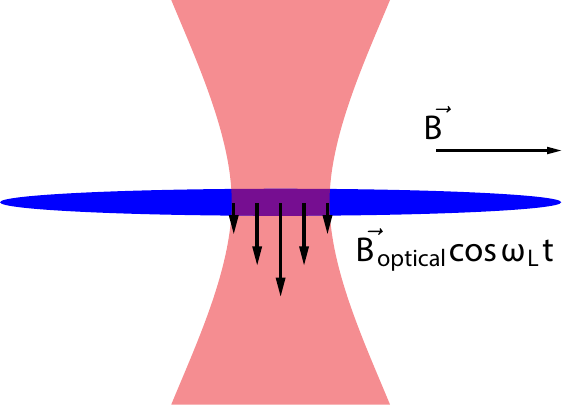}
	\captionsetup{skip=3pt}
	\caption{\label{fig:optical_zeeman_cartoon1}}
\end{subfigure}
\end{center}

\begin{center}
\begin{subfigure}[t]{.95\textwidth}
	\centering\includegraphics[scale=0.8]{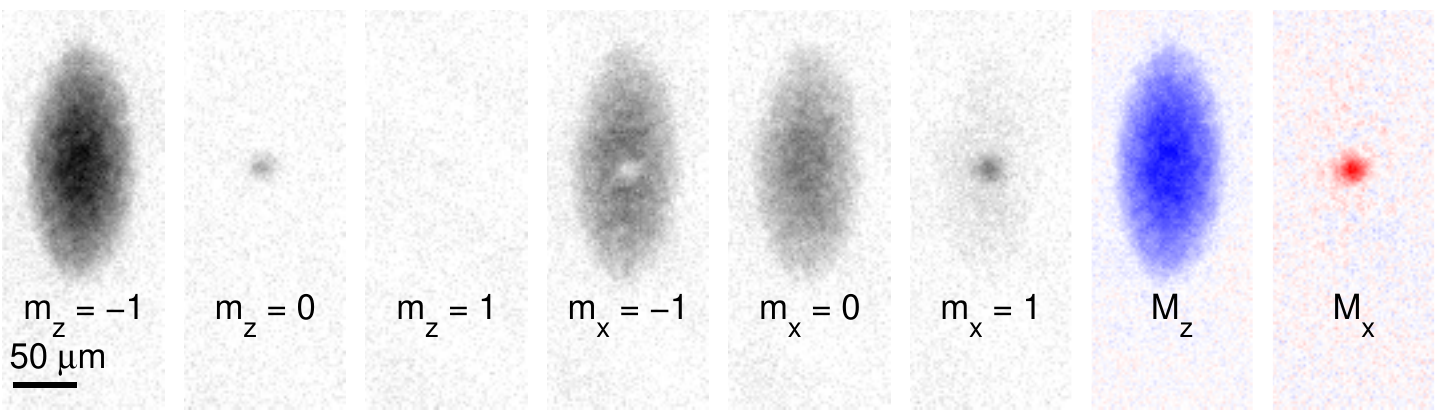}
	\captionsetup{skip=3pt}
	\caption{\label{fig:magnon_weak}}
\end{subfigure}
\end{center}
\caption[AC Stark shift and suppression of phonons at the magic-zero wavelength.]{(a) AC Stark shift and suppression of phonons at the magic-zero wavelength. Calculated ac Stark shift of the $F=1$ states for circularly polarized light. The ac Stark shift of the $m_F=0$ state vanishes at 790.034(7) nm \cite{aror11tuneout}. (b) A focused light beam (red hyperbola) introduces an effective transverse magnetic field proportional to the local intensity. As the laser amplitude is modulated at the Larmor frequency, the magnetization of the atoms (blue trap) is rotated about a transverse spin axis. (c) A rotation of the longitudinal magnetization can be measured as small wavepacket in the $m_F=0$ population with ASSISI.  Here six image frames (first six, shown in gray scale) are combined to measure the column magnetizations $\tilde{M}_z$ and $\tilde{M}_x$ (last two images, in color)}
\end{figure}

In alkali atoms, hyperfine states can be coupled through the electronic excited states. Fig.~\ref{fig:Stark-Shift} shows the calculated ac Stark shift of the three $F=1$ Zeeman states in the presence of a circularly polarized laser. At the ``magic-zero'' or ``tune-out'' wavelength 790.03 nm, the ac Stark shift vanishes for atoms in the  $|m_F=0\rangle$ ground state \cite{aror11tuneout}. Like the linear Zeeman effect of a magnetic field, the Stark shift at this wavelength is proportional to $m_F$. The strength of the effective magnetic field is given by the two-photon Rabi frequency $\Omega$, which is proportional to the laser intensity. A laser tuned to this wavelength cannot excite phonons in an $m_F=0$ condensate (Fig.~\ref{fig:Stark-Shift}(b)), but can create magnetic excitations.

When the propagation axis of the light is perpendicular to the magnetic field $\mathbf B$, the light can induce a spatially and temporally-varying transverse magnetic field $H_\mathrm{laser}(\mathbf r, t) = F_y \Omega(\mathbf r, t)$, where $\Omega(\mathbf r, t)$ is proportional to the local optical intensity $I(\mathbf r, t)$. Like a ``real'' magnetic field, the light drives transitions between Zeeman states when its intensity is driven resonantly, giving $\Omega(\mathbf r, t) = \frac{1}{2}\Omega(\mathbf r)(1-\cos \omega_L t)$, where $\omega_L=\mu |\mathbf B|/\hbar F$ is the Larmor frequency.

One simple structure is a near-Gaussian wavepacket in the middle of the condensate. This structure can be made by focusing a light beam through the center of the cloud (Fig.~\ref{fig:optical_zeeman_cartoon1}). The effective magnetic field is now inhomogeneous, with $\Omega(\mathbf r, t) = \frac{1}{2} \Omega_0 (1- \cos (\omega_L t))\,e^{-|\mathbf r|^2/(2 \sigma^2)}$. After a time $\tau$, the light is extinguished and the magnetization ends up having been rotated by an angle $\theta(\mathbf r) = \theta_0 e^{-|\mathbf r|^2/(2 \sigma^2)}$, where $\theta_0 = \Omega_0 \tau \ll 1$. For experimental convenience, the atomic spin is initially oriented along $-\mathbf z$, with a wavefunction of $\psi_\mathrm{ground}=\sqrt{n}(0,0,1)$, where $n$ is the local density of the condensate. After rotation, the wavefunction now acquires a population in the $|m_z=0\rangle$ state (see Eq.~\ref{eqn:magnon_solution}):
\begin{equation}
\psi(\mathbf r) = e^{-i F_y \theta(\mathbf r)/\hbar} \sqrt{n} \left( \begin{array}{c} 0 \\ 0 \\ 1 \end{array} \right) = \sqrt{n} \left( \begin{array}{c} O(\theta_0^2) \\ -\frac{1}{\sqrt 2} \theta_0 e^{-|\mathbf r|^2/(2 \sigma^2)} \\ 1 - O(\theta_0^2) \end{array} \right)
\end{equation}

An image of the resulting structure measured by the ASSISI method is shown in Fig.~\ref{fig:magnon_weak}. To monitor dynamics and the evolution of magnon textures, it is often sufficient to monitor the evolution of the population in the $|m_z=0\rangle$ state, $|\psi_{m_F=0}(t)|^2=\frac{1}{2} n \theta(\mathbf r)^2$, as will be done in Sec.~\ref{sec:magnon_propagation}. As discussed in Sec.~\ref{sec:ferromagnetic_magnons}, the wavefunction of the $|m_F=0\rangle$ component is a good description of magnons in an $|m_z=+1\rangle$ (or $|m_z=-1\rangle$) condensate, and so images of $|\psi_{m_F=0}|^2$ can be thought of as direct images of the magnon population.

\begin{figure}[t]
\centering
\begin{subfigure}[t]{0.44\textwidth}
	\centering
	\includegraphics[scale=0.9]{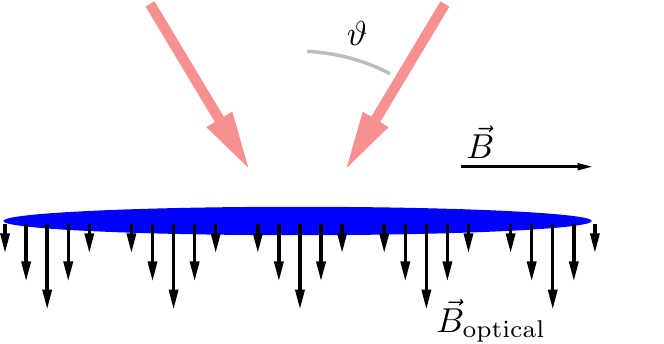}
	\caption{\label{fig:optical_zeeman_cartoon2}}
\end{subfigure}
\begin{subfigure}[t]{.45\textwidth}
	\centering
	\includegraphics[scale=0.9]{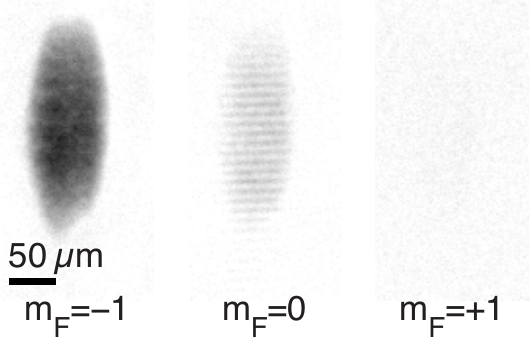}
	\captionsetup{skip=3pt}
	\caption{\label{fig:magnon_assisi}}
\end{subfigure}
\caption[Standing wave of magnons]{Standing wave of magnons. (a) A standing wave of magnons is generated by illuminating a longitudinally magnetized spinor gas with two coherent plane waves of light that intersect at a half-angle $\vartheta$.  With the light tuned to the proper wavevector ($k_L=2\pi/(790.03\;\mathrm{nm})$ for $^{87}$Rb), the ac Stark shift of the light is equivalent to an effective magnetic field $B_\mathrm{optical}$ that is proportional to the local intensity of the optical standing wave.  This light rotates the magnetization, creating magnons at $\mathbf{0}$, $-\mathbf{k}$, and $-\mathbf{k}$, where $|\mathbf{k}| = 2 k_L\sin\vartheta$. (b) ASSISI imaging of standing wave of magnons with a $10\;\mu\mathrm{m}$ wavelength and $\theta_\mathrm{avg} = 14^\circ$. Note the high visibility fringes in the $m_F=0$ image. \label{fig:interferometer-cartoon}}
\end{figure}

Another important spin texture is a standing wave of magnons. This structure is created by two plane waves of light, both at the same wavelength that produces a vector ac Stark shift, intersecting at a half-angle $\vartheta$ (Fig.~\ref{fig:optical_zeeman_cartoon2}). Now, $\theta(\mathbf r) = \theta_\mathrm{avg} (1+\cos \mathbf k\cdot \mathbf r)$ with $|\mathbf{k}| = 2 k_L \sin\vartheta$ with $k_L$ being the optical wavenumber, and $\theta_\mathrm{avg} = \Omega_0 \tau/2$. For small angles ($\theta_\mathrm{avg}\ll 1$), we can approximate the perturbed wavefunction as
\begin{equation}
\psi(\mathbf r) = e^{-i F_y \theta(\mathbf r)/\hbar} \sqrt{n} \left( \begin{array}{c} 0 \\ 0 \\ 1 \end{array} \right) = \sqrt{n} \left( \begin{array}{c} O(\theta_0^2) \\ -\frac{1}{\sqrt 2} \theta_\mathrm{avg} (1+\cos \mathbf k\cdot \mathbf r) \\ 1 - O(\theta_0^2) \end{array} \right)
\end{equation}
Expanding this expression to order $\theta_\mathrm{avg}^2$, we have
\begin{align}
\left( \begin{array}{c}
0 \\[0.1cm]
-\frac{1}{\sqrt 2} \theta_\mathrm{avg} (1 + \cos \mathbf k \cdot \mathbf r) \\[0.1cm]
1
\end{array} \right)
 = \nonumber\\
\left( \begin{array}{c} 0 \\[0.1cm] 0 \\[0.1cm] 1 \end{array} \right)
- \frac{1}{\sqrt 2} \theta_\mathrm{avg} \left( \begin{array}{c} 0 \\[0.1cm] 1 \\[0.1cm] 0 \end{array} \right)
&- \frac{1}{2\sqrt 2} \theta_\mathrm{avg} \left( \begin{array}{c} 0 \\[0.1cm] e^{-i \mathbf k \cdot \mathbf r} \\[0.1cm] 0 \end{array} \right)
- \frac{1}{2\sqrt 2} \theta_\mathrm{avg} \left( \begin{array}{c} 0 \\[0.1cm] e^{i \mathbf k \cdot \mathbf r} \\[0.1cm] 0 \end{array} \right)\label{eqn:magnon_plane_waves}
\end{align}
In Sec.~\ref{sec:magnon_interferometry}, this initial state will serve as the basis for magnon interferometry.

An alternative approach to creating coherent magnon wavepackets is to use light at constant intensity, rather than modulating the intensity at the Larmor frequency.  When the circular polarization of the light has a component along the magnetic field, then it also introduces a Hamiltonian $H(\mathbf r) = \hbar \Omega(\mathbf r,t) F_z$.  Such a static Hamiltonian creates magnon excitations within a transversely oriented ferromagnet.  This method was used to write spin textures that served as test targets to evaluate the performance of a spinor-gas magnetometer  \cite{veng07mag}.

Raman transitions can also be used to create excitations that cannot be generated with an effective magnetic field. Magnetic fields drive transitions with $\Delta m = 0, \pm 1$, whereas a two-photon Raman transition can also drive $\Delta m = \pm 2$ transitions. These were used in a gas of $F=2$ ${}^{87}$Rb atoms to generate superpositions of only the even $m_F$ states ($|m_F=\{-2, 0, 2\}\rangle$) \cite{lesl09skyrmion}.

\subsection{Magnon propagation\label{sec:magnon_propagation}}
In experiments, spinor condensates are trapped in an inhomogeneous trapping potential. Surprisingly, however, magnons propagate through the volume of an inhomogeneous spinor condensate as if they were free particles in a uniform potential. To see this, we add an additional state-independent potential $V(\mathbf r)$ to the Hamiltonian and derive a slightly different linearized equation for magnons:
\begin{align}\label{eqn:magnon_effective_potential}
i \hbar \dot \xi = {}& - \frac{\hbar^2}{2m} \nabla^2 \xi + \left[ \left(c_0^{(1)} + c_1^{(1)}\right) n(\mathbf r) + V(\mathbf r) - \mu \right] \xi \\\nonumber
= {}& - \frac{\hbar^2}{2m} \nabla^2 \xi + V_\mathrm{eff}(\mathbf r) \xi
\end{align}
with
\begin{equation}
V_\mathrm{eff}(\mathbf r) = \left(c_0^{(1)} + c_1^{(1)}\right) n(\mathbf r) + V(\mathbf r) - \mu
\end{equation}
Here, $n(\mathbf r)$ is the nonuniform density of the condensate (the majority spin state). If $V(\mathbf r)$ varies slowly and $\xi$ is small, then the condensate ``fills'' its container such that $n(\mathbf r) = \left(\mu - V(\mathbf r) \right) / \left( c_0^{(1)} + c_1^{(1)}\right)$, at least everywhere  $\mu > V(\mathbf{r})$ and the condensate density $n$ is nonzero. Thus, throughout the volume where the majority-spin condensate has nonzero density, $V_\mathrm{eff}(\mathbf r)$ is precisely zero and the magnons can propagate through the volume of a nonuniform condensate like noninteracting particles through vacuum.

\begin{figure}[t]
\centering
\begin{subfigure}[c]{.6\textwidth}
	\centering
	\includegraphics[scale=0.85]{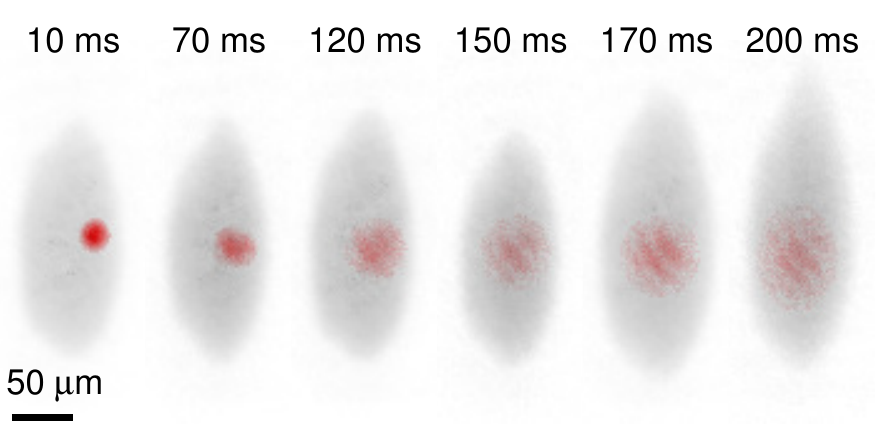}
	\caption{\label{fig:magnon_expansion_images}}
\end{subfigure}
\begin{subfigure}[c]{.38\textwidth}
	\centering
	\includegraphics[scale=0.9]{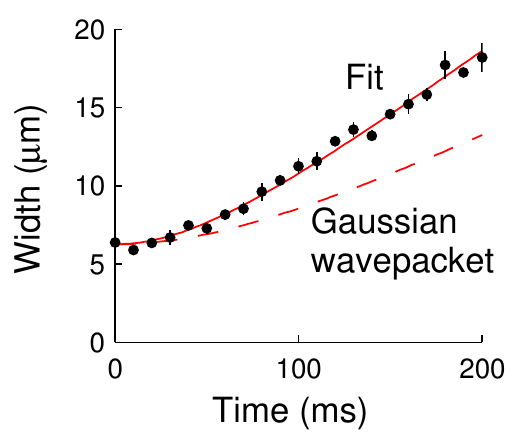}
	\caption{\label{fig:magnon_expansionB}}
\end{subfigure}
\caption[Expansion of a magnon wavepacket]{Expansion of a magnon wavepacket. A Gaussian wavepacket of spin is created with a focused optical beam that applies a nonuniform spin rotation. (a) Expansion of the $m_F=0$ atoms (red) in the presence of a weak magnetic field gradient over 200 ms. The condensate profile is shown in light gray. (b) An average of the width over time for three runs (black circles).  The data are fitted to a hyperbola (red line), as expected for the evolution of a free Gaussian wavepacket in time, but the expansion appears faster than is expected for a Heisenberg-limited wavepacket (red dashed line). This discrepancy is most likely because the light intensity is not exactly Gaussian, as suggested by the structure seen after 120 ms.\label{fig:magnon_expansion}}
\end{figure}

This free propagation of magnons is illustrated qualitatively in the images of Fig.\ \ref{fig:magnon_expansion}, where we create a localized near-Gaussian wavepacket of magnons within a ferromagnetic spinor Bose-Einstein condensate and observe its propagation in time.  The wavepacket is observed to expand slowly over 100's of ms, in spite of being confined (along with the rest of the spinor gas) within the optical potential of a focused light-beam trap and the pull of gravity.

What if the potential were spin-dependent, such as would be created by an inhomogeneous magnetic field? In this case we would consider a potential $V_{-1}(\mathbf r)$ for the majority spin state and a slightly different potential $V_{0}(\mathbf r)$ for the $m_F=0$ component. The effective potential governing magnon propagation, appearing in Eq.~\ref{eqn:magnon_effective_potential}, would instead be written as
\begin{equation}
V_\mathrm{eff}(\mathbf r) = V_0(\mathbf x) - \left(c_0^{(1)} + c_1^{(1)}\right) n(\mathbf r) - \mu = V_0(\mathbf r) - V_{-1}(\mathbf r).
\end{equation}

\begin{figure}[t]
\centering
\includegraphics[width=\textwidth]{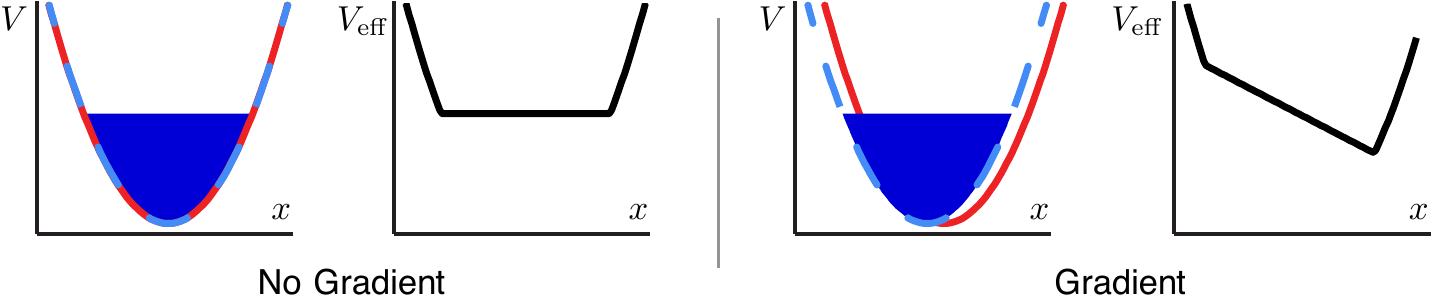}
\caption[Effective potential of a magnon in a gradient.]{Effective potential of a magnon in a gradient. (Left) Without a magnetic field gradient, the potential $V_{-1}$ for $m_F=-1$ atoms (dashed blue line) is the same as the potential $V_{0}$  for $m_F=0$ atoms (solid red line). The effective potential for magnons $V_{-1}-V_{0}$ is a combination of the attractive optical trap and repulsive interaction energy of the $m_F=-1$ condensate (blue region). These contribute to an effectively flat trap where the local chemical potential is nonzero (solid black line). (Right) In the presence of a magnetic field gradient, the density of the $m_F=-1$ condensate is shifted (dark blue region) while the potential for $m_F=0$ (red line) is unchanged. The effective potential (black) for $m_F=0$ is the sum, which contains the gradient. \label{fig:acceleration_cartoon}}
\end{figure}

\begin{figure}[!ht]
\centering
\begin{subfigure}[t]{.78 \textwidth}
	\centering
	\includegraphics[scale=1]{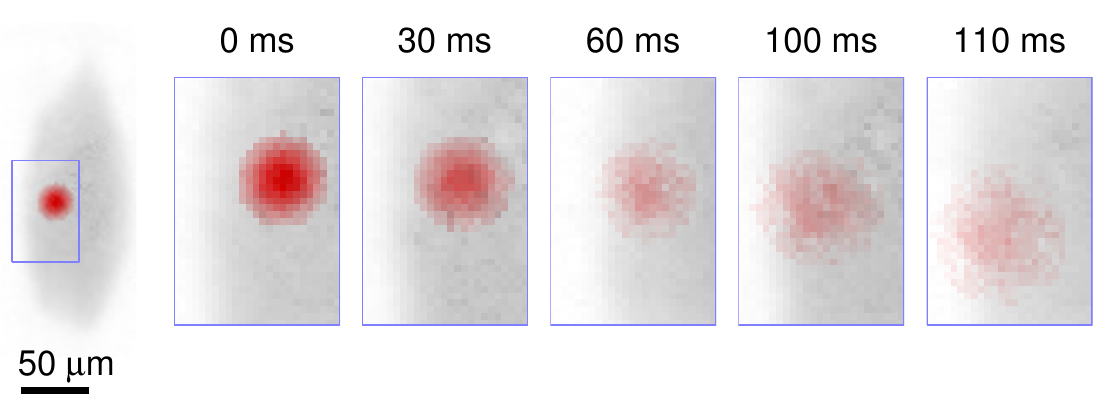}
	\caption{\label{magnetic_moment_images}}
\end{subfigure}
\begin{subfigure}[t]{.48\textwidth}
	\centering
	\includegraphics[scale=1]{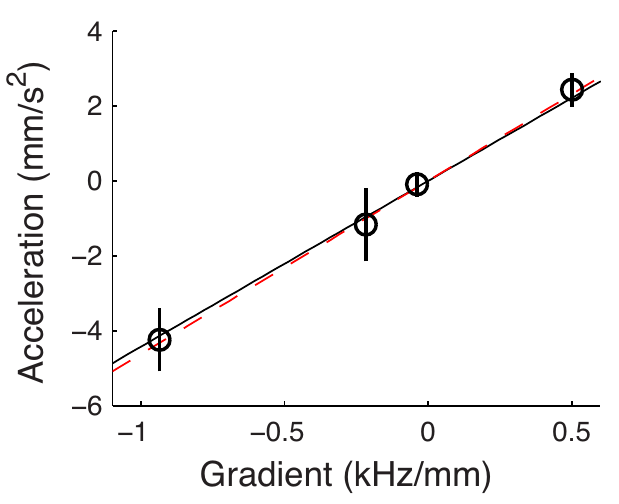}
	\caption{\label{fig:magnetic_moment2}}
\end{subfigure}
\caption[Magnetic moment of a magnon]{Magnetic moment of a magnon. A Gaussian wavepacket is created in one location. (a) The wavepacket is accelerated in the direction of the magnetic field gradient. The region-of-interest (blue box) is 50 by 75 $\mu\mathrm{m}$. (b) We fit the position versus time to a quadratic function to extract the acceleration (black circles). The magnon accelerates towards higher field with an effective magnetic moment $\mu^\ast = -1.04(8) \mu_\mathrm{condensate}$ (solid black line), consistent with the prediction of $\mu^\ast = -\mu_\mathrm{condensate}$ (red dashed line).
\label{fig:magnetic_moment}}
\end{figure}

One origin for such a state-dependent potential is an inhomogeneous magnetic field.  In particular, a magnetic field gradient produces an effective force on magnons that is uniform.  We consider just the adiabatic magnetic potential produced by an inhomogeneous magnetic field $\mathbf{B}(\mathbf{r})$, of the form $V_B(\mathbf{r}) = -(\mu/F) |\mathbf{B}(\mathbf{r})| m_F$.  Letting $|\mathbf{B}|$ vary just linearly with a gradient $B^\prime$ in the $\mathbf{x}$ direction, we have simply $V_0(\mathbf r) - V_{-1}(\mathbf r) = \mu B' x$. The state-independent component is exactly zero and only the state-dependent component remains, resulting in a uniform force proportional to the magnetic field gradient (Fig.~\ref{fig:acceleration_cartoon}). The resulting acceleration $a = \mu^\ast B'/m^\ast$ depends on the effective magnetic moment $\mu^\ast$ and effective mass $m^\ast$ of a magnon. The acceleration can be directly seen by the displacement of a magnon wavepacket over time (Fig.~\ref{fig:magnetic_moment}), and from this we can determine the effective magnon magnetic moment.  From the data of Fig.\ \ref{fig:magnetic_moment}, and as described in Ref.\ \cite{mart14magnon}, we found $\mu^\ast = -1.04(8) \mu_{-1}$ which has the same magnitude as but the opposite sign of the magnetic moment of the atoms that make up the condensate. It may be interesting to note that the magnetic moment of the magnon is nonzero even though its constituent particle, an $|F=1,m_F=0\rangle$ ${}^{87}$Rb atom, has zero magnetic moment.

\subsection{Magnon contrast interferometry and recoil frequency\label{sec:magnon_interferometry}}
How do we experimentally measure the predicted dispersion relation of a magnon in the ferromagnetic condensate, $E(\mathbf k) = \hbar^2 |\mathbf k|^2/2m$? One powerful answer is atom interferometry. We implemented a specific form interferometry, contrast interferometry, by manipulating the spin and momentum of the condensate. Every atom of the condensate is placed initially in a superposition of mostly the $|m_F = -1\rangle$ state with momentum $\mathbf 0$, and a small admixture of the $|m_F = 0\rangle$ state in the discrete momentum states $\mathbf 0$, $\hbar \mathbf k$, and $-\hbar \mathbf k$. This superposition can be described magnons with momenta $\mathbf 0$, $\hbar \mathbf k$, and $-\hbar \mathbf k$ moving through the condensate. Each magnon momentum state evolves in time $t$ with a phase $\phi(\mathbf k) = E(\mathbf k) t/\hbar$. Interference between different momentum states causes a standing wave in the spin density, with an amplitude that oscillates in time with the frequency $\left[E(\mathbf k) +E(-\mathbf k) - 2E(\mathbf 0)\right]/\hbar$.
%Physicists understands superpositions, waves, propagation, and interference. But she wants to learn the details because they're so exciting!
%
%\cite{Phuc2013}
%
%\cite{Saito2015}

Such a magnon contrast interferometer was demonstrated in Ref.\ \cite{mart14magnon}, inspired by the application of contrast interferometry with density excitations in a sodium condensate \cite{gupt02inter}.  The superposition of magnon waves described above was created using the optical methods detailed in Sec.\ \ref{sec:making_detecting_magnons}.  This scheme can also be described in terms of Raman scattering, in which atoms in the $m_F = -1$ condensate absorb a photon from one light beam and then transition to the $|m_F = 0\rangle$ state by emitting another photon into either of the light beams illuminating the gas.  The Zeeman energy difference between these internal states is made up by the frequency difference between the two photons. Recall that in our optical method of imprinting magnons we utilize light that is amplitude modulated at the Larmor frequency, so that the spectrum of the light incident upon the atoms contains frequency components suited for such stimulated Raman transitions.  Different choices of which light field is absorbed and which light field is emitted provide the Raman scattering pathways to final states with magnon momenta $\mathbf{0}$ and $\pm \hbar \mathbf{k}$.  The coherence between these different transition amplitudes establishes the initial high-contrast interference of the magnon waves.

%\subsection{Description of experiment: spin wave and contrast interferometry\label{sec:basic_magnon_interferometry}}
\begin{figure}[!t]
\centering
\begin{subfigure}[t]{0.44\textwidth}
	\centering
	\includegraphics[scale=0.9]{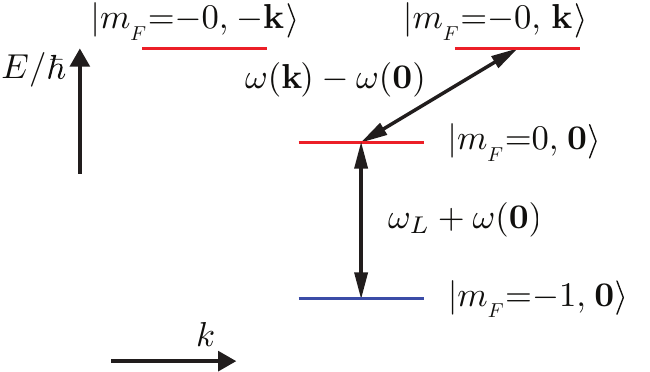}
	\caption{}
\end{subfigure}
\begin{subfigure}[t]{0.44\textwidth}
	\centering
	\includegraphics[scale=0.85]{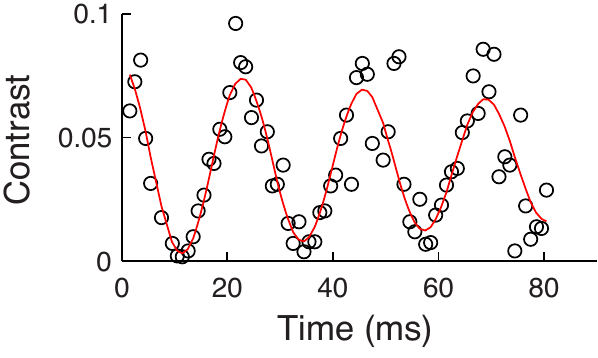}
	\caption{}
\end{subfigure}
\caption[Contrast interferometer scheme]{Contrast interferometer scheme. (a) Every atom in the ground state $|m_F={-}1, \mathbf{0}\rangle$ level is excited into a superposition with a small admixture of the $|m_F=0,\mathbf{0}\rangle$, $|m_F=0,-\mathbf{k}\rangle$, and $|m_F=0,\mathbf{k}\rangle$  states (the states are labeled within the ket by their wavevector).  We define the dispersion relation $\omega(\mathbf k) = E(\mathbf k)/\hbar$ as offset by the Larmor frequency $\omega_L\mu B/\hbar$, the energy shift induced by the magnetic field. (b) The contrast of the $\cos(\mathbf{k}\cdot\mathbf{r})$ modulation of the magnon density oscillates in time at a frequency $(E(\mathbf k) + E(-\mathbf k) - 2 E(\mathbf 0))/\hbar = \hbar |\mathbf k|^2/m^\ast$. These data show the temporal evolution of the pattern imaged in Fig.~\ref{fig:magnon_assisi}.}
\end{figure}

%As discussed in Sec.~\ref{sec:making_detecting_magnons}, plane waves of magnons can be generated with the optical Zeeman effect by crossing two laser beams with a half-angle $\vartheta$. Adjusting $\vartheta$ changes the momentum transfer, $|\mathbf k| = 2 |\mathbf k_1| \sin(\vartheta/2)$.

\begin{figure}[t]
\centering
	\includegraphics[scale=1]{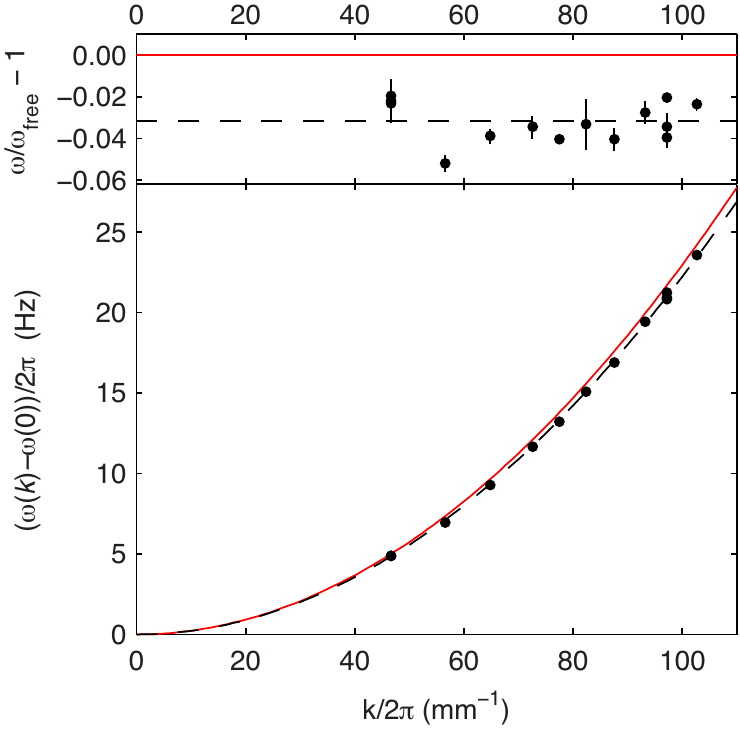}
\caption[Dispersion relation of a magnon]{Our final data for the dispersion relation of a magnon. As expected, the dispersion relation is very close to a quadratic result $\omega \propto |\mathbf k|^2$ (black dashed line). The surprising result is that the frequencies are systematically shifted down, corresponding to a magnon mass of $3.3\%$ heavier than the bare rubidium mass (solid red line) \cite{mart14magnon}.\label{fig:dispersion_relation}}
\end{figure}

For the structure described in Eq.~\ref{eqn:magnon_plane_waves}, we assume that each momentum component evolves in time with a phase evolution $e^{-i E(\mathbf k) t/\hbar}$:
\begin{equation}
\psi_{m_F=0}(t) = -\frac{1}{2\sqrt 2} \theta_\mathrm{avg}\sqrt{n}e^{-i \mu t/\hbar} \left[ e^{-i E(\mathbf k) t/\hbar} e^{-i \mathbf k \cdot \mathbf r} + e^{-i E(-\mathbf k) t/\hbar} e^{i \mathbf k \cdot \mathbf r} + 2 e^{-i E(\mathbf 0) t/\hbar} \right]
\end{equation}
The interference between different momentum states is directly visible in the density of the $m_F=0$ component (Sec.~\ref{sec:assisi}):
\begin{align}\label{eqn:oscillating_density}
|\psi_{m_F=0}(t)|^2 = n(\mathbf r)\,\theta_\mathrm{avg}^2 \Bigg[& \frac{3}{4} + \cos \mathbf k \cdot \mathbf r \, \cos \left( \frac{E(\mathbf k) + E(-\mathbf k) - 2 E(\mathbf 0)}{2\hbar}t \right) \nonumber\\
&+ \frac{1}{4} \cos \left( 2 \mathbf k \cdot \mathbf r + \frac{E(\mathbf k) - E(-\mathbf k)}{\hbar} t \right) \Bigg]
\end{align}
The signal (contrast in the standing wave in density) oscillates at a frequency of $(E(\mathbf k)+E(-\mathbf k)-2 E(\mathbf 0))/2\hbar = \hbar |\mathbf k|^2/m^\ast$.  With good knowledge of $\mathbf k$, we can determine the effective mass $m^\ast$ (see Fig.~\ref{fig:dispersion_relation}).

One feature of contrast interferometry is that it is insensitive to velocity, acceleration, and constant forces.   A constant force, over time, would add a constant $\hbar \delta \mathbf k$ to the momenta of each of the magnon waves.  The same alteration results if the condensate in which we imprint magnons is moving initially with uniform wavevector $\delta \mathbf{k}$.  In either case, this boost does not change the \emph{contrast} of the magnon interference pattern, since
\[ E(\mathbf k)+E(-\mathbf k)-2 E(\mathbf 0) = E(\mathbf k + \delta \mathbf k)+E(-\mathbf k + \delta \mathbf k)-2 E(\delta \mathbf k)\,.
\]

\section{Conclusion}

%- history
%- non-interacting sample, phase transition, magnetism
%- weakly interacting: competition of ferromagnetism and polar states
%- imaging: seeing order, correlations
%- spin dynamics: large-scale changes; far out-of-equilibrium, low damping, squeezing
%- magnetic excitations: small scale fluctuations, quasiparticles, emergence

As we have discussed, the mechanism that underlies magnetism in a spinor Bose gas is different from that in solid-state materials.  Magnetism in solid state magnets may come about from the short range interaction between neighboring magnetic moments that are confined to sites of a lattice. The interaction is quantified by an exchange energy $J$, and magnetic order might appear at temperatures for which $k_B T < z J$ where $z$ is a number of order unity that is a measure of the coordination within the crystal lattice.  In other situations, magnetism may emerge from the interaction between itinerant fermions.  In this instance, one itinerant particle interacts with all others, leading to magnetic order only when the interaction defeats the demagnetizing effect of Fermi statistics (as in the Stoner theory).

In contrast to the case of itinerant fermions, within a spinor Bose gas, the emergence of magnetic order is \emph{strengthened}, rather than \emph{suppressed}, by quantum statistics.  Like itinerant fermions but unlike in spin-lattice systems, the ``coordination number'' is very large, being essentially the macroscopic number of atoms $N$ in the gas.  The ``exchange energy,'' $J$, defined as the strength of interactions between just two atoms, is small, scaling inversely with the volume.  It is the quantity $N J$ which defines the spin-dependent energy per particle.  Unlike in the spin-lattice examples of magnetism, this spin-dependent energy need not compete with the thermal energy $k_B T$ if the spinor Bose gas undergoes Bose-Einstein condensation.  Rather, this spin-dependent energy competes against other influences, such as homogeneous or inhomogeneous magnetic fields and dipolar interactions, in dictating the magnetic order of a gas that already, through Bose-Einstein condensation, has a strong predilection to become ordered.

Given these profound differences in the origin of magnetism, it is remarkable how many generic features of magnetically ordered materials are exhibited in both the solid state and in spinor Bose gases.  Like in their solid-state ``counterparts,'' spinor Bose gases manifest magnetic phase transitions with spontaneous symmetry breaking, magnon excitations, persistent spin textures and other topological excitations, thermodynamic effects of demagnetization, non-trivial ordering in the case of antiferromagnetic interactions, and more.

The similarity between the systems is an illustration of emergence, in which the large scale structure of different complex systems can be quite similar even when the microscopic descriptions of those systems are dissimilar.  Given this emergent commonality of materials, we are led to the conclusion that if we want to study in detail the emergent, long-range/low-energy properties of materials, it behooves us to focus on the simplest systems in which these properties emerge.  Their simplicity gives us a chance to understand the properties of a many-body system -- and also the methods by which we probe these properties -- from first principles, and then to test our understanding with quantitative rigor.  Spinor Bose-Einstein gases serve as such quantitative test cases, ``gedanken materials,'' for understanding better the topic of magnetism.

Moreover, spinor Bose gases realize phenomena that are either inaccessible in solid-state materials, or even do not have a solid-state analogue. As we saw in Sec.\ \ref{sec:dynamics}, these gases can be prepared far out of equilibrium, even in high-lying magnetic states, with spin dynamics that  take the many-body quantum system through a large portion of its phase space. Ground states may be highly correlated and require descriptions more complicated than single-particle order parameters and standard Bose-Einstein condensation (Sec.\ \ref{sec:manybody}).  High-spin models are being realized in highly dipolar gases, in which the anisotropic interactions may dramatically change the thermodynamics and ground states.

Not only are spinor Bose gases  a ``material'' that can be used toward the ends pursued by condensed-matter physicists and materials scientists, but also they are a ``medium'' that can be utilized to further the aims of precision measurement and sensing which are common pursuits among atomic physicists.  Spinor Bose-Einstein condensates have been demonstrated to serve as magnetic field sensors, excelling particularly in the mapping of inhomogeneous magnetic fields \cite{veng07mag}.  The distinct properties of these gases that we have discussed in this document, e.g.\ their low energies, the coexistence of magnetic order and superfluidity, and the rotational symmetry of interactions, all contribute to improving the sensitivity and resolution, and to reducing systematic bias, of such a magnetometer.  Spatially resolved magnetometry using scalar Bose-Einstein condensates with a non-zero magnetic moment has also been demonstrated \cite{wild05magnature,aign08} and is being developed further \cite{naid13}.  In comparison with these, the sensitivity of the spinor-gas magnetometer is much better because of the ability to detect very small Larmor frequency variations over the very long spin coherence time (demonstrated in our laboratory to be on the order of seconds).  We have applied the spinor-gas magnetometry concept in using a rubidium spinor gas to measure the $360\;\mathrm{pT}$ magnetic field generated by the gas itself \cite{mart14magnon}.  It will be interesting to find practical ways of applying spinor-gas magnetometry to the measurement of fields from other sources as well; some ideas about this possibility are presented in Ref.\ \cite{stam15seeing}.

Coherent magnon optics offer another potential resource for precise sensing using spinor Bose gases.  Magnons propagate within a ferromagnetic spinor condensate in a manner that is nearly equivalent to the motion of a free particles in a potential-free volume \cite{mart14magnon}.  Magnon thermalization and decoherence times can be extremely long, particularly at magnon velocities well below the speed of sound \cite{chik00}, allowing for coherent optics and interferometry of magnetic excitations for extended periods of time. This approach has been used to measure the dispersion relation of magnetic excitations. Future experiments may use magnon interferometry to measure nonlinearities of quasiparticles, the role of dipole-dipole coupling, or transport phenomena of spin structures. As we improve our understanding of spinor condensates, it may be possible to use magnon interferometry to measure parameters of the outside world, such as inertial forces or short-range gravity.

Much of the discussion in this manuscript, like much of the experimental work in the field of spinor Bose gases, has focused on alkali gases.  In the coming years, we expect to see greater focus on high-spin spinor gases.  Experiments on chromium gases have already shown signatures of spinor-gas thermodynamics and its application to cooling, and the beginnings of research on the magnetic order favored by spin-dependent interactions \cite{pasq11crspinor,pasq12thermo,nayl15demag}.  That work highlights the need for exceptional magnetic field control in order to study gaseous spin mixtures in the presence of rapid dipolar relaxation \cite{pasq10dipolar,burd15fermionic}.  In the case of the lanthanide elements, the application of spin-dependent direct imaging methods, following the examples discussed in Sec.\ \ref{sec:imaging}, and of optical imprinting methods to create coherent magnetic excitations, as described in Sec.\ \ref{sec:excitations}, is likely to be very effective owing to the non-zero electronic orbital angular momentum of the ground state.  The rich array of magnetic orders that may be realized in these cases, along with the strong role of dipolar interactions, present compelling targets for such methods.

\acknowledgments

D.M.S.-K.\ thanks the organizers of the Enrico Fermi summer school on \emph{Quantum Matter at Ultralow Temperatures}, Massimo Inguscio, Wolfgang Ketterle, Sandro Stringari, and also Giacomo Roati, for allowing him the honor of attending the school and presenting the lectures recapitulated in this manuscript, and also thanks the staff of the Societ\`{a} Italiana di Fisica for administering the school so effectively and for their patience with this manuscript.  The authors are indebted to our coworkers at Berkeley with whom the ideas described in this manuscript were derived and explored.  In particular, we thank the ``E4 crew,'' including Eric Copenhaver, Fang Fang, Holger Kadau, Sean Lourette, Andrew MacRae, Thomas Mittiga, Shun Wu, and particularly Ryan Olf, with whom we played around with ideas about spin-dependent imaging, skyrmion spin textures, optical imprinting of magnons, magnon interferometry, and magnon thermodynamics.  Their collective effort yielded many of the methods and results described in Secs.\ \ref{sec:imaging} and \ref{sec:excitations}.  We thank also the ``E1/E5 crew,'' including Ananth Chikkatur, Jennie Guzman, James Higbie, Shin Inouye, Gyu-Boong Jo, Sabrina Leslie, Kater Murch, Lorraine Sadler, Veronique Savalli, Jay Sau, Claire Thomas, Friedhelm Serwane, Mukund Vengalattore, and Andre Wenz,  for their inspiring work on dispersive birefringent imaging, quantum quenches and spin dynamics, magnetic ordering and coarsening, and spinor-gas magnetometry, which is highlighted throughout this document.

We are grateful for recent financial support for our research on magnetism in quantum gases and on coherent magnon optics and interferometry from AFOSR through the MURI program, NASA, NSF, and DTRA.  G.E.M.\ thanks the Hertz Foundation for support during his time at Berkeley, and the NRC for support at his present institution.

\bibliographystyle{varenna}
%\bibliography{allrefs_x2,main}
\bibliography{allrefs_x2}
%\begin{thebibliography}{0}
%\bibitem{ref:apo} \BY{Einstein A. \atque Fermi E.}
%  \IN{Phys. Rev. A}{13}{1999}{12};
%  \SAME{69}{999}{1666}.
%\bibitem{ref:pul} \BY{Newton I.}
%  preprint INFN 8181.
%\bibitem{ref:bra} \BY{Bragg~B.}
%  \TITLE{Complete Works}, in \TITLE{Workers Playtime}, edited by \NAME{Tizio A. \atque Caio B.} (Unexeditor, Bologna) 1997, pp.~1-10.
%
%
%
%\end{thebibliography}

\end{document}